%% file: Main.tex
\documentclass[journal]{IEEEtran}

\usepackage{xcolor,soul,framed} 

\colorlet{shadecolor}{yellow}
\usepackage[pdftex]{graphicx}
\graphicspath{{figures/}}
\DeclareGraphicsExtensions{.pdf,.jpeg,.png}

\usepackage[cmex10]{amsmath}
\usepackage{mdwmath}
\usepackage{mdwtab}
\usepackage{eqparbox}
\usepackage{url}

\usepackage{algorithm}
\usepackage{algpseudocode}
\usepackage{amssymb}
\usepackage{multirow}
\usepackage{svg}
\usepackage{pgfplots}
\usepackage{physics}
\usepackage{amsmath}
\usepackage{tikz}
\usepackage{mathdots}
\usepackage{yhmath}
\usepackage{cancel}
\usepackage{color}
\usepackage{siunitx}
\usepackage{array}
\usepackage{multirow}
\usepackage{amssymb}
\usepackage{gensymb}
\usepackage{tabularx}
\usepackage{extarrows}
\usepackage{adjustbox}
\usepackage{booktabs}
\usetikzlibrary{fadings}
\usetikzlibrary{patterns}
\usetikzlibrary{shadows.blur}
\usetikzlibrary{shapes}

\usepackage{tikz}
\usepackage{amsmath}

\newcommand{\fancy}[1]{\mathbb{#1}}
\newcommand{\prob}[2]{\fancy{P}_{#1}\left[#2\right]}

\DeclareMathOperator*{\argmax}{arg\,max}
\DeclareMathOperator*{\argmin}{arg\,min}

\renewcommand{\baselinestretch}{1}\normalsize
\renewcommand{\paragraph}[1]{\textbf{#1}:$\:$}

\begin{document}
    \title{AudioFool: Fast, Universal and synchronization-free Cross-Domain Attack on Speech Recognition}
  \author{Mohamad Fakih, Rouwaida Kanj, Fadi Kurdahi
      and~Mohammed~E.~Fouda

  \thanks{Manuscript received xxx xx, xxx.}
  \thanks{M. Fakih and R. Kanj are  withElectrical and Computer Engineering Dept., American University of Beirut, Lebanon, 1107 202.}
  \thanks{F. Kurdahi is with Center for Embedded \& Cyber-physical Systems, University of California-Irvine, Irvine, CA, USA 92697-2625.}%
  \thanks{M. Fouda was with Center for Embedded \& Cyber-physical Systems, University of California-Irvine, Irvine, CA, USA 92697-2625 and is currently with Rain Neuromorphics, Inc (e-mail: foudam@uci.edu).}
  }


\maketitle

\begin{abstract}
Automatic Speech Recognition systems have been shown to be vulnerable to adversarial attacks that manipulate the command executed on the device. Recent research has focused on exploring methods to create such attacks, however, some issues relating to Over-The-Air (OTA) attacks have not been properly addressed. In our work, we examine the needed properties of robust attacks compatible with the OTA model, and we design a method of generating attacks with arbitrary such desired properties, namely the invariance to synchronization, and the robustness to filtering: this allows a Denial-of-Service (DoS) attack against ASR systems. We achieve these characteristics by constructing attacks in a modified frequency domain through an inverse Fourier transform. We evaluate our method on standard keyword classification tasks and analyze it in OTA, and we analyze the properties of the cross-domain attacks to explain the efficiency of the approach.
\end{abstract}


%
\IEEEpeerreviewmaketitle

\section{Introduction}
\label{sec:intro}
\IEEEPARstart{A}{utomatic} Speech Recognition (ASR) Systems are widely used in virtual assistant applications, such as Android’s Google Assistant\cite{gass}, Apple’s Siri\cite{siri} and Microsoft’s Cortana\cite{cortana}. The ASR system is responsible for identifying the user’s speech and converting it into text. ASR systems are also used in critical operations such as air traffic control, surveillance, monitoring, and many other applications where voice-to-text could be beneficial. The widespread use of such systems motivated researchers to improve the accuracy of speech recognition, as well as investigate the robustness of the systems to different environments and noise conditions. This also introduces a vulnerability allowing some attackers to inject some imperceptible command to gain unauthorized access to sensitive information, hijack accounts or even infiltrate critical systems.
    
Deep Neural Networks (DNNs) have recently made significant improvements in speech recognition \cite{audionet,deepspeech}. The success of the DNNs in speech recognition depends on the pre-processing of the raw audio signal, which is crucial for the system to learn the features of the audio signal. Most DNNs use Mel Frequency Cepstral Coefficients (MFCCs) as the audio features \cite{deepspeech}. MFCCs are obtained by performing an FFT on the raw audio signal, and then computing relevant coefficients of the power spectrum. However, other methods \cite{audionet} opt to operate directly on the raw audio signal, which also produces good results.
    
DNNs have been repeatedly shown to be vulnerable to adversarial perturbations that can compromise the performance of neural networks \cite{ian}. Such adversarial attacks are imperceptible noise examples that cause networks to misclassify or fail to classify their inputs and are therefore a source of concern for machine learning systems operating in the real world. Adversarial attacks are commonly considered in the context of white-box attacks where the attacker has access to the model's architecture and parameters. The earliest attack methods considered the case of untargeted attacks where the perturbation would change the classification of the model to any other class than the original prediction. However, recent works focus on targeted attacks where the attack is designed for making the classifier output a certain class.

    Another class of studied attacks is Universal Perturbations \cite{uap} which can be applied to any data sample and would fool the classifier. Such attacks are expensive to construct but are highly sought after given the simplicity of deploying them.
    
\paragraph{Motivation}
    To properly deploy attacks against ASR systems in the wild, the method needs to have certain properties that relate to the fact that the attacker cannot predict the signal, the channel, or the receiver's relative position. Such properties are summarized as:
    \begin{itemize}
    \item The user's speech can not be anticipated which necessitates making the perturbation universal.
    \item Furthermore, the attack should work regardless of synchronization: in simulated experiments, the perturbation and the signal are known ahead of time and are made to originate at the same time sample. In practice, this is not feasible since the attacker does not know when the user will speak.
    \item Lastly, since the attacker does not control the attack's perceived power at the attacked device, the attack vector should work for a range of relative power.
    \end{itemize}
    
    These issues are further aggravated by the delay introduced during the physical propagation inside the channel (eg: a distance of 10 meters is enough to introduce 233 timesteps of latency for an 8kHz signal). Previous works either infinitely repeat the same perturbation disregarding the synchronization issue \cite{realtime}, or design short pulses that are synchronization-free due to the distribution they optimize over \cite{advpulse}. We solve the synchronization problem by designing signals that directly verify invariance to shift through generating attacks by finding a power spectral density that is constant through time, this has the added benefit of converging much faster since iterating in the search space is more efficient than the data space.

\paragraph{Contributions}
We summarize our novelties as follows:
\begin{itemize}
    \item We propose solving the synchronization problem by designing attacks that inherently require no such alignment.
    \item We propose constructing attacks in a designed co-domain, different than the signal domain, which allows the construction of the attack signals with properties guaranteed by the mapping function.
    \item We extensively study the designed attacks on software emulated as well as physical OTA setups, as well as on convolutional and recurrent networks to demonstrate the cross-architecture and cross-model transferability of the attacks. To the best of our knowledge, this is the first work that showcases performance across architectures.
    \item We identify a new method to study the two search spaces allowing us to explain the efficiency of building attacks in the co-domain instead of the main domain.
\end{itemize}

\section{RELATED WORK}

\paragraph{Attacks against classifiers} The concept of adversarial attacks has been extensively researched following the work by Szegedy et Al. \cite{ian} which first described a gradient-based algorithm to construct attacks that fool a neural network classifier on a certain input, followed by the Fast-Gradient-Sign-Method by Goodfellow et al. \cite{ianog} which iteratively takes the same step size towards a vector that minimizes the true output of the classifier. Different classes of algorithms have emerged, each focusing on some particular objective that aims to be minimized. Most gradient methods aim to minimize some $l_p$ norm of the attack vector given some constraints on the values present, such as to change the output of the classifier. Namely, the work by Moosavi-Dezfooli et Al. \cite{deepfool} minimizes the $L_2$ norm of the perturbation by iteratively linearizing the local decision boundary of the classifier around a certain data input. Alternatively, the work by Carlini and Wagner proposes a more general algorithm that works with $L_0$, $L_1$, $L_2$, or $L_{\infty}$ norms by choosing an appropriate objective function that minimizes both the $l_p$ norm of the attack vector, as well as maximizes the difference between the logit values of the top fake choice and the true label of the data point chosen. Particular algorithms exist for $L_1$ based attacks, commonly referred to as 'sparse' attacks: SparseFool \cite{sparsefool} by Moosavi-Dezfooli et al. iteratively aggregates $L_2$ attacks generated using DeepFool \cite{deepfool} by identifying and accumulating the best dimension at each step to approach the closest decision boundary of the classifier around a certain data point. Other adversarial methods exist such as the Jacobian-based Saliency Map Attack (JSMA)\cite{JSMA} which identifies how changes to a particular vector dimension will affect all output logits. These different algorithms construct the attack in the same domain of the data, we aim to explore the cross-domain construction of attacks by choosing an appropriate domain basis that guarantees certain properties.

\paragraph{Attacks on Audio Classifiers}Some works study the feasibility and effectiveness of attacks on the audio sensor (microphone) \cite{dolphin, jamming}, mainly by jamming the sensor by injecting audio that lies outside the hearing domain of the human ear. These attacks, although provably efficient, require extensive preparation and uncommon equipment on the part of the attacker. Thus we focus on adversarial attacks that build audio signals that can be played back using normal speaker drivers.
CommanderSong \cite{CommanderSong} is one of the earliest works towards Adversarial Attacks for ASR systems, where Yuan et. Al propose hijacking a song by inserting a designed perturbation stemming from a given keyword which the ASR system should interpret as a command while remaining imperceptible to the average human listener. They use a simple gradient descent approach with a constraint on the norm of the imposed perturbation.
Carlini and Wagner's work \cite{carlini2} on adversarial attacks for ASR systems target full Text-To-Speech (TTS) systems by maximizing the Connectionist Temporal Classification (CTC) Loss \cite {CTC} which measures the accuracy of TTS systems on audio recordings paired with transcripts without any timing information.
Neither of these two works addresses the issue of generalizing the attack for any user input. Universal Adversarial Attacks on ASR systems leverage the work done by Moosavi-Dezfooli et al. \cite{uap} which proposes aggregating per-image attacks iteratively until a certain desired performance on the entire dataset is achieved: Projected Gradient Descent is used to guarantee the $L_2$ norm constraint by scaling the aggregated perturbation if its norm exceeds the constraint.
The work by Abdoli et al. \cite {uapa2} directly modifies the original UAP algorithm \cite{uap} by using aggregated updates through mini-batches from the set instead of stochastic sampling, and the perturbation on each iteration is computed using a simpler method similar to FGSM \cite {ianog}.
Alternatively, the work by Neekhara et al. \cite{uapa1} modifies the UAP algorithm by iteratively aggregating perturbations that maximize the error rate of a TTS system.
Xie et al. \cite{idk} propose a universal attack against speaker recognition systems by crafting short attacks that are either cropped or repeated to cover the data signal length; they also propose overcoming the Over-The-Air channel by simulating the attack signal through Room Impulse Response (RIR) \cite{RIR} and their attack construction uses a simple projected gradient descent.
The work on generative models by Xie et al. \cite{gen} leverages the Wave-U-Net generative architecture \cite{unet} to craft an adversarial vector against any input, although the attack vector itself is not universal, the model queried on construction time is pretrained and therefore lightweight enough to be considered real-time. The generative U-Net model is trained to minimize the accuracy of the audio keyword classifier on the entire dataset.
The invariance to synchronization and timing has also been investigated by constructing attacks robust to such cyclical shifting. Namely, the work by Li et al. \cite {advpulse} uses short sub-second perturbations that are not repeated over the speech signal length, instead they design the perturbation by minimizing the expected accuracy over the distribution of possible start offsets by using an aggregating algorithm similar to UAP \cite {uap}.
Finally, the work by Gong et al. \cite{audidos} uses a modified Fast Feature Fool algorithm \cite{FFF} that constructs an attack through maximally exciting particular neurons in the neural network; the modifications proposed include the expectation over transformation trick over the distribution of cyclical shifts.

\section{BACKGROUND}
In this section, we overview the methods and formulations used as building blocks in our method, as well as some motivation for our design.
\label{sec:background}
\subsection{Adversarial attacks formulation}
    Adversarial attacks were first introduced in \cite{ian} as a way to change the output of a classifier on a certain image. Consider a classifier $\mathbb{C}: \mathcal{X} \rightarrow \left\{1, ..., k\right\}$ applied on some data in the domain $\mathcal{X} = \mathbb{R}^m$ and producing one of $k$ classes. Internally, the classifier first produces how likely each of the $k$ classes is, then yields the class $\hat{k}$ corresponding to the maximal value. Consider $f:\mathcal{X}\rightarrow\mathbb{R}^k$ to be the mentioned distribution function then we can express $\mathbb{C}$ as follows:
    
\begin{equation}
        \begin{aligned}
            \mathbb{C}\left(x\right) &= \argmax_k f_k\left(x\right)
        \end{aligned}
        \label{eq:class}
    \end{equation}

Adversarial attacks are small perturbations $r \in \mathcal{X}$ such that $\hat{x} = x + r$ for some data $x$ will have a different predicted class that $x$. The perturbations $r$ are designed to be "small" for some p-norm. Formally, this is equivalent to finding $r$ using the formulation:
\begin{equation}
        \begin{aligned}
            \min_r \quad & ||r||_{p}\\
            \textrm{s.t.}\quad &  \mathbb{C}(\hat{x}) \neq \mathbb{C}(x)
        \end{aligned}
        \label{eq:adv}
    \end{equation}
This formulation finds the smallest perturbation that changes the output of the classifier. Other formulations exist to have the best perturbation given some norm budget $\epsilon$. Given the initial output of the classifier $k_0 = \mathbb{C}(x)$, the 'best' perturbation is the one that moves the classifier as far away as possible from the original output $k_0$:
\begin{equation}
    \begin{aligned}
        \min_r \quad & f_{k_0}(\hat{x})\\
        \textrm{s.t.}\quad &  ||r||_{p} \leq \epsilon
    \end{aligned}
    \label{eq:adv2}
\end{equation}
Many methods \cite{ian, pgd, deepfool} exist to solve such problems for the $L_2$ norm, as well as the $L_1$ norm \cite{onepixel, sparsefool}.

\subsection{Universal Attack algorithms}
Typical formulations of Universal Adversarial Perturbations \cite{uap} aim to generate a single attack vector $V$ that is able to fool the classifier $ \mathbb{C} $ on "most" of the inputs $x$ that are sampled from the data $D$. Such formulations solve a minimization problem of the form:
    
\begin{equation}
\centering
\begin{aligned}
\min_V \quad & ||V||_{p}\\
\textrm{s.t.}\quad & \prob{x \sim D}{\mathbb{C}(x + V) = \mathbb{C}(x)}\leq 1 - \epsilon
\end{aligned}
\label{eq:UAP}
\end{equation}
where $\prob{x \sim D}{.}$ is the probability when sampling $x$ from the dataset $D$, and the parameter $\epsilon$ controls the target fool rate of the perturbation $V$. The fool rate can be formally defined in this case as $FR(V) = \prob{x \sim D}{\mathbb{C}(x + V) \neq \mathbb{C}(x)}$. In practice, this is found by tracking the ratio of test samples that change classification after the perturbation.

\subsection{Audio Signals}
Since our method is focused on the audio domain, there are a few essential signal concepts we use throughout our work. In all that follows, an audio signal has some duration $T$ (in seconds) and a sampling rate $s$, yielding data $x \in \mathbb{R}^{T * s}$ that has length $N = T * s$. We use two types of transformations to operate one of the audio signals. The first manipulation is the shift operation that rotates the signal in a cycle. For $x' = shift\left(x, \tau\right)$ The operation is defined as:
    \begin{equation}
        \begin{aligned}
            x'_i &= x_{(i+\tau) mod\:N}
        \end{aligned}
        \label{eq:shift}
    \end{equation}
where $\tau$ is a parameter controlling the shift introduced to the signal in number of samples.
    The second operation we use is the single-sided Fourier transform and its inverse. The FFT (Fast Fourier Transform) is very commonly used in signal processing applications. The implementation of the FFT involves the use of sine and cosine functions to extract spectral components across all the signals. The operation normally yields complex-valued numbers however for practical applications, only the magnitude of these numbers is considered.

\section{AUDIOFOOL SETUP}
\subsection{Setup}
    Our designed method aims to circumvent the need for cost terms related to desired properties in the minimization problem: Consider inputs to the classifiers $x \in \mathcal{X}$, instead of looking for perturbations $V$ in the entire input space $\mathcal{X}$ and enforcing a cost $\mathcal{L}\left(V\right) $ for elements that do not verify the desired property, we propose looking in a space $z \in \mathcal{Z}$ that we feed to a mapping function $g$ that guarantees that $g\left(z\right)$ verifies the property needed.
    
    \begin{equation}
        \begin{aligned}
            g &: \mathcal{Z}  \rightarrow \mathcal{X} \\
            \forall z \in \mathcal{Z} &: \mathcal{L}\left(g(z)\right)  = 0
        \end{aligned}
    \end{equation}
    For example, if we desire that the attack to be 
    Additionally, we can design such functions $f$ that introduce invariance to certain transformations. Consider some transformation $T\left(x\right)$, we attempt to design a mapping $f_T\left(z\right)$ such that it its images are (semi-)invariant to that transform.
    
    \begin{equation}
        \begin{aligned}
            T\left(x\right)&: \mathcal{X} \rightarrow \mathcal{X}\\
            \forall z \in \mathcal{Z} &:T\left(g_T\left(z\right)\right) \approx g_T\left(z\right)
        \end{aligned}
    \end{equation}
    
    We assume that any such designed function is differentiable in the rest of our work. Our approach differs from the traditional method of minimizing the expectation over transformation \cite{advpulse} by guaranteeing the needed (semi-)invariance to some transformation instead of sampling from a transformation distribution.

\subsection{Spectral-Domain Attacks}

    Using our setup, let the function $g_F(z)$ be the inverse-FFT operation on zero-phased spectrums followed by an (optional) padding operation: $g_F(z)$ takes the domain $\mathcal{Z}$ as the set of of real valued signals having length $N_T$, and phase $= 0$. E.g for signals of length 8000, $\mathcal{Z} \equiv \left\{c \in \mathbb{C}^{4001} s.t. \angle c = 0\right\}\equiv \mathbb{R}^{4001}$. The padding operation repeats the resulting real-time signal so as to become in the same domain $\mathcal{X}$ as our waveform data. Smaller values of $N_T$ make the attack more robust to rotation since it sets the upper bound to the rotation cycle (i.e: $shift(V, N_T) = V$). However, bigger $N_T$ values allow for more degrees of freedom in constructing the attack. $N_T$ is chosen such as to make $g_F(z)$ semi-invariant to rotation/shifting within 30 milliseconds which is the propagation delay at 10 meters of distance.
    
    The designed function $g_F(z)$ is fully differentiable, and it verifies the invariance to the rotation transformation. We use it to first construct stationary attacks on Classification DNNs.
    
\begin{figure}[t!]
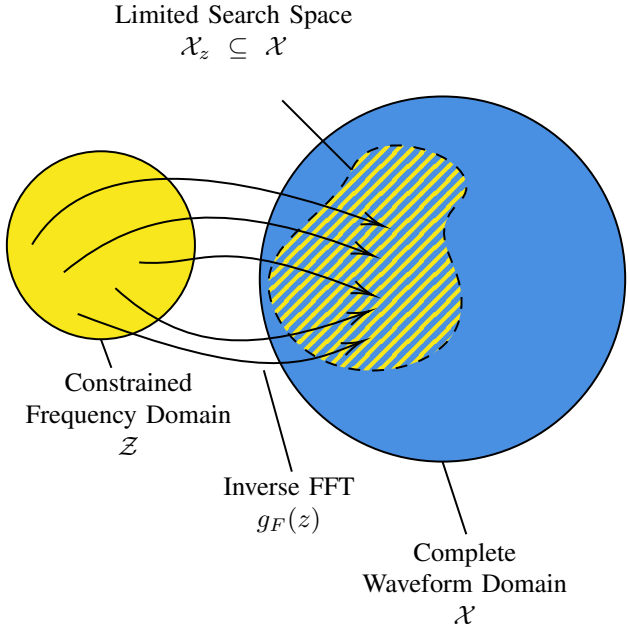

    \centering
    \include{figures/link}
    \vspace{-0.4in}
    \caption{Domain relationship overview}
    \label{fig:domains}
    \vspace{-0.1in}
\end{figure}

\subsection{Formulation}
To achieve the needed properties of shift-invariance and robustness to perfect filtering, previous methods \cite{advpulse} minimize the expectation of having a correct classification over many distributions of transformations and the distribution of the data. Namely, the transformations here are the shift and scaling of the attack. Let $s$ be the shift parameter and $\alpha$ the attenuation parameter, and $x$ the signal to be perturbed. Typically chosen distributions uniformly cover some arbitrary range (for attenuation) or the entire span of the parameter (for shift), therefore sampling $s$ from $\mathcal{U}( 0;1)$ and $\alpha$ from $\mathcal{U}( 0;4)$, in addition to sampling the data $x$ from the dataset $D$. The attack $V$ is taken to cover all the signal space $\mathcal{X}$, which in the case of $N$ second recordings with a sample rate of $T_s$ is $\mathcal{X} \equiv \mathbb{R}^{N * T_s}$. The formulation minimizes the accuracy of the classifier over these distributions as follows:

\begin{equation}
    \argmin_{V \in \mathcal{X}} \mathbb{E}_{x\sim D, \alpha \sim \mathcal{U}( 0;4) , s\sim \mathcal{U}( 0;1)}[\mathcal{C} (x)  = \mathcal{C}( x + \alpha  * shift( V, s )  )]
\end{equation}

However, as we previously discussed this is neither sufficient nor efficient for constructing attacks that verify the properties needed. We introduce the transformation-invariant domains to the formulation. We use the mentioned $g_F(z)$ mapping with $z$ being from $\mathcal{Z}$, the set of Fourier transforms of signals semi-invariant to shifting. Furthermore, since the attacker does not control the distance to the victim, we constrain the construction with a limit on the attack signal power up to a parameter that the attacker controls; this parameter would dictate the perceptibility as well as the range of effectiveness of the attack. We study the effect of choosing this parameter in our experiments. Our proposed optimization problem is therefore the following:

\begin{equation}
    \begin{aligned}
    \argmin_{z \in \mathcal{Z}} &\; \mathbb{E}_{x\sim D}[\mathcal{C} (x)  = \mathcal{C}( x + g_F(z)  )] \\
    s.t. &\; ||g_F(z)||_2^2 \leq \epsilon
    \end{aligned}
\end{equation}

\subsection{Universal Attacks}
    To construct the universal perturbations, we opt to use a modified formulation of the Universal Adversarial Perturbation (UAP) Algorithm \cite{uap}. The original UAP algorithm aggregates individual attacks on samples from the dataset and enforces some norm constraints by projecting into an $L_p$ ball of known radius. The first modification is that instead of updating the perturbation in a fashion close to Stochastic Gradient Descent where the candidate perturbation is updated for every sample in the dataset until convergence, we use mini-batches with momentum-based updates to have a stronger direction which negates the effect of unhelpful directions. Additionally, we only compute perturbations for samples in the batch that were not fooled by the current candidate, and we only aggregate the successful attacks (since most attack methods do not guarantee convergence). The second modification relates to the target norm of the perturbation: our method computes the average power of a signal from the dataset and designs a perturbation such as not exceeding some given SNR limit. Our algorithm uses the DeepFool algorithm by Moosavi-Dezfooli et. al \cite{deepfool} as the core L2 attack for each iteration of the AudioFool algorithm. DeepFool finds an adversarial attack by minimizing the L2 norm of the perturbation such that the perturbation fools the classifier. Note that the formulation and implementation of the AudioFool algorithm allow for plug-and-play configurations with other Adversarial Perturbation algorithms such as the Carlini and Wagner \cite{carlini} attack or the FGSM attack \cite{ian}. Our choice of the DeepFool algorithm centers around the fact that this attack method provided the most efficient as well as the perturbations with the smallest L2 norm. The resulting AudioFool algorithm is shown in Algorithm \ref{alg:AudioFool}, where we show the iterative aggregation of attacks on previously unfooled examples from the dataset using a momentum update and a projection back into the space that verfies the norm constraint.

\begin{figure}[t!]
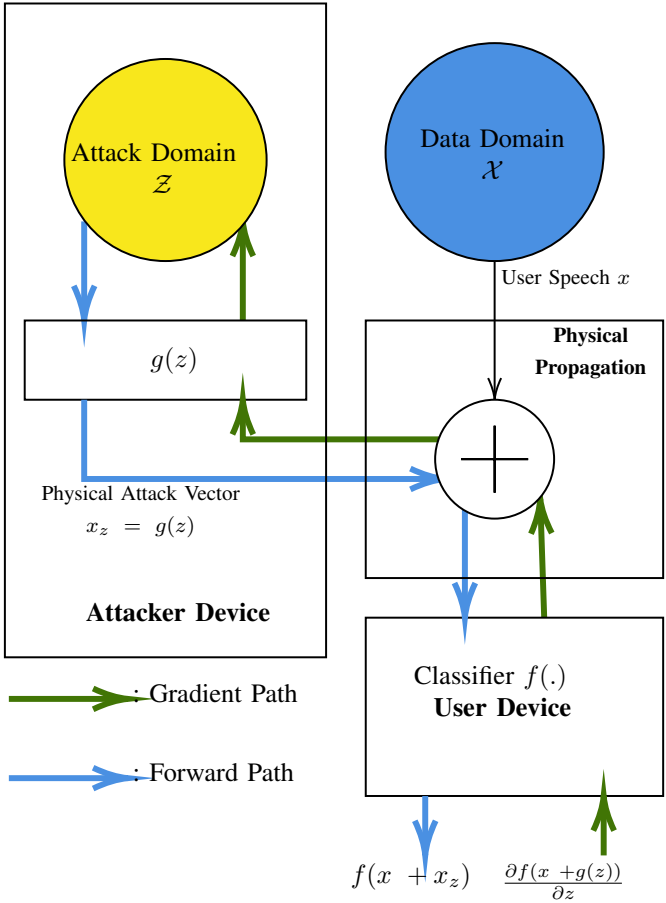

\centering
\vspace{-0.1in}    \hspace{-5cm}\include{figures/diagram2}
\vspace{-0.4in}
\caption{Attack framework overview}
\label{fig:framework}
\vspace{-0.4in}
\end{figure}

\begin{algorithm}[t!]
\caption{AudioFool}\label{alg:AudioFool}
\begin{algorithmic}
\Require Desired Fooling Rate $\epsilon$, Classifier $f$, max iter $I_M$, dataset $S$, $SNR$, Learning Rate $LR$, Decay Rate $\alpha$, Domain function $g$
\Ensure Universal Perturbation $U$
\State $i \gets 0$
\State $fooling\_rate \gets 0$
\State $U \gets 0$
\State $\Delta U \gets 0$
\State $l2target \gets E_{v \sim S}[|v|_2] * 10^{-\frac{SNR}{10}}$ \Comment{Average L2 norm of data in S}
\While{$fooling\_rate < \epsilon$ and $i < I_M$}
    \State {Get Batch $B$ from $S$ s.t $\forall$ $X \in B, f(X) = f(X + g(U))$}
    \State $R \gets 0$
    \State $c \gets 0$
    \For {$W \in B$}
        \State $Candidate \gets fool(f, W, g)$
        \If {$f(W) = f(W + g(Candidate))$}
            \State $Discard \: Candidate$
        \Else
            \State $R \gets R + Candidate$
            \State $c \gets c + 1$
        \EndIf
    \EndFor
    \State $R \gets R / c$ \Comment{Get the average computed perturbation}
    \State $\Delta U \gets \alpha * \Delta U + LR * R$
    \State $U \gets U + \Delta U$
    \State $U \gets Q_g(U, l2target)$ \Comment{Project into L2 ball with radius l2target}
    \State Update $fooling\_rate$ using the entirety of S 
    \State $i \gets i + 1$
\EndWhile
\end{algorithmic}
\end{algorithm}

\section{THREAT MODEL}

We assume that the attacker can install a speaker in the vicinity of the victim device(s) if the victim is physically stationary, or that the attacker can keep a mobile speaker within a valid range of the mobile victim device if they are moving. The attack requires no further knowledge, information, or preparation on the part of the attacker since the attack vector is already present on the speaker device. Typical information about the user's speech, the environment, and the timing is not necessary here:
\begin{itemize}
    \item \textbf{Synchronization}: The attack is invariant to user timing, as well as the propagation delay usually incurred when playing back any attack signal through the channel. The attacker does not control the user's keyword initiation and cannot predict the start of the phrase. Even when overcoming this causality issue, the attacker cannot be expected to maintain the exact distance within a range of 4 centimeters (propagation delay for 1/8000th of a second). This invariance is essential for a true DoS-style attack.
    \item \textbf{Signal Content}: The attack is also invariant to the content of the signal (i.e: which words the user will speak), which is also essential since the attacker is not able to anticipate most of the user's speech since the entropy of human speech is quite high in most cases: the attacker might anticipate some common sequence of words or phrases but in the majority of scenarios this is not possible. The designed attack is universal
\end{itemize}

This attack approach can therefore be used in simultaneous jamming of multiple user applications without being perceptible to the human ear if designed as such.

\begin{figure}[t!]
        \vspace{-0.7in}
    \includegraphics[width=1.5\columnwidth]{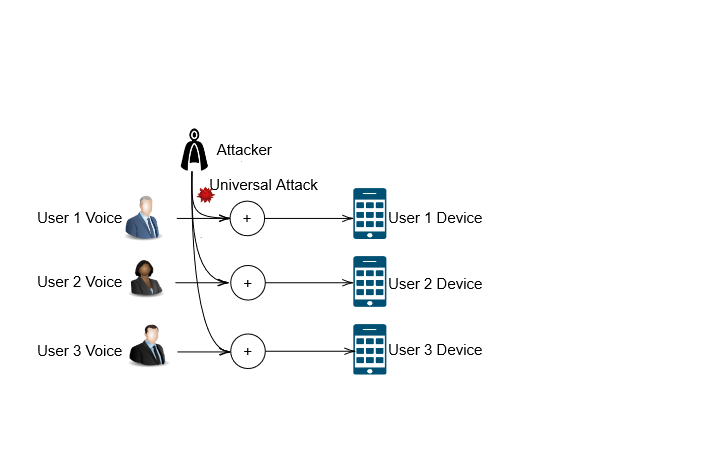}
        \vspace{-0.7in}
    \caption{The setup of the real-world attack}
    \label{fig:setup}
\end{figure}

\section{Attack Evalutation}
\subsection{Experimental Setup}
\paragraph{Task}
To measure the effectiveness of our proposed attack method, we target popular ASR applications that have become standard in testing such attacks. The SpeechCommands \cite{speechcommands} dataset consists of a limited vocabulary corpus of one-second utterances produced by various speakers. The dataset contains 35 distinct words, as well as explicit background noise that can be used to augment the data. We train a classifier to predict one of the 36 possible classes. The noise class is left as a choice to study the effect of drowning the speech signal with noise. The dataset was downsampled from 16kHz to 8kHz which is a common pre-processing step in audio applications on edge devices which are the focus of our attack. The dataset was divided into a 95/5 train/test split, with a batch size of 64 samples.

\paragraph{Architecture}
    We use two different models to study the effect of cross-architecture transferability, as well as the effect of having certain pre-processing steps and recurrent units.
    
    The first architecture is the well-studied AudioNet \cite{audionet} architecture that takes as input raw audio signals and uses a series of 1D convolutions, BatchNormalization, and max pooling, with a final dense layer to extract needed features. This architecture is the simplest yet efficient form of convolutional classifiers, as we consider our attack performance against this architecture to be representative of the performance against convolutional models that are similar in depth and size (number of channels per layer). 
    The second model mimics the advanced DeepSpeech architecture\cite{deepspeech} by having MFCC feature extraction, followed by two 2D convolution layers, and finally an LSTM layer over the timesteps. The architecture is dubbed "SpecCRNN". We use the same parameters as the original architecture: 40 mfcc bins, and 128 channels for the inner convolutions, with the 'same' padding.
    The two models were trained using AdaDelta optimizers until satisfactory and comparable performance is reached. The test accuracies for the AudioNet and SpecCRNN are  92.28\% and  92.68\%, respectively.

\paragraph{Over-The-Air}
    We measure the performance of the proposed algorithm on Over-The-Air audio by crafting a data superset that contains the software test set recorded over two different channels for two different environments: A calm residential suburb and a busy university common space. The data was collected by playing the audio at full volume using 'device A' (2013 MacBook Pro), and recorded using 'device B' (Galaxy Tab S6 Lite); Additionally, for half of the test set in each scenario, 'device C' (Wired Speakers) was playing back a previously designed attack on the software-trained models. In total, this approach yielded around 90 minutes of unperturbed audio, and 90 minutes of attacked audio. To address the drop in accuracy induced by introducing the channel, we finetune each of the models previously mentioned on the unperturbed OTA audio until the same accuracy is reached as the software case. (92.28\% for AudioNet, 92.68\% for SpecCRNN). In the following experiments we refer to the models trained on the data through the channel as "OTA", and the initial software only models as "SW" or "Software models". Attacks built in the data domain (Waveform domain) might be referred to as "wav attacks", and the Attacks in the Co-Domain (Frequency Domain) might be referred to as freq attacks".

\subsection{Results}
\begin{figure}
\centering
\includegraphics{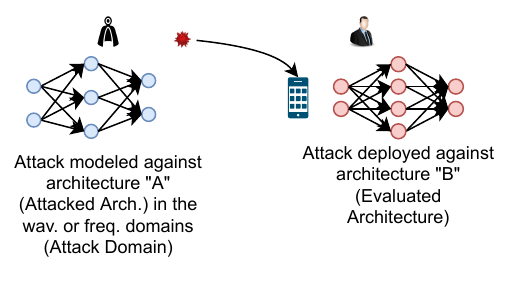}
\vspace{-0.3in}
\caption{Overview of the transferability experiment}
\label{fig:explainTransfer}
\end{figure}

\paragraph{Cross-Architecture Transferability}
We test transferring attacks between architectures by constructing different attacks for different (model, domain) pairs, yielding four attacks that we evaluate on both architectures. We show this method's overview in figure \ref{fig:explainTransfer}. We report the Fool Rates in Table \ref{fig:foursc}. The attack evaluation show that the currently used algorithm produces attacks in both the waveform and the modified frequency domains that can be transferred between two widely different architectures. This finding allows us to build later experiment on either of the two architectures, since the attacks have largely similar performance.

\begin{figure}[t!]
\centering
\vspace{-0.1in}
\begin{adjustbox}{width=\columnwidth, clip,trim=0 0 0 0.8cm}
\input{figures/acc_OTA.pgf}
\end{adjustbox}
\begin{adjustbox}{width=\columnwidth, clip,trim=0 0 0 0.8cm}
\input{figures/acc_SW.pgf}
\end{adjustbox}
\vspace{-0.15in}
\caption{SNR (dB) versus fool rate over the air (top) and on the software (bottom).}
\label{fig:SNR}
\vspace{-0.1in}
\end{figure}
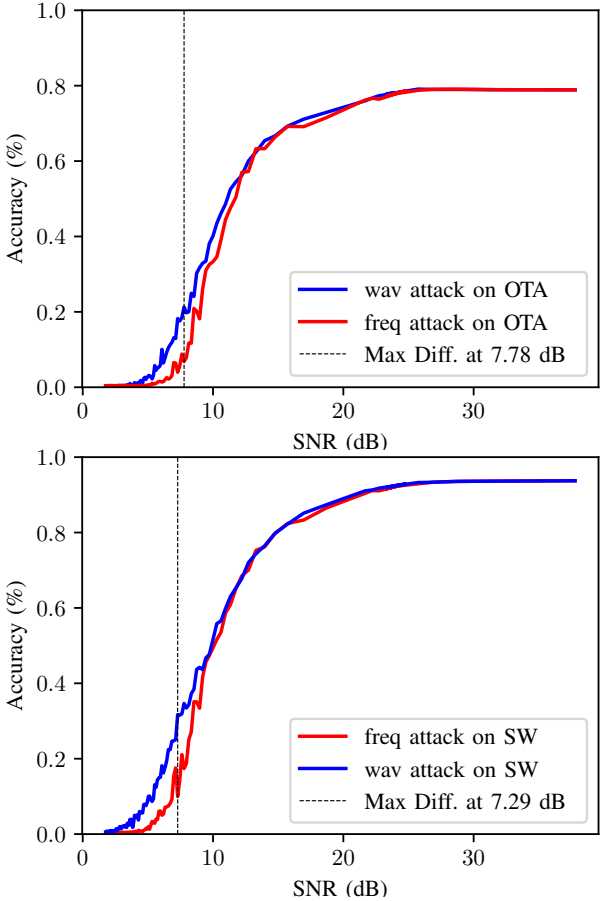

\begin{table}[t!]
\centering
\label{fig:foursc}
\caption{Attack performance in cross-domain and cross-architecture scenarios}
\begin{tabular}{cc|cccc|}
\cline{3-6}
\multicolumn{2}{c|}{}            & \multicolumn{4}{c|}{Evaluated Architecture}               \\ \cline{3-6} 
\multicolumn{1}{l}{}                                                                            & \multicolumn{1}{l|}{} & \multicolumn{2}{c|}{SpecCRNN}                          & \multicolumn{2}{c|}{AudioNet}               \\ \hline
\multicolumn{2}{|c|}{Attack Domain}                                                                                     & \multicolumn{1}{c|}{Wav.} & \multicolumn{1}{c|}{Freq.} & \multicolumn{1}{c|}{Wav.} & Freq.           \\ \hline
\multicolumn{1}{|c|}{\multirow{2}{*}{\begin{tabular}[c]{@{}c@{}}Attacked\\ Arch.\end{tabular}}} & SpecCRNN              & \multicolumn{1}{c|}{74\%} & \multicolumn{1}{c|}{53\%}  & \multicolumn{1}{c|}{78\%} & 95.4\%          \\ \cline{2-6} 
\multicolumn{1}{|c|}{}                                                                          & AudioNet              & \multicolumn{1}{c|}{60\%} & \multicolumn{1}{c|}{51\%}  & \multicolumn{1}{c|}{73\%} & \textbf{95.7\%} \\ \hline
\end{tabular}
\vspace{-0.4in}
\end{table}
\paragraph{Robustness to attenuation} We test the effect of attenuating the attack relative to the signal to simulate the attacker getting closer/further to the device of the user or the effect of targeted filtering of the perturbation. We report the accuracy for the different attacks on the AudioNet architecture under the SW scenario as well as the OTA scenario against the SNR ratio in figure \ref{fig:SNR}. Three distinct regions are identifiable: Region A (SNR $<$ 4dB), where both the attacks are too loud relative to the data, which drowns the signal in the noise and the classification model cannot perform well. Region B (4dB $<$ SNR $<$ 15dB) where there exists a substantial disparity between the attack domains reaching a 12.7\% advantage in the OTA scenario, and 20.2\% in the SW scenario. Finally, Region C (15dB $<$ SNR) where both attacks perform largely similarly.

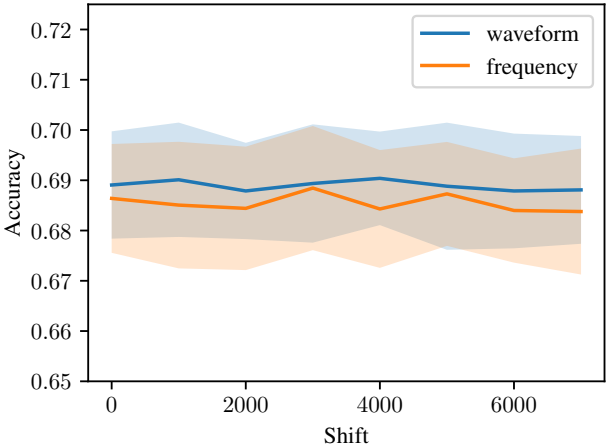
\begin{figure}[H]
\centering
\vspace{-0.1in}
\begin{adjustbox}{width=\columnwidth, clip,trim=0 0 0 0.8cm}
\input{figures/foolrate_shift.pgf}
\end{adjustbox}
\vspace{-0.1in}
\caption{Impact of the noise shift on the Foolrate.}
\label{fig:shift}
\vspace{-0.1in}
\end{figure}

\paragraph{Synchronization} We test the effect of cyclically shifting the attack relative to the beginning of the signal to simulate the user talking at random offsets relative to the attack. We report the fool rate against the number of timestep samples shifted in figure \ref{fig:shift}. We repeat the experiment on 100 different attack vectors in each domain which allows us to extract the statistical deviation which is shown alongside the Accuracy mean for each shift value. The frequency-based attacks are on average more robust to shifting than the waveform-based attacks.

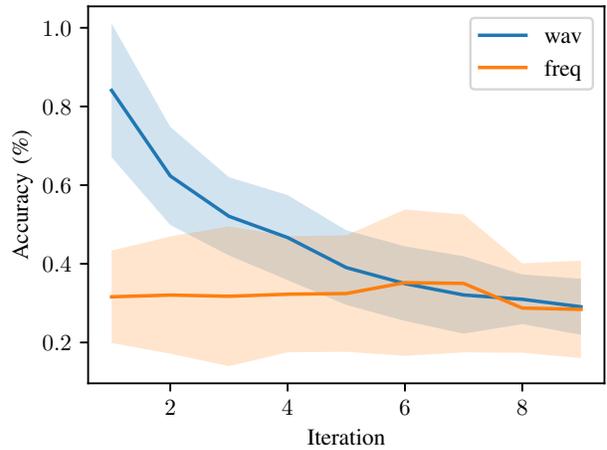
\begin{figure}
\centering
\begin{adjustbox}{width=\columnwidth,clip,trim=0 0 0 0.8cm}
\input{figures/accuracy_vs_iterations.pgf}
\end{adjustbox}
\vspace{-0.1in}
\caption{The convergence of the two attack domains over iterations for SpecCRNN.}
\label{fig:ratio}
\end{figure}

\paragraph{Efficiency} For the constructed attacks on the AudioNet model, we track the Accuracy of the model across iterations of the AudioFool algorithm, and we compare the mean performance and the deviation of the two attack domains in figure \ref{fig:ratio}. Attacks in the frequency domain converge on a good attack from the first iteration, while waveform attacks need upwards of 5 iterations on average to achieve the same performance. This disparity in efficiency is furthermore studied in the Discussion section.

\section{Discussion}

\paragraph{Designed Constraints}
The results showcased the validity of the attack approach: An attacker with access to publicly available resources can quickly mount a robust attack that can be shaped with arbitrary properties as long as the attacker can design a function that generates the needed constrained signal. This can be used to dodge defenses if the attacker has further knowledge of the system. For example, if the system under attack is known to have some band-pass filter, the attacker can then limit their search space to the relevant space without needing to penalize the attack construction if frequencies outside the space are chosen. Such capability might also help in constructing attacks in the real-time as a means to adapt to the environment between the attacker and the attacked system: by projecting some form of pre-constructed general 'meta-attack' that resides in the unconstrained search space into the newly acquired environment-specific search space, the attacker will have a better starting attack vector that they can fine-tune in a short time. For example, if the attacker's model includes reverberation modeling, the information on the current room's Room-Impulse-Response (RIR) can be approximated and used alongside the positioning of both the attacker and the victim devices to construct an attack that is either invariant to such reverberation, or that might use the reverberations to its advantage after fine-tuning. Removing the need for distributions of transformations and having to sample different transformation combinations at construction time will lead to much faster construction time and much more adaptable attacks.

\begin{figure}
\centering
\begin{adjustbox}{width=\columnwidth, clip,trim=0 0 0 0.8cm}
\input{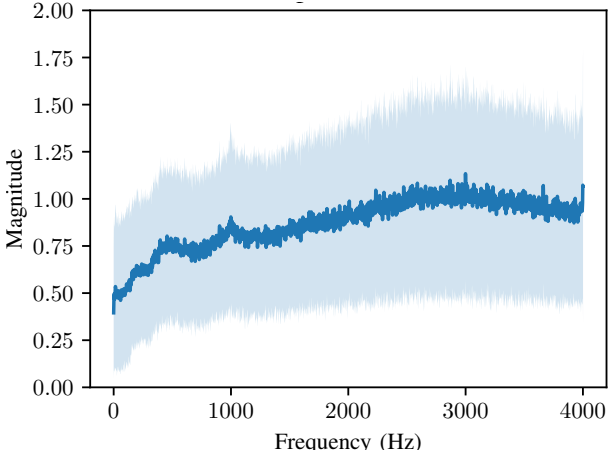}
\end{adjustbox}
\vspace{-0.22in}
\caption{Frequency composition of all generated attacks, Smoothed with a Gaussian kernel of size 5}
\label{fig:freq}
\end{figure}

\paragraph{Defenses}
The choice of the modified frequency domain for solving the synchronization problem aims to use the fact that the phase is zero for all frequencies, while still maintaining a value at all timesteps. Other choices might be designed for the fundamental basis instead of the frequency domain to achieve different characteristics. For example, one might use the latent space of some generator function that uniquely generates bird sounds to build a perturbation that mimics the sound of a bird (More accurately phrased, the algorithm would find the possible bird sound output that highly perturbs the keyword classifier). Alternatively and more relevant to our research, a function that generates non-stationary signals can be used to dodge band-stop filters that are tuned to the top frequencies detected from our attacks. Although we have already studied the effect of perfect filtering on the attacks, targeted filtering might behave differently. To better understand this possibility, we analyze the frequency content of the attacks and extract relevant statistics shown in figure \ref{fig:freq}. The composition clearly shows peaks at around 1000Hz and 500Hz, which coincides with the frequency peaks of the human voice.

\begin{figure}[t!]
\centering
\includegraphics[scale=0.3,clip,trim=2cm 2cm 4cm 4cm]{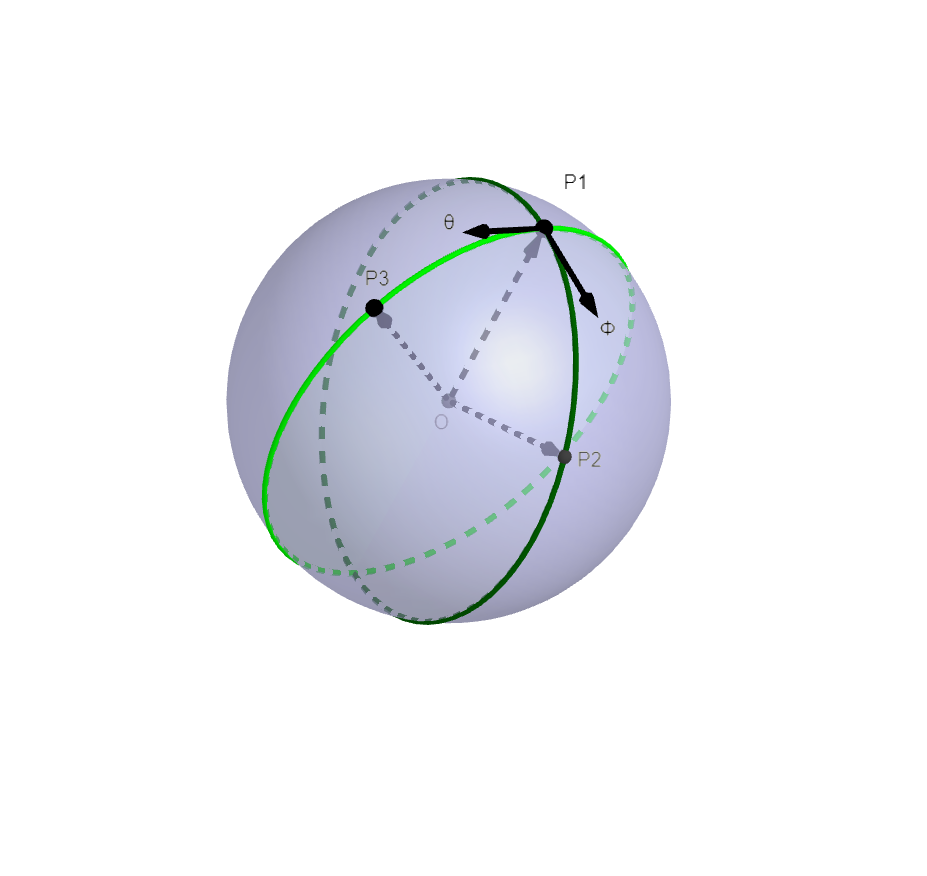}
\vspace{-2cm}
\caption{Visualization of surface studied}
\label{fig:hpplane}
\end{figure}

\paragraph{Attack search spaces}
Perhaps the most surprising result is the difference in attack construction efficiency between two domains (that are equivalent), or when the attack is more potent overall if constructed in a more constrained domain. Theoretically, the choice of space only consists of some transformation of the original bases, and the co-domain might have bases that stretch the space in the direction that provides the largest surface for the algorithm, therefore granting larger ease for iterative updates since our approach aggregates attacks that are built on gradient information. We study here the effect of having this transformation of search space by exploring the difference between the search surface of the frequency domain and that of the waveform domain for the same signal powers (i.e: ball of the same radius in both domains). We generate 3 attacks for each of the domains with a different initial point to achieve different final attacks with the same SNR level. Since all of the attack vectors generated to have the same magnitude, they form a unique sphere around the origin, which we unroll into a $\left(\Phi, \Theta\right)$ space, and study the classifier's accuracy when adding perturbations from this surface to the test set. This allows us to measure the geometry of the search spaces involved in the attack construction. A visualization of the method is shown in figure \ref{fig:hpplane}. Formally, we choose the points P1, P2, and P3 from outputs of three distinct runs of the AudioFool Algorithm, and we measure the accuracy of the model when adding $R_{\left(\Phi, \Theta\right)}$ to the data by sweeping over the pair $\left(\Phi \in \left[0 - 360\right], \Theta\in \left[0 - 360\right]\right)$, with $R_{\left(\Phi, \Theta\right)}$ being:
\begin{equation}
R_{\left(\Phi, \Theta\right)} = sin(\Phi)cos(\Theta)P3 + sin(\Phi) sin(\Theta)P2 + cos(\Phi)P1 
\end{equation}

\begin{figure}[t!]
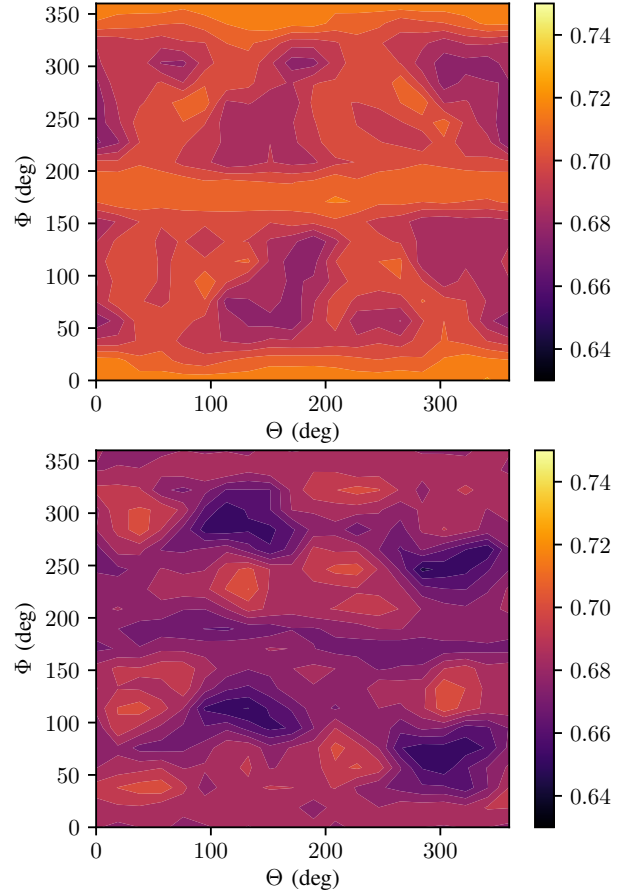

\centering
\begin{adjustbox}{width=\columnwidth, clip,trim=0 0 0 0.8cm}
\input{figures/w_accuracy_vs_theta_phi.pgf}
\end{adjustbox}
\begin{adjustbox}{width=\columnwidth, clip,trim=0 0 0 0.8cm}
\input{figures/f_accuracy_vs_theta_phi.pgf}
\end{adjustbox}
\vspace{-0.1in}
\caption{Search space vs the accuracy for waveform-doamin attack (top) and for frequency-domain attack (bottom).}
\label{fig:surface0}
\end{figure}

\begin{figure}[t!]
\centering
\begin{adjustbox}{width=\columnwidth, clip,trim=0 0 0 0.8cm}
\input{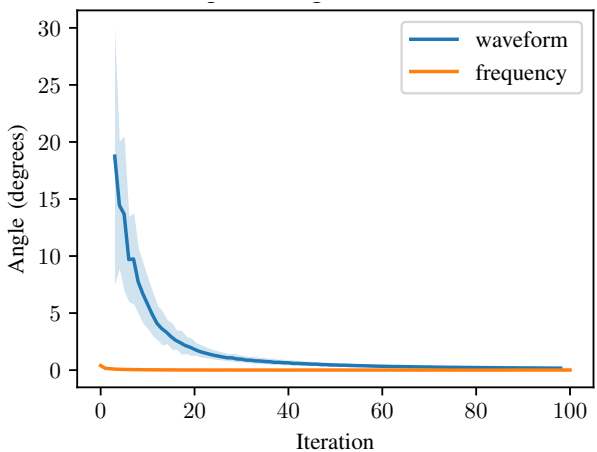}
\end{adjustbox}
\vspace{-0.15in}
\caption{Update angles for each of the domains}
\label{fig:seq}
\vspace{-0.15in}
\end{figure}

The resulting surfaces of $f(x + R_{\left(\Phi, \Theta\right)})$ are shown in figure \ref{fig:surface0}. The difference in geometry between the two surfaces highlights the different search space of the algorithm in the respective domains: The frequency domain has fewer local minima, as well as wider minima loci, from any starting point on this surface, and the path to the nearest optimum is straighter than the respective path in the waveform domain. To verify this hypothesis, we track the gradient updates through iterations and compute the relative angle between successive updates to the perturbation vector $U$. Formally, at each iteration $i$, we compute:
\begin{align*}
    \nabla U_i &= \Delta U_i - \Delta U_{i-1} \\
    \theta_i &= arccos\left(\frac{<\nabla U_i, \nabla U_{i-1}>}{||\nabla U_i||* ||\nabla U_{i-1}||}\right)
\end{align*}
Where $<., .>$ is the inner product. We plot the $\theta$ sequences for 100 runs for each of the frequency and waveform domains and we extract relevant statistics in figure \ref{fig:seq}. Updates in the waveform domain, on average, take around 20 iterations to solidify into the final direction, pointing less than 2 degrees towards the local minimum, with most of the variance happening in this phase of iteration. Updates in the frequency domain, on the other hand, start with update angles significantly less than the waveform domain: Even from the first iteration, the angle is less than 1 degree, signifying that the algorithm does indeed find the best direction towards the local minimum much faster than the waveform based algorithm. It is worth noting that the variance in this case mainly happens in the second phase of the iterations; i.e: when the algorithm starts to yield close enough points to the optimum.

\section{Conclusion}
This paper proposed a frequency-domain attack which is more potent and more efficient than waveform perturbations. The proposed method is designed to generate a synchronization-free attacks which are very desirable for speech attacks.Our formulation allows for dodging stationarity detecting circuits, as well as the design of arbitrary properties for attacks. We demonstrate the validity of the DoS attack against OTA systems, the transferability of attacks across model architectures, and the effect of perfect filtering on the attack. Furthermore, we analyse the fundamental differences between the domains proposed to explain the disparity in results observed by showing that the search space in the frequency domain is inherently more efficient than the waveform domain. In the future works, the audioFool will be tested and evaluated on other speech models and datasets.


\ifCLASSOPTIONcaptionsoff
  \newpage
\fi


\end{document}

%% file: figures/link.tex
 
\tikzset{
pattern size/.store in=\mcSize, 
pattern size = 5pt,
pattern thickness/.store in=\mcThickness, 
pattern thickness = 0.3pt,
pattern radius/.store in=\mcRadius, 
pattern radius = 1pt}
\makeatletter
\pgfutil@ifundefined{pgf@pattern@name@_2oy728jlj}{
\pgfdeclarepatternformonly[\mcThickness,\mcSize]{_2oy728jlj}
{\pgfqpoint{0pt}{0pt}}
{\pgfpoint{\mcSize+\mcThickness}{\mcSize+\mcThickness}}
{\pgfpoint{\mcSize}{\mcSize}}
{
\pgfsetcolor{\tikz@pattern@color}
\pgfsetlinewidth{\mcThickness}
\pgfpathmoveto{\pgfqpoint{0pt}{0pt}}
\pgfpathlineto{\pgfpoint{\mcSize+\mcThickness}{\mcSize+\mcThickness}}
\pgfusepath{stroke}
}}
\makeatother
\tikzset{every picture/.style={line width=0.75pt}} 

\begin{tikzpicture}[x=0.75pt,y=0.75pt,yscale=-1,xscale=1]

\draw  [fill={rgb, 255:red, 74; green, 144; blue, 226 }  ,fill opacity=1 ] (132.03,147.82) .. controls (132.03,96.92) and (173.29,55.67) .. (224.18,55.67) .. controls (275.08,55.67) and (316.33,96.92) .. (316.33,147.82) .. controls (316.33,198.71) and (275.08,239.97) .. (224.18,239.97) .. controls (173.29,239.97) and (132.03,198.71) .. (132.03,147.82) -- cycle ;
\draw  [pattern=_2oy728jlj,pattern size=5.1pt,pattern thickness=1.5pt,pattern radius=0pt, pattern color={rgb, 255:red, 248; green, 231; blue, 28}][dash pattern={on 4.5pt off 4.5pt}] (178.33,92.63) .. controls (190.33,65.33) and (252.33,89.33) .. (232.33,109.33) .. controls (212.33,129.33) and (240.33,141.63) .. (232.33,169.33) .. controls (224.33,197.03) and (164.33,207.03) .. (142.33,169.33) .. controls (120.33,131.63) and (166.33,119.93) .. (178.33,92.63) -- cycle ;
\draw  [fill={rgb, 255:red, 248; green, 231; blue, 28 }  ,fill opacity=1 ] (4.47,130.77) .. controls (4.47,104.57) and (25.7,83.33) .. (51.9,83.33) .. controls (78.1,83.33) and (99.33,104.57) .. (99.33,130.77) .. controls (99.33,156.96) and (78.1,178.2) .. (51.9,178.2) .. controls (25.7,178.2) and (4.47,156.96) .. (4.47,130.77) -- cycle ;
\draw    (40.2,165.43) .. controls (132.32,200.71) and (154.33,191.43) .. (182.47,179.75) ;
\draw [shift={(184.2,179.03)}, rotate = 157.52] [color={rgb, 255:red, 0; green, 0; blue, 0 }  ][line width=0.75]    (10.93,-3.29) .. controls (6.95,-1.4) and (3.31,-0.3) .. (0,0) .. controls (3.31,0.3) and (6.95,1.4) .. (10.93,3.29)   ;
\draw    (17.2,130.43) .. controls (46.06,82.67) and (118.47,92.93) .. (193.08,120.61) ;
\draw [shift={(194.2,121.03)}, rotate = 200.47] [color={rgb, 255:red, 0; green, 0; blue, 0 }  ][line width=0.75]    (10.93,-3.29) .. controls (6.95,-1.4) and (3.31,-0.3) .. (0,0) .. controls (3.31,0.3) and (6.95,1.4) .. (10.93,3.29)   ;
\draw    (71.2,139.43) .. controls (106.03,142.02) and (107.19,123.22) .. (187.98,155.54) ;
\draw [shift={(189.2,156.03)}, rotate = 201.92] [color={rgb, 255:red, 0; green, 0; blue, 0 }  ][line width=0.75]    (10.93,-3.29) .. controls (6.95,-1.4) and (3.31,-0.3) .. (0,0) .. controls (3.31,0.3) and (6.95,1.4) .. (10.93,3.29)   ;
\draw    (33.2,144.43) .. controls (74,110.6) and (118.75,108.06) .. (188.15,135.62) ;
\draw [shift={(189.2,136.03)}, rotate = 201.8] [color={rgb, 255:red, 0; green, 0; blue, 0 }  ][line width=0.75]    (10.93,-3.29) .. controls (6.95,-1.4) and (3.31,-0.3) .. (0,0) .. controls (3.31,0.3) and (6.95,1.4) .. (10.93,3.29)   ;
\draw    (59.2,152.43) .. controls (93.85,184.11) and (128.5,186) .. (185.47,164.69) ;
\draw [shift={(187.2,164.03)}, rotate = 159.23] [color={rgb, 255:red, 0; green, 0; blue, 0 }  ][line width=0.75]    (10.93,-3.29) .. controls (6.95,-1.4) and (3.31,-0.3) .. (0,0) .. controls (3.31,0.3) and (6.95,1.4) .. (10.93,3.29)   ;
\draw [line width=0.75]    (134.07,193.87) -- (148.2,245.03) ;
\draw    (57.2,192.63) -- (51.9,178.2) ;
\draw    (236.2,276.63) -- (224.18,239.97) ;
\draw    (143.33,57.63) -- (178.33,92.63) ;

\draw (106.8,246.67) node [anchor=north west][inner sep=0.75pt]   [align=left] {\begin{minipage}[lt]{57.72pt}\setlength\topsep{0pt}
\begin{center}
Inverse FFT\\$\displaystyle g_{F}( z)$
\end{center}

\end{minipage}};
\draw (5.07,193.8) node [anchor=north west][inner sep=0.75pt]   [align=left] {\begin{minipage}[lt]{88.94pt}\setlength\topsep{0pt}
\begin{center}
Constrained\\Frequency Domain\\$\displaystyle \mathcal{Z}$
\end{center}

\end{minipage}};
\draw (176.07,279.4) node [anchor=north west][inner sep=0.75pt]   [align=left] {\begin{minipage}[lt]{86.84pt}\setlength\topsep{0pt}
\begin{center}
Complete\\Waveform Domain\\$\displaystyle \mathcal{X}$
\end{center}

\end{minipage}};
\draw (49.07,7.93) node [anchor=north west][inner sep=0.75pt]   [align=left] {\begin{minipage}[lt]{102.56pt}\setlength\topsep{0pt}
\begin{center}
Limited Search Space\\$\displaystyle \mathcal{X}_{z} \ \subseteq \ \mathcal{X}$
\end{center}

\end{minipage}};

\end{tikzpicture}

%% file: figures/diagram2.tex
\tikzset{every picture/.style={line width=0.75pt}} 

\begin{tikzpicture}[x=0.75pt,y=0.75pt,yscale=-1,xscale=1]

\draw [color={rgb, 255:red, 65; green, 117; blue, 5 }  ,draw opacity=1 ][line width=2.25]    (280,320) -- (278.13,264) ;
\draw [shift={(278,260)}, rotate = 88.09] [color={rgb, 255:red, 65; green, 117; blue, 5 }  ,draw opacity=1 ][line width=2.25]    (17.49,-5.26) .. controls (11.12,-2.23) and (5.29,-0.48) .. (0,0) .. controls (5.29,0.48) and (11.12,2.23) .. (17.49,5.26)   ;
\draw [color={rgb, 255:red, 65; green, 117; blue, 5 }  ,draw opacity=1 ][line width=2.25]    (128,170) -- (128,124) ;
\draw [shift={(128,120)}, rotate = 90] [color={rgb, 255:red, 65; green, 117; blue, 5 }  ,draw opacity=1 ][line width=2.25]    (17.49,-5.26) .. controls (11.12,-2.23) and (5.29,-0.48) .. (0,0) .. controls (5.29,0.48) and (11.12,2.23) .. (17.49,5.26)   ;
\draw [color={rgb, 255:red, 74; green, 144; blue, 226 }  ,draw opacity=1 ][line width=2.25]    (48,120) -- (48,166) ;
\draw [shift={(48,170)}, rotate = 270] [color={rgb, 255:red, 74; green, 144; blue, 226 }  ,draw opacity=1 ][line width=2.25]    (17.49,-5.26) .. controls (11.12,-2.23) and (5.29,-0.48) .. (0,0) .. controls (5.29,0.48) and (11.12,2.23) .. (17.49,5.26)   ;
\draw  [fill={rgb, 255:red, 248; green, 231; blue, 28 }  ,fill opacity=1 ] (38,89) .. controls (38,60.83) and (60.83,38) .. (89,38) .. controls (117.17,38) and (140,60.83) .. (140,89) .. controls (140,117.17) and (117.17,140) .. (89,140) .. controls (60.83,140) and (38,117.17) .. (38,89) -- cycle ;
\draw  [color={rgb, 255:red, 0; green, 0; blue, 0 }  ,draw opacity=1 ] (18,170) -- (160,170) -- (160,210) -- (18,210) -- cycle ;
\draw  [fill={rgb, 255:red, 74; green, 144; blue, 226 }  ,fill opacity=1 ] (200,85) .. controls (200,54.62) and (224.62,30) .. (255,30) .. controls (285.38,30) and (310,54.62) .. (310,85) .. controls (310,115.38) and (285.38,140) .. (255,140) .. controls (224.62,140) and (200,115.38) .. (200,85) -- cycle ;
\draw   (190,320) -- (340,320) -- (340,410) -- (190,410) -- cycle ;
\draw [color={rgb, 255:red, 74; green, 144; blue, 226 }  ,draw opacity=1 ][line width=2.25]    (240,260) -- (240,316) ;
\draw [shift={(240,320)}, rotate = 270] [color={rgb, 255:red, 74; green, 144; blue, 226 }  ,draw opacity=1 ][line width=2.25]    (17.49,-5.26) .. controls (11.12,-2.23) and (5.29,-0.48) .. (0,0) .. controls (5.29,0.48) and (11.12,2.23) .. (17.49,5.26)   ;
\draw    (255,140) -- (255,208) ;
\draw [shift={(255,210)}, rotate = 270] [color={rgb, 255:red, 0; green, 0; blue, 0 }  ][line width=0.75]    (10.93,-3.29) .. controls (6.95,-1.4) and (3.31,-0.3) .. (0,0) .. controls (3.31,0.3) and (6.95,1.4) .. (10.93,3.29)   ;
\draw [color={rgb, 255:red, 74; green, 144; blue, 226 }  ,draw opacity=1 ][line width=2.25]    (48,210) -- (48,250) -- (226,250) ;
\draw [shift={(230,250)}, rotate = 180] [color={rgb, 255:red, 74; green, 144; blue, 226 }  ,draw opacity=1 ][line width=2.25]    (17.49,-5.26) .. controls (11.12,-2.23) and (5.29,-0.48) .. (0,0) .. controls (5.29,0.48) and (11.12,2.23) .. (17.49,5.26)   ;
\draw [color={rgb, 255:red, 74; green, 144; blue, 226 }  ,draw opacity=1 ][line width=2.25]    (220,410) -- (220,430.23) -- (220,436) ;
\draw [shift={(220,440)}, rotate = 270] [color={rgb, 255:red, 74; green, 144; blue, 226 }  ,draw opacity=1 ][line width=2.25]    (17.49,-5.26) .. controls (11.12,-2.23) and (5.29,-0.48) .. (0,0) .. controls (5.29,0.48) and (11.12,2.23) .. (17.49,5.26)   ;
\draw [color={rgb, 255:red, 65; green, 117; blue, 5 }  ,draw opacity=1 ][line width=2.25]    (310,440) -- (310,414) ;
\draw [shift={(310,410)}, rotate = 90] [color={rgb, 255:red, 65; green, 117; blue, 5 }  ,draw opacity=1 ][line width=2.25]    (17.49,-5.26) .. controls (11.12,-2.23) and (5.29,-0.48) .. (0,0) .. controls (5.29,0.48) and (11.12,2.23) .. (17.49,5.26)   ;
\draw [color={rgb, 255:red, 65; green, 117; blue, 5 }  ,draw opacity=1 ][line width=2.25]    (230,230) -- (128,230) -- (128,214) ;
\draw [shift={(128,210)}, rotate = 90] [color={rgb, 255:red, 65; green, 117; blue, 5 }  ,draw opacity=1 ][line width=2.25]    (17.49,-5.26) .. controls (11.12,-2.23) and (5.29,-0.48) .. (0,0) .. controls (5.29,0.48) and (11.12,2.23) .. (17.49,5.26)   ;
\draw   (8,10) -- (170,10) -- (170,340) -- (8,340) -- cycle ;
\draw   (190,170) -- (340,170) -- (340,300) -- (190,300) -- cycle ;
\draw [color={rgb, 255:red, 65; green, 117; blue, 5 }  ,draw opacity=1 ][line width=2.25]    (10,361) -- (66,361) ;
\draw [shift={(70,361)}, rotate = 180] [color={rgb, 255:red, 65; green, 117; blue, 5 }  ,draw opacity=1 ][line width=2.25]    (17.49,-5.26) .. controls (11.12,-2.23) and (5.29,-0.48) .. (0,0) .. controls (5.29,0.48) and (11.12,2.23) .. (17.49,5.26)   ;
\draw [color={rgb, 255:red, 74; green, 144; blue, 226 }  ,draw opacity=1 ][line width=2.25]    (10,401) -- (50,401) -- (66,401) ;
\draw [shift={(70,401)}, rotate = 180] [color={rgb, 255:red, 74; green, 144; blue, 226 }  ,draw opacity=1 ][line width=2.25]    (17.49,-5.26) .. controls (11.12,-2.23) and (5.29,-0.48) .. (0,0) .. controls (5.29,0.48) and (11.12,2.23) .. (17.49,5.26)   ;
\draw  [fill={rgb, 255:red, 255; green, 255; blue, 255 }  ,fill opacity=1 ] (225,240) .. controls (225,223.43) and (238.43,210) .. (255,210) .. controls (271.57,210) and (285,223.43) .. (285,240) .. controls (285,256.57) and (271.57,270) .. (255,270) .. controls (238.43,270) and (225,256.57) .. (225,240) -- cycle ;
\draw   (255,223.09) -- (255,223.09) -- (255,240) -- (271.91,240) -- (271.91,240) -- (255,240) -- (255,256.91) -- (255,256.91) -- (255,240) -- (238.09,240) -- (238.09,240) -- (255,240) -- cycle ;

\draw (40,79) node [anchor=north west][inner sep=0.75pt]   [align=left] {\begin{minipage}[lt]{69.07pt}\setlength\topsep{0pt}
Attack Domain
\begin{center}
 $\displaystyle \mathcal{Z}$
\end{center}

\end{minipage}};
\draw (80,182) node [anchor=north west][inner sep=0.75pt]  [color={rgb, 255:red, 0; green, 0; blue, 0 }  ,opacity=1 ] [align=left] {$\displaystyle g( z)$};
\draw (213,342) node [anchor=north west][inner sep=0.75pt]   [align=left] {\begin{minipage}[lt]{66.74pt}\setlength\topsep{0pt}
Classifier $\displaystyle f( .)$
\begin{center}
\textbf{User Device}
\end{center}

\end{minipage}};
\draw (181,442.4) node [anchor=north west][inner sep=0.75pt]    {$f( x\ +x_{z})$};
\draw (257,442.4) node [anchor=north west][inner sep=0.75pt]    {$\frac{\partial f( x\ +g( z))}{\partial z}$};
\draw (41,311) node [anchor=north west][inner sep=0.75pt]   [align=left] {\begin{minipage}[lt]{79.31pt}\setlength\topsep{0pt}
\begin{center}
\textbf{Attacker Device}
\end{center}

\end{minipage}};
\draw (257,143) node [anchor=north west][inner sep=0.75pt]   [align=left] {{\footnotesize User Speech $\displaystyle x$}};
\draw (21,252) node [anchor=north west][inner sep=0.75pt]   [align=left] {\begin{minipage}[lt]{81.06pt}\setlength\topsep{0pt}
\begin{center}
{\footnotesize Physical Attack Vector}\\{\footnotesize  $\displaystyle x_{z} \ =\ g( z)$}
\end{center}

\end{minipage}};
\draw (269,172) node [anchor=north west][inner sep=0.75pt]   [align=left] {\begin{minipage}[lt]{49.5pt}\setlength\topsep{0pt}
\begin{center}
{\footnotesize \textbf{Physical}}\\{\footnotesize \textbf{Propagation}}
\end{center}

\end{minipage}};
\draw (71,352) node [anchor=north west][inner sep=0.75pt]   [align=left] {: Gradient Path};
\draw (71,392) node [anchor=north west][inner sep=0.75pt]   [align=left] {: Forward Path};
\draw (211,72) node [anchor=north west][inner sep=0.75pt]   [align=left] {\begin{minipage}[lt]{62.28pt}\setlength\topsep{0pt}
\begin{center}
Data Domain\\$\displaystyle \mathcal{X}$
\end{center}

\end{minipage}};

\end{tikzpicture}

%% file: figures/acc_OTA.pgf
\begingroup%
\makeatletter%
\begin{pgfpicture}%
\pgfpathrectangle{\pgfpointorigin}{\pgfqpoint{4.000000in}{3.000000in}}%
\pgfusepath{use as bounding box, clip}%
\begin{pgfscope}%
\pgfsetbuttcap%
\pgfsetmiterjoin%
\pgfsetlinewidth{0.000000pt}%
\definecolor{currentstroke}{rgb}{1.000000,1.000000,1.000000}%
\pgfsetstrokecolor{currentstroke}%
\pgfsetstrokeopacity{0.000000}%
\pgfsetdash{}{0pt}%
\pgfpathmoveto{\pgfqpoint{0.000000in}{0.000000in}}%
\pgfpathlineto{\pgfqpoint{4.000000in}{0.000000in}}%
\pgfpathlineto{\pgfqpoint{4.000000in}{3.000000in}}%
\pgfpathlineto{\pgfqpoint{0.000000in}{3.000000in}}%
\pgfpathclose%
\pgfusepath{}%
\end{pgfscope}%
\begin{pgfscope}%
\pgfsetbuttcap%
\pgfsetmiterjoin%
\definecolor{currentfill}{rgb}{1.000000,1.000000,1.000000}%
\pgfsetfillcolor{currentfill}%
\pgfsetlinewidth{0.000000pt}%
\definecolor{currentstroke}{rgb}{0.000000,0.000000,0.000000}%
\pgfsetstrokecolor{currentstroke}%
\pgfsetstrokeopacity{0.000000}%
\pgfsetdash{}{0pt}%
\pgfpathmoveto{\pgfqpoint{0.500000in}{0.375000in}}%
\pgfpathlineto{\pgfqpoint{3.600000in}{0.375000in}}%
\pgfpathlineto{\pgfqpoint{3.600000in}{2.640000in}}%
\pgfpathlineto{\pgfqpoint{0.500000in}{2.640000in}}%
\pgfpathclose%
\pgfusepath{fill}%
\end{pgfscope}%
\begin{pgfscope}%
\pgfsetbuttcap%
\pgfsetroundjoin%
\definecolor{currentfill}{rgb}{0.000000,0.000000,0.000000}%
\pgfsetfillcolor{currentfill}%
\pgfsetlinewidth{0.803000pt}%
\definecolor{currentstroke}{rgb}{0.000000,0.000000,0.000000}%
\pgfsetstrokecolor{currentstroke}%
\pgfsetdash{}{0pt}%
\pgfsys@defobject{currentmarker}{\pgfqpoint{0.000000in}{-0.048611in}}{\pgfqpoint{0.000000in}{0.000000in}}{%
\pgfpathmoveto{\pgfqpoint{0.000000in}{0.000000in}}%
\pgfpathlineto{\pgfqpoint{0.000000in}{-0.048611in}}%
\pgfusepath{stroke,fill}%
}%
\begin{pgfscope}%
\pgfsys@transformshift{0.503228in}{0.375000in}%
\pgfsys@useobject{currentmarker}{}%
\end{pgfscope}%
\end{pgfscope}%
\begin{pgfscope}%
\definecolor{textcolor}{rgb}{0.000000,0.000000,0.000000}%
\pgfsetstrokecolor{textcolor}%
\pgfsetfillcolor{textcolor}%
\pgftext[x=0.503228in,y=0.277778in,,top]{\color{textcolor}\rmfamily\fontsize{10.000000}{12.000000}\selectfont \(\displaystyle {0}\)}%
\end{pgfscope}%
\begin{pgfscope}%
\pgfsetbuttcap%
\pgfsetroundjoin%
\definecolor{currentfill}{rgb}{0.000000,0.000000,0.000000}%
\pgfsetfillcolor{currentfill}%
\pgfsetlinewidth{0.803000pt}%
\definecolor{currentstroke}{rgb}{0.000000,0.000000,0.000000}%
\pgfsetstrokecolor{currentstroke}%
\pgfsetdash{}{0pt}%
\pgfsys@defobject{currentmarker}{\pgfqpoint{0.000000in}{-0.048611in}}{\pgfqpoint{0.000000in}{0.000000in}}{%
\pgfpathmoveto{\pgfqpoint{0.000000in}{0.000000in}}%
\pgfpathlineto{\pgfqpoint{0.000000in}{-0.048611in}}%
\pgfusepath{stroke,fill}%
}%
\begin{pgfscope}%
\pgfsys@transformshift{1.285585in}{0.375000in}%
\pgfsys@useobject{currentmarker}{}%
\end{pgfscope}%
\end{pgfscope}%
\begin{pgfscope}%
\definecolor{textcolor}{rgb}{0.000000,0.000000,0.000000}%
\pgfsetstrokecolor{textcolor}%
\pgfsetfillcolor{textcolor}%
\pgftext[x=1.285585in,y=0.277778in,,top]{\color{textcolor}\rmfamily\fontsize{10.000000}{12.000000}\selectfont \(\displaystyle {10}\)}%
\end{pgfscope}%
\begin{pgfscope}%
\pgfsetbuttcap%
\pgfsetroundjoin%
\definecolor{currentfill}{rgb}{0.000000,0.000000,0.000000}%
\pgfsetfillcolor{currentfill}%
\pgfsetlinewidth{0.803000pt}%
\definecolor{currentstroke}{rgb}{0.000000,0.000000,0.000000}%
\pgfsetstrokecolor{currentstroke}%
\pgfsetdash{}{0pt}%
\pgfsys@defobject{currentmarker}{\pgfqpoint{0.000000in}{-0.048611in}}{\pgfqpoint{0.000000in}{0.000000in}}{%
\pgfpathmoveto{\pgfqpoint{0.000000in}{0.000000in}}%
\pgfpathlineto{\pgfqpoint{0.000000in}{-0.048611in}}%
\pgfusepath{stroke,fill}%
}%
\begin{pgfscope}%
\pgfsys@transformshift{2.067942in}{0.375000in}%
\pgfsys@useobject{currentmarker}{}%
\end{pgfscope}%
\end{pgfscope}%
\begin{pgfscope}%
\definecolor{textcolor}{rgb}{0.000000,0.000000,0.000000}%
\pgfsetstrokecolor{textcolor}%
\pgfsetfillcolor{textcolor}%
\pgftext[x=2.067942in,y=0.277778in,,top]{\color{textcolor}\rmfamily\fontsize{10.000000}{12.000000}\selectfont \(\displaystyle {20}\)}%
\end{pgfscope}%
\begin{pgfscope}%
\pgfsetbuttcap%
\pgfsetroundjoin%
\definecolor{currentfill}{rgb}{0.000000,0.000000,0.000000}%
\pgfsetfillcolor{currentfill}%
\pgfsetlinewidth{0.803000pt}%
\definecolor{currentstroke}{rgb}{0.000000,0.000000,0.000000}%
\pgfsetstrokecolor{currentstroke}%
\pgfsetdash{}{0pt}%
\pgfsys@defobject{currentmarker}{\pgfqpoint{0.000000in}{-0.048611in}}{\pgfqpoint{0.000000in}{0.000000in}}{%
\pgfpathmoveto{\pgfqpoint{0.000000in}{0.000000in}}%
\pgfpathlineto{\pgfqpoint{0.000000in}{-0.048611in}}%
\pgfusepath{stroke,fill}%
}%
\begin{pgfscope}%
\pgfsys@transformshift{2.850299in}{0.375000in}%
\pgfsys@useobject{currentmarker}{}%
\end{pgfscope}%
\end{pgfscope}%
\begin{pgfscope}%
\definecolor{textcolor}{rgb}{0.000000,0.000000,0.000000}%
\pgfsetstrokecolor{textcolor}%
\pgfsetfillcolor{textcolor}%
\pgftext[x=2.850299in,y=0.277778in,,top]{\color{textcolor}\rmfamily\fontsize{10.000000}{12.000000}\selectfont \(\displaystyle {30}\)}%
\end{pgfscope}%
\begin{pgfscope}%
\definecolor{textcolor}{rgb}{0.000000,0.000000,0.000000}%
\pgfsetstrokecolor{textcolor}%
\pgfsetfillcolor{textcolor}%
\pgftext[x=2.050000in,y=0.098766in,,top]{\color{textcolor}\rmfamily\fontsize{10.000000}{12.000000}\selectfont SNR (dB)}%
\end{pgfscope}%
\begin{pgfscope}%
\pgfsetbuttcap%
\pgfsetroundjoin%
\definecolor{currentfill}{rgb}{0.000000,0.000000,0.000000}%
\pgfsetfillcolor{currentfill}%
\pgfsetlinewidth{0.803000pt}%
\definecolor{currentstroke}{rgb}{0.000000,0.000000,0.000000}%
\pgfsetstrokecolor{currentstroke}%
\pgfsetdash{}{0pt}%
\pgfsys@defobject{currentmarker}{\pgfqpoint{-0.048611in}{0.000000in}}{\pgfqpoint{-0.000000in}{0.000000in}}{%
\pgfpathmoveto{\pgfqpoint{-0.000000in}{0.000000in}}%
\pgfpathlineto{\pgfqpoint{-0.048611in}{0.000000in}}%
\pgfusepath{stroke,fill}%
}%
\begin{pgfscope}%
\pgfsys@transformshift{0.500000in}{0.375000in}%
\pgfsys@useobject{currentmarker}{}%
\end{pgfscope}%
\end{pgfscope}%
\begin{pgfscope}%
\definecolor{textcolor}{rgb}{0.000000,0.000000,0.000000}%
\pgfsetstrokecolor{textcolor}%
\pgfsetfillcolor{textcolor}%
\pgftext[x=0.225308in, y=0.326775in, left, base]{\color{textcolor}\rmfamily\fontsize{10.000000}{12.000000}\selectfont \(\displaystyle {0.0}\)}%
\end{pgfscope}%
\begin{pgfscope}%
\pgfsetbuttcap%
\pgfsetroundjoin%
\definecolor{currentfill}{rgb}{0.000000,0.000000,0.000000}%
\pgfsetfillcolor{currentfill}%
\pgfsetlinewidth{0.803000pt}%
\definecolor{currentstroke}{rgb}{0.000000,0.000000,0.000000}%
\pgfsetstrokecolor{currentstroke}%
\pgfsetdash{}{0pt}%
\pgfsys@defobject{currentmarker}{\pgfqpoint{-0.048611in}{0.000000in}}{\pgfqpoint{-0.000000in}{0.000000in}}{%
\pgfpathmoveto{\pgfqpoint{-0.000000in}{0.000000in}}%
\pgfpathlineto{\pgfqpoint{-0.048611in}{0.000000in}}%
\pgfusepath{stroke,fill}%
}%
\begin{pgfscope}%
\pgfsys@transformshift{0.500000in}{0.828000in}%
\pgfsys@useobject{currentmarker}{}%
\end{pgfscope}%
\end{pgfscope}%
\begin{pgfscope}%
\definecolor{textcolor}{rgb}{0.000000,0.000000,0.000000}%
\pgfsetstrokecolor{textcolor}%
\pgfsetfillcolor{textcolor}%
\pgftext[x=0.225308in, y=0.779775in, left, base]{\color{textcolor}\rmfamily\fontsize{10.000000}{12.000000}\selectfont \(\displaystyle {0.2}\)}%
\end{pgfscope}%
\begin{pgfscope}%
\pgfsetbuttcap%
\pgfsetroundjoin%
\definecolor{currentfill}{rgb}{0.000000,0.000000,0.000000}%
\pgfsetfillcolor{currentfill}%
\pgfsetlinewidth{0.803000pt}%
\definecolor{currentstroke}{rgb}{0.000000,0.000000,0.000000}%
\pgfsetstrokecolor{currentstroke}%
\pgfsetdash{}{0pt}%
\pgfsys@defobject{currentmarker}{\pgfqpoint{-0.048611in}{0.000000in}}{\pgfqpoint{-0.000000in}{0.000000in}}{%
\pgfpathmoveto{\pgfqpoint{-0.000000in}{0.000000in}}%
\pgfpathlineto{\pgfqpoint{-0.048611in}{0.000000in}}%
\pgfusepath{stroke,fill}%
}%
\begin{pgfscope}%
\pgfsys@transformshift{0.500000in}{1.281000in}%
\pgfsys@useobject{currentmarker}{}%
\end{pgfscope}%
\end{pgfscope}%
\begin{pgfscope}%
\definecolor{textcolor}{rgb}{0.000000,0.000000,0.000000}%
\pgfsetstrokecolor{textcolor}%
\pgfsetfillcolor{textcolor}%
\pgftext[x=0.225308in, y=1.232775in, left, base]{\color{textcolor}\rmfamily\fontsize{10.000000}{12.000000}\selectfont \(\displaystyle {0.4}\)}%
\end{pgfscope}%
\begin{pgfscope}%
\pgfsetbuttcap%
\pgfsetroundjoin%
\definecolor{currentfill}{rgb}{0.000000,0.000000,0.000000}%
\pgfsetfillcolor{currentfill}%
\pgfsetlinewidth{0.803000pt}%
\definecolor{currentstroke}{rgb}{0.000000,0.000000,0.000000}%
\pgfsetstrokecolor{currentstroke}%
\pgfsetdash{}{0pt}%
\pgfsys@defobject{currentmarker}{\pgfqpoint{-0.048611in}{0.000000in}}{\pgfqpoint{-0.000000in}{0.000000in}}{%
\pgfpathmoveto{\pgfqpoint{-0.000000in}{0.000000in}}%
\pgfpathlineto{\pgfqpoint{-0.048611in}{0.000000in}}%
\pgfusepath{stroke,fill}%
}%
\begin{pgfscope}%
\pgfsys@transformshift{0.500000in}{1.734000in}%
\pgfsys@useobject{currentmarker}{}%
\end{pgfscope}%
\end{pgfscope}%
\begin{pgfscope}%
\definecolor{textcolor}{rgb}{0.000000,0.000000,0.000000}%
\pgfsetstrokecolor{textcolor}%
\pgfsetfillcolor{textcolor}%
\pgftext[x=0.225308in, y=1.685775in, left, base]{\color{textcolor}\rmfamily\fontsize{10.000000}{12.000000}\selectfont \(\displaystyle {0.6}\)}%
\end{pgfscope}%
\begin{pgfscope}%
\pgfsetbuttcap%
\pgfsetroundjoin%
\definecolor{currentfill}{rgb}{0.000000,0.000000,0.000000}%
\pgfsetfillcolor{currentfill}%
\pgfsetlinewidth{0.803000pt}%
\definecolor{currentstroke}{rgb}{0.000000,0.000000,0.000000}%
\pgfsetstrokecolor{currentstroke}%
\pgfsetdash{}{0pt}%
\pgfsys@defobject{currentmarker}{\pgfqpoint{-0.048611in}{0.000000in}}{\pgfqpoint{-0.000000in}{0.000000in}}{%
\pgfpathmoveto{\pgfqpoint{-0.000000in}{0.000000in}}%
\pgfpathlineto{\pgfqpoint{-0.048611in}{0.000000in}}%
\pgfusepath{stroke,fill}%
}%
\begin{pgfscope}%
\pgfsys@transformshift{0.500000in}{2.187000in}%
\pgfsys@useobject{currentmarker}{}%
\end{pgfscope}%
\end{pgfscope}%
\begin{pgfscope}%
\definecolor{textcolor}{rgb}{0.000000,0.000000,0.000000}%
\pgfsetstrokecolor{textcolor}%
\pgfsetfillcolor{textcolor}%
\pgftext[x=0.225308in, y=2.138775in, left, base]{\color{textcolor}\rmfamily\fontsize{10.000000}{12.000000}\selectfont \(\displaystyle {0.8}\)}%
\end{pgfscope}%
\begin{pgfscope}%
\pgfsetbuttcap%
\pgfsetroundjoin%
\definecolor{currentfill}{rgb}{0.000000,0.000000,0.000000}%
\pgfsetfillcolor{currentfill}%
\pgfsetlinewidth{0.803000pt}%
\definecolor{currentstroke}{rgb}{0.000000,0.000000,0.000000}%
\pgfsetstrokecolor{currentstroke}%
\pgfsetdash{}{0pt}%
\pgfsys@defobject{currentmarker}{\pgfqpoint{-0.048611in}{0.000000in}}{\pgfqpoint{-0.000000in}{0.000000in}}{%
\pgfpathmoveto{\pgfqpoint{-0.000000in}{0.000000in}}%
\pgfpathlineto{\pgfqpoint{-0.048611in}{0.000000in}}%
\pgfusepath{stroke,fill}%
}%
\begin{pgfscope}%
\pgfsys@transformshift{0.500000in}{2.640000in}%
\pgfsys@useobject{currentmarker}{}%
\end{pgfscope}%
\end{pgfscope}%
\begin{pgfscope}%
\definecolor{textcolor}{rgb}{0.000000,0.000000,0.000000}%
\pgfsetstrokecolor{textcolor}%
\pgfsetfillcolor{textcolor}%
\pgftext[x=0.225308in, y=2.591775in, left, base]{\color{textcolor}\rmfamily\fontsize{10.000000}{12.000000}\selectfont \(\displaystyle {1.0}\)}%
\end{pgfscope}%
\begin{pgfscope}%
\definecolor{textcolor}{rgb}{0.000000,0.000000,0.000000}%
\pgfsetstrokecolor{textcolor}%
\pgfsetfillcolor{textcolor}%
\pgftext[x=0.169753in,y=1.507500in,,bottom,rotate=90.000000]{\color{textcolor}\rmfamily\fontsize{10.000000}{12.000000}\selectfont Accuracy (\%)}%
\end{pgfscope}%
\begin{pgfscope}%
\pgfpathrectangle{\pgfqpoint{0.500000in}{0.375000in}}{\pgfqpoint{3.100000in}{2.265000in}}%
\pgfusepath{clip}%
\pgfsetrectcap%
\pgfsetroundjoin%
\pgfsetlinewidth{1.505625pt}%
\definecolor{currentstroke}{rgb}{0.000000,0.000000,1.000000}%
\pgfsetstrokecolor{currentstroke}%
\pgfsetdash{}{0pt}%
\pgfpathmoveto{\pgfqpoint{3.459091in}{2.161833in}}%
\pgfpathlineto{\pgfqpoint{2.988065in}{2.163092in}}%
\pgfpathlineto{\pgfqpoint{2.752552in}{2.164350in}}%
\pgfpathlineto{\pgfqpoint{2.614786in}{2.164350in}}%
\pgfpathlineto{\pgfqpoint{2.517039in}{2.166447in}}%
\pgfpathlineto{\pgfqpoint{2.379273in}{2.144217in}}%
\pgfpathlineto{\pgfqpoint{2.326897in}{2.137506in}}%
\pgfpathlineto{\pgfqpoint{2.441221in}{2.150928in}}%
\pgfpathlineto{\pgfqpoint{2.281526in}{2.126600in}}%
\pgfpathlineto{\pgfqpoint{2.241507in}{2.112758in}}%
\pgfpathlineto{\pgfqpoint{2.197318in}{2.095561in}}%
\pgfpathlineto{\pgfqpoint{1.965974in}{2.026353in}}%
\pgfpathlineto{\pgfqpoint{1.829609in}{1.985667in}}%
\pgfpathlineto{\pgfqpoint{1.732565in}{1.943303in}}%
\pgfpathlineto{\pgfqpoint{1.657169in}{1.885000in}}%
\pgfpathlineto{\pgfqpoint{1.595503in}{1.856478in}}%
\pgfpathlineto{\pgfqpoint{1.543328in}{1.788108in}}%
\pgfpathlineto{\pgfqpoint{1.498109in}{1.734839in}}%
\pgfpathlineto{\pgfqpoint{1.458207in}{1.647594in}}%
\pgfpathlineto{\pgfqpoint{1.422503in}{1.608586in}}%
\pgfpathlineto{\pgfqpoint{1.390196in}{1.565383in}}%
\pgfpathlineto{\pgfqpoint{1.360696in}{1.476042in}}%
\pgfpathlineto{\pgfqpoint{1.333554in}{1.420256in}}%
\pgfpathlineto{\pgfqpoint{1.308421in}{1.360275in}}%
\pgfpathlineto{\pgfqpoint{1.285019in}{1.282678in}}%
\pgfpathlineto{\pgfqpoint{1.263126in}{1.238217in}}%
\pgfpathlineto{\pgfqpoint{1.242558in}{1.132517in}}%
\pgfpathlineto{\pgfqpoint{1.223165in}{1.120353in}}%
\pgfpathlineto{\pgfqpoint{1.204819in}{1.093508in}}%
\pgfpathlineto{\pgfqpoint{1.187414in}{1.060372in}}%
\pgfpathlineto{\pgfqpoint{1.170856in}{0.921117in}}%
\pgfpathlineto{\pgfqpoint{1.155068in}{0.939572in}}%
\pgfpathlineto{\pgfqpoint{1.139982in}{0.828419in}}%
\pgfpathlineto{\pgfqpoint{1.125536in}{0.822128in}}%
\pgfpathlineto{\pgfqpoint{1.111680in}{0.856522in}}%
\pgfpathlineto{\pgfqpoint{1.098367in}{0.793606in}}%
\pgfpathlineto{\pgfqpoint{1.085556in}{0.773053in}}%
\pgfpathlineto{\pgfqpoint{1.073211in}{0.787733in}}%
\pgfpathlineto{\pgfqpoint{1.061298in}{0.667772in}}%
\pgfpathlineto{\pgfqpoint{1.049789in}{0.672806in}}%
\pgfpathlineto{\pgfqpoint{1.038657in}{0.643864in}}%
\pgfpathlineto{\pgfqpoint{1.027878in}{0.629603in}}%
\pgfpathlineto{\pgfqpoint{1.017431in}{0.611147in}}%
\pgfpathlineto{\pgfqpoint{1.007295in}{0.585142in}}%
\pgfpathlineto{\pgfqpoint{0.997453in}{0.550747in}}%
\pgfpathlineto{\pgfqpoint{0.987888in}{0.520967in}}%
\pgfpathlineto{\pgfqpoint{0.978585in}{0.602339in}}%
\pgfpathlineto{\pgfqpoint{0.969530in}{0.508383in}}%
\pgfpathlineto{\pgfqpoint{0.960710in}{0.495381in}}%
\pgfpathlineto{\pgfqpoint{0.952113in}{0.491606in}}%
\pgfpathlineto{\pgfqpoint{0.943728in}{0.476506in}}%
\pgfpathlineto{\pgfqpoint{0.935545in}{0.505028in}}%
\pgfpathlineto{\pgfqpoint{0.927555in}{0.444628in}}%
\pgfpathlineto{\pgfqpoint{0.919748in}{0.425333in}}%
\pgfpathlineto{\pgfqpoint{0.912117in}{0.437917in}}%
\pgfpathlineto{\pgfqpoint{0.904653in}{0.445467in}}%
\pgfpathlineto{\pgfqpoint{0.897350in}{0.444208in}}%
\pgfpathlineto{\pgfqpoint{0.890200in}{0.433722in}}%
\pgfpathlineto{\pgfqpoint{0.883198in}{0.422397in}}%
\pgfpathlineto{\pgfqpoint{0.876337in}{0.434561in}}%
\pgfpathlineto{\pgfqpoint{0.869612in}{0.428269in}}%
\pgfpathlineto{\pgfqpoint{0.863017in}{0.402683in}}%
\pgfpathlineto{\pgfqpoint{0.856548in}{0.408136in}}%
\pgfpathlineto{\pgfqpoint{0.850200in}{0.415267in}}%
\pgfpathlineto{\pgfqpoint{0.843969in}{0.403103in}}%
\pgfpathlineto{\pgfqpoint{0.837849in}{0.396392in}}%
\pgfpathlineto{\pgfqpoint{0.831838in}{0.401844in}}%
\pgfpathlineto{\pgfqpoint{0.825931in}{0.394714in}}%
\pgfpathlineto{\pgfqpoint{0.820126in}{0.396811in}}%
\pgfpathlineto{\pgfqpoint{0.814417in}{0.402683in}}%
\pgfpathlineto{\pgfqpoint{0.808804in}{0.389681in}}%
\pgfpathlineto{\pgfqpoint{0.803281in}{0.385906in}}%
\pgfpathlineto{\pgfqpoint{0.797847in}{0.388842in}}%
\pgfpathlineto{\pgfqpoint{0.792498in}{0.386744in}}%
\pgfpathlineto{\pgfqpoint{0.787232in}{0.397650in}}%
\pgfpathlineto{\pgfqpoint{0.782046in}{0.385906in}}%
\pgfpathlineto{\pgfqpoint{0.776939in}{0.390100in}}%
\pgfpathlineto{\pgfqpoint{0.771907in}{0.388842in}}%
\pgfpathlineto{\pgfqpoint{0.766948in}{0.384647in}}%
\pgfpathlineto{\pgfqpoint{0.762061in}{0.387164in}}%
\pgfpathlineto{\pgfqpoint{0.757243in}{0.386744in}}%
\pgfpathlineto{\pgfqpoint{0.752493in}{0.385486in}}%
\pgfpathlineto{\pgfqpoint{0.747808in}{0.385906in}}%
\pgfpathlineto{\pgfqpoint{0.743187in}{0.385906in}}%
\pgfpathlineto{\pgfqpoint{0.738627in}{0.386744in}}%
\pgfpathlineto{\pgfqpoint{0.734129in}{0.385067in}}%
\pgfpathlineto{\pgfqpoint{0.729689in}{0.384647in}}%
\pgfpathlineto{\pgfqpoint{0.725306in}{0.386744in}}%
\pgfpathlineto{\pgfqpoint{0.720979in}{0.385486in}}%
\pgfpathlineto{\pgfqpoint{0.716706in}{0.385906in}}%
\pgfpathlineto{\pgfqpoint{0.712487in}{0.384228in}}%
\pgfpathlineto{\pgfqpoint{0.708319in}{0.385067in}}%
\pgfpathlineto{\pgfqpoint{0.704202in}{0.385067in}}%
\pgfpathlineto{\pgfqpoint{0.700134in}{0.384228in}}%
\pgfpathlineto{\pgfqpoint{0.696114in}{0.384228in}}%
\pgfpathlineto{\pgfqpoint{0.692141in}{0.384228in}}%
\pgfpathlineto{\pgfqpoint{0.688214in}{0.384228in}}%
\pgfpathlineto{\pgfqpoint{0.684332in}{0.384228in}}%
\pgfpathlineto{\pgfqpoint{0.680494in}{0.385067in}}%
\pgfpathlineto{\pgfqpoint{0.676698in}{0.384647in}}%
\pgfpathlineto{\pgfqpoint{0.672945in}{0.385067in}}%
\pgfpathlineto{\pgfqpoint{0.669233in}{0.385486in}}%
\pgfpathlineto{\pgfqpoint{0.665560in}{0.384647in}}%
\pgfpathlineto{\pgfqpoint{0.661927in}{0.384647in}}%
\pgfpathlineto{\pgfqpoint{0.658333in}{0.384647in}}%
\pgfpathlineto{\pgfqpoint{0.654776in}{0.384228in}}%
\pgfpathlineto{\pgfqpoint{0.651256in}{0.384228in}}%
\pgfpathlineto{\pgfqpoint{0.647772in}{0.384647in}}%
\pgfpathlineto{\pgfqpoint{0.644323in}{0.384228in}}%
\pgfpathlineto{\pgfqpoint{0.640909in}{0.384228in}}%
\pgfusepath{stroke}%
\end{pgfscope}%
\begin{pgfscope}%
\pgfpathrectangle{\pgfqpoint{0.500000in}{0.375000in}}{\pgfqpoint{3.100000in}{2.265000in}}%
\pgfusepath{clip}%
\pgfsetrectcap%
\pgfsetroundjoin%
\pgfsetlinewidth{1.505625pt}%
\definecolor{currentstroke}{rgb}{1.000000,0.000000,0.000000}%
\pgfsetstrokecolor{currentstroke}%
\pgfsetdash{}{0pt}%
\pgfpathmoveto{\pgfqpoint{3.459091in}{2.162253in}}%
\pgfpathlineto{\pgfqpoint{2.988065in}{2.162672in}}%
\pgfpathlineto{\pgfqpoint{2.752552in}{2.166028in}}%
\pgfpathlineto{\pgfqpoint{2.614786in}{2.165189in}}%
\pgfpathlineto{\pgfqpoint{2.517039in}{2.159317in}}%
\pgfpathlineto{\pgfqpoint{2.379273in}{2.138764in}}%
\pgfpathlineto{\pgfqpoint{2.326897in}{2.124922in}}%
\pgfpathlineto{\pgfqpoint{2.441221in}{2.156381in}}%
\pgfpathlineto{\pgfqpoint{2.281526in}{2.105628in}}%
\pgfpathlineto{\pgfqpoint{2.241507in}{2.111500in}}%
\pgfpathlineto{\pgfqpoint{2.197318in}{2.100175in}}%
\pgfpathlineto{\pgfqpoint{1.965974in}{1.994894in}}%
\pgfpathlineto{\pgfqpoint{1.829609in}{1.940786in}}%
\pgfpathlineto{\pgfqpoint{1.732565in}{1.942464in}}%
\pgfpathlineto{\pgfqpoint{1.657169in}{1.877031in}}%
\pgfpathlineto{\pgfqpoint{1.595503in}{1.808242in}}%
\pgfpathlineto{\pgfqpoint{1.543328in}{1.808242in}}%
\pgfpathlineto{\pgfqpoint{1.498109in}{1.671503in}}%
\pgfpathlineto{\pgfqpoint{1.458207in}{1.665631in}}%
\pgfpathlineto{\pgfqpoint{1.422503in}{1.515889in}}%
\pgfpathlineto{\pgfqpoint{1.390196in}{1.452553in}}%
\pgfpathlineto{\pgfqpoint{1.360696in}{1.383344in}}%
\pgfpathlineto{\pgfqpoint{1.333554in}{1.262125in}}%
\pgfpathlineto{\pgfqpoint{1.308421in}{1.158522in}}%
\pgfpathlineto{\pgfqpoint{1.285019in}{1.129161in}}%
\pgfpathlineto{\pgfqpoint{1.263126in}{1.112383in}}%
\pgfpathlineto{\pgfqpoint{1.242558in}{1.077150in}}%
\pgfpathlineto{\pgfqpoint{1.223165in}{0.965578in}}%
\pgfpathlineto{\pgfqpoint{1.204819in}{0.786056in}}%
\pgfpathlineto{\pgfqpoint{1.187414in}{0.832194in}}%
\pgfpathlineto{\pgfqpoint{1.170856in}{0.848553in}}%
\pgfpathlineto{\pgfqpoint{1.155068in}{0.639669in}}%
\pgfpathlineto{\pgfqpoint{1.139982in}{0.640508in}}%
\pgfpathlineto{\pgfqpoint{1.125536in}{0.552425in}}%
\pgfpathlineto{\pgfqpoint{1.111680in}{0.529775in}}%
\pgfpathlineto{\pgfqpoint{1.098367in}{0.574236in}}%
\pgfpathlineto{\pgfqpoint{1.085556in}{0.501672in}}%
\pgfpathlineto{\pgfqpoint{1.073211in}{0.464761in}}%
\pgfpathlineto{\pgfqpoint{1.061298in}{0.520128in}}%
\pgfpathlineto{\pgfqpoint{1.049789in}{0.523064in}}%
\pgfpathlineto{\pgfqpoint{1.038657in}{0.444208in}}%
\pgfpathlineto{\pgfqpoint{1.027878in}{0.429947in}}%
\pgfpathlineto{\pgfqpoint{1.017431in}{0.423656in}}%
\pgfpathlineto{\pgfqpoint{1.007295in}{0.433303in}}%
\pgfpathlineto{\pgfqpoint{0.997453in}{0.420719in}}%
\pgfpathlineto{\pgfqpoint{0.987888in}{0.409394in}}%
\pgfpathlineto{\pgfqpoint{0.978585in}{0.406039in}}%
\pgfpathlineto{\pgfqpoint{0.969530in}{0.409394in}}%
\pgfpathlineto{\pgfqpoint{0.960710in}{0.409814in}}%
\pgfpathlineto{\pgfqpoint{0.952113in}{0.405200in}}%
\pgfpathlineto{\pgfqpoint{0.943728in}{0.398069in}}%
\pgfpathlineto{\pgfqpoint{0.935545in}{0.395553in}}%
\pgfpathlineto{\pgfqpoint{0.927555in}{0.393456in}}%
\pgfpathlineto{\pgfqpoint{0.919748in}{0.398908in}}%
\pgfpathlineto{\pgfqpoint{0.912117in}{0.390939in}}%
\pgfpathlineto{\pgfqpoint{0.904653in}{0.386744in}}%
\pgfpathlineto{\pgfqpoint{0.897350in}{0.389681in}}%
\pgfpathlineto{\pgfqpoint{0.890200in}{0.387583in}}%
\pgfpathlineto{\pgfqpoint{0.883198in}{0.385906in}}%
\pgfpathlineto{\pgfqpoint{0.876337in}{0.386744in}}%
\pgfpathlineto{\pgfqpoint{0.869612in}{0.385067in}}%
\pgfpathlineto{\pgfqpoint{0.863017in}{0.384228in}}%
\pgfpathlineto{\pgfqpoint{0.856548in}{0.385067in}}%
\pgfpathlineto{\pgfqpoint{0.850200in}{0.385067in}}%
\pgfpathlineto{\pgfqpoint{0.843969in}{0.384647in}}%
\pgfpathlineto{\pgfqpoint{0.837849in}{0.385486in}}%
\pgfpathlineto{\pgfqpoint{0.831838in}{0.384228in}}%
\pgfpathlineto{\pgfqpoint{0.825931in}{0.384228in}}%
\pgfpathlineto{\pgfqpoint{0.820126in}{0.385067in}}%
\pgfpathlineto{\pgfqpoint{0.814417in}{0.384228in}}%
\pgfpathlineto{\pgfqpoint{0.808804in}{0.385067in}}%
\pgfpathlineto{\pgfqpoint{0.803281in}{0.384228in}}%
\pgfpathlineto{\pgfqpoint{0.797847in}{0.384228in}}%
\pgfpathlineto{\pgfqpoint{0.792498in}{0.384228in}}%
\pgfpathlineto{\pgfqpoint{0.787232in}{0.384228in}}%
\pgfpathlineto{\pgfqpoint{0.782046in}{0.384228in}}%
\pgfpathlineto{\pgfqpoint{0.776939in}{0.384228in}}%
\pgfpathlineto{\pgfqpoint{0.771907in}{0.384228in}}%
\pgfpathlineto{\pgfqpoint{0.766948in}{0.384228in}}%
\pgfpathlineto{\pgfqpoint{0.762061in}{0.384228in}}%
\pgfpathlineto{\pgfqpoint{0.757243in}{0.384228in}}%
\pgfpathlineto{\pgfqpoint{0.752493in}{0.384228in}}%
\pgfpathlineto{\pgfqpoint{0.747808in}{0.384228in}}%
\pgfpathlineto{\pgfqpoint{0.743187in}{0.384228in}}%
\pgfpathlineto{\pgfqpoint{0.738627in}{0.384228in}}%
\pgfpathlineto{\pgfqpoint{0.734129in}{0.384228in}}%
\pgfpathlineto{\pgfqpoint{0.729689in}{0.384228in}}%
\pgfpathlineto{\pgfqpoint{0.725306in}{0.384228in}}%
\pgfpathlineto{\pgfqpoint{0.720979in}{0.384228in}}%
\pgfpathlineto{\pgfqpoint{0.716706in}{0.384228in}}%
\pgfpathlineto{\pgfqpoint{0.712487in}{0.384228in}}%
\pgfpathlineto{\pgfqpoint{0.708319in}{0.384228in}}%
\pgfpathlineto{\pgfqpoint{0.704202in}{0.384228in}}%
\pgfpathlineto{\pgfqpoint{0.700134in}{0.384228in}}%
\pgfpathlineto{\pgfqpoint{0.696114in}{0.384228in}}%
\pgfpathlineto{\pgfqpoint{0.692141in}{0.384228in}}%
\pgfpathlineto{\pgfqpoint{0.688214in}{0.384228in}}%
\pgfpathlineto{\pgfqpoint{0.684332in}{0.384228in}}%
\pgfpathlineto{\pgfqpoint{0.680494in}{0.384228in}}%
\pgfpathlineto{\pgfqpoint{0.676698in}{0.384228in}}%
\pgfpathlineto{\pgfqpoint{0.672945in}{0.384228in}}%
\pgfpathlineto{\pgfqpoint{0.669233in}{0.384228in}}%
\pgfpathlineto{\pgfqpoint{0.665560in}{0.384228in}}%
\pgfpathlineto{\pgfqpoint{0.661927in}{0.384228in}}%
\pgfpathlineto{\pgfqpoint{0.658333in}{0.384228in}}%
\pgfpathlineto{\pgfqpoint{0.654776in}{0.384228in}}%
\pgfpathlineto{\pgfqpoint{0.651256in}{0.384228in}}%
\pgfpathlineto{\pgfqpoint{0.647772in}{0.384228in}}%
\pgfpathlineto{\pgfqpoint{0.644323in}{0.384228in}}%
\pgfpathlineto{\pgfqpoint{0.640909in}{0.384228in}}%
\pgfusepath{stroke}%
\end{pgfscope}%
\begin{pgfscope}%
\pgfpathrectangle{\pgfqpoint{0.500000in}{0.375000in}}{\pgfqpoint{3.100000in}{2.265000in}}%
\pgfusepath{clip}%
\pgfsetbuttcap%
\pgfsetroundjoin%
\pgfsetlinewidth{0.501875pt}%
\definecolor{currentstroke}{rgb}{0.000000,0.000000,0.000000}%
\pgfsetstrokecolor{currentstroke}%
\pgfsetdash{{1.850000pt}{0.800000pt}}{0.000000pt}%
\pgfpathmoveto{\pgfqpoint{1.111680in}{0.375000in}}%
\pgfpathlineto{\pgfqpoint{1.111680in}{2.653889in}}%
\pgfusepath{stroke}%
\end{pgfscope}%
\begin{pgfscope}%
\pgfsetrectcap%
\pgfsetmiterjoin%
\pgfsetlinewidth{0.803000pt}%
\definecolor{currentstroke}{rgb}{0.000000,0.000000,0.000000}%
\pgfsetstrokecolor{currentstroke}%
\pgfsetdash{}{0pt}%
\pgfpathmoveto{\pgfqpoint{0.500000in}{0.375000in}}%
\pgfpathlineto{\pgfqpoint{0.500000in}{2.640000in}}%
\pgfusepath{stroke}%
\end{pgfscope}%
\begin{pgfscope}%
\pgfsetrectcap%
\pgfsetmiterjoin%
\pgfsetlinewidth{0.803000pt}%
\definecolor{currentstroke}{rgb}{0.000000,0.000000,0.000000}%
\pgfsetstrokecolor{currentstroke}%
\pgfsetdash{}{0pt}%
\pgfpathmoveto{\pgfqpoint{3.600000in}{0.375000in}}%
\pgfpathlineto{\pgfqpoint{3.600000in}{2.640000in}}%
\pgfusepath{stroke}%
\end{pgfscope}%
\begin{pgfscope}%
\pgfsetrectcap%
\pgfsetmiterjoin%
\pgfsetlinewidth{0.803000pt}%
\definecolor{currentstroke}{rgb}{0.000000,0.000000,0.000000}%
\pgfsetstrokecolor{currentstroke}%
\pgfsetdash{}{0pt}%
\pgfpathmoveto{\pgfqpoint{0.500000in}{0.375000in}}%
\pgfpathlineto{\pgfqpoint{3.600000in}{0.375000in}}%
\pgfusepath{stroke}%
\end{pgfscope}%
\begin{pgfscope}%
\pgfsetrectcap%
\pgfsetmiterjoin%
\pgfsetlinewidth{0.803000pt}%
\definecolor{currentstroke}{rgb}{0.000000,0.000000,0.000000}%
\pgfsetstrokecolor{currentstroke}%
\pgfsetdash{}{0pt}%
\pgfpathmoveto{\pgfqpoint{0.500000in}{2.640000in}}%
\pgfpathlineto{\pgfqpoint{3.600000in}{2.640000in}}%
\pgfusepath{stroke}%
\end{pgfscope}%
\begin{pgfscope}%
\pgfsetbuttcap%
\pgfsetmiterjoin%
\definecolor{currentfill}{rgb}{1.000000,1.000000,1.000000}%
\pgfsetfillcolor{currentfill}%
\pgfsetfillopacity{0.800000}%
\pgfsetlinewidth{1.003750pt}%
\definecolor{currentstroke}{rgb}{0.800000,0.800000,0.800000}%
\pgfsetstrokecolor{currentstroke}%
\pgfsetstrokeopacity{0.800000}%
\pgfsetdash{}{0pt}%
\pgfpathmoveto{\pgfqpoint{1.777466in}{0.444444in}}%
\pgfpathlineto{\pgfqpoint{3.502778in}{0.444444in}}%
\pgfpathquadraticcurveto{\pgfqpoint{3.530556in}{0.444444in}}{\pgfqpoint{3.530556in}{0.472222in}}%
\pgfpathlineto{\pgfqpoint{3.530556in}{1.039352in}}%
\pgfpathquadraticcurveto{\pgfqpoint{3.530556in}{1.067129in}}{\pgfqpoint{3.502778in}{1.067129in}}%
\pgfpathlineto{\pgfqpoint{1.777466in}{1.067129in}}%
\pgfpathquadraticcurveto{\pgfqpoint{1.749688in}{1.067129in}}{\pgfqpoint{1.749688in}{1.039352in}}%
\pgfpathlineto{\pgfqpoint{1.749688in}{0.472222in}}%
\pgfpathquadraticcurveto{\pgfqpoint{1.749688in}{0.444444in}}{\pgfqpoint{1.777466in}{0.444444in}}%
\pgfpathclose%
\pgfusepath{stroke,fill}%
\end{pgfscope}%
\begin{pgfscope}%
\pgfsetrectcap%
\pgfsetroundjoin%
\pgfsetlinewidth{1.505625pt}%
\definecolor{currentstroke}{rgb}{0.000000,0.000000,1.000000}%
\pgfsetstrokecolor{currentstroke}%
\pgfsetdash{}{0pt}%
\pgfpathmoveto{\pgfqpoint{1.805244in}{0.962963in}}%
\pgfpathlineto{\pgfqpoint{2.083022in}{0.962963in}}%
\pgfusepath{stroke}%
\end{pgfscope}%
\begin{pgfscope}%
\definecolor{textcolor}{rgb}{0.000000,0.000000,0.000000}%
\pgfsetstrokecolor{textcolor}%
\pgfsetfillcolor{textcolor}%
\pgftext[x=2.194133in,y=0.914352in,left,base]{\color{textcolor}\rmfamily\fontsize{10.000000}{12.000000}\selectfont wav attack on OTA}%
\end{pgfscope}%
\begin{pgfscope}%
\pgfsetrectcap%
\pgfsetroundjoin%
\pgfsetlinewidth{1.505625pt}%
\definecolor{currentstroke}{rgb}{1.000000,0.000000,0.000000}%
\pgfsetstrokecolor{currentstroke}%
\pgfsetdash{}{0pt}%
\pgfpathmoveto{\pgfqpoint{1.805244in}{0.769290in}}%
\pgfpathlineto{\pgfqpoint{2.083022in}{0.769290in}}%
\pgfusepath{stroke}%
\end{pgfscope}%
\begin{pgfscope}%
\definecolor{textcolor}{rgb}{0.000000,0.000000,0.000000}%
\pgfsetstrokecolor{textcolor}%
\pgfsetfillcolor{textcolor}%
\pgftext[x=2.194133in,y=0.720679in,left,base]{\color{textcolor}\rmfamily\fontsize{10.000000}{12.000000}\selectfont freq attack on OTA}%
\end{pgfscope}%
\begin{pgfscope}%
\pgfsetbuttcap%
\pgfsetroundjoin%
\pgfsetlinewidth{0.501875pt}%
\definecolor{currentstroke}{rgb}{0.000000,0.000000,0.000000}%
\pgfsetstrokecolor{currentstroke}%
\pgfsetdash{{1.850000pt}{0.800000pt}}{0.000000pt}%
\pgfpathmoveto{\pgfqpoint{1.805244in}{0.575617in}}%
\pgfpathlineto{\pgfqpoint{2.083022in}{0.575617in}}%
\pgfusepath{stroke}%
\end{pgfscope}%
\begin{pgfscope}%
\definecolor{textcolor}{rgb}{0.000000,0.000000,0.000000}%
\pgfsetstrokecolor{textcolor}%
\pgfsetfillcolor{textcolor}%
\pgftext[x=2.194133in,y=0.527006in,left,base]{\color{textcolor}\rmfamily\fontsize{10.000000}{12.000000}\selectfont Max Diff. at 7.78 dB}%
\end{pgfscope}%
\end{pgfpicture}%
\makeatother%
\endgroup%

%% file: figures/acc_SW.pgf
\begingroup%
\makeatletter%
\begin{pgfpicture}%
\pgfpathrectangle{\pgfpointorigin}{\pgfqpoint{4.000000in}{3.000000in}}%
\pgfusepath{use as bounding box, clip}%
\begin{pgfscope}%
\pgfsetbuttcap%
\pgfsetmiterjoin%
\pgfsetlinewidth{0.000000pt}%
\definecolor{currentstroke}{rgb}{1.000000,1.000000,1.000000}%
\pgfsetstrokecolor{currentstroke}%
\pgfsetstrokeopacity{0.000000}%
\pgfsetdash{}{0pt}%
\pgfpathmoveto{\pgfqpoint{0.000000in}{0.000000in}}%
\pgfpathlineto{\pgfqpoint{4.000000in}{0.000000in}}%
\pgfpathlineto{\pgfqpoint{4.000000in}{3.000000in}}%
\pgfpathlineto{\pgfqpoint{0.000000in}{3.000000in}}%
\pgfpathclose%
\pgfusepath{}%
\end{pgfscope}%
\begin{pgfscope}%
\pgfsetbuttcap%
\pgfsetmiterjoin%
\definecolor{currentfill}{rgb}{1.000000,1.000000,1.000000}%
\pgfsetfillcolor{currentfill}%
\pgfsetlinewidth{0.000000pt}%
\definecolor{currentstroke}{rgb}{0.000000,0.000000,0.000000}%
\pgfsetstrokecolor{currentstroke}%
\pgfsetstrokeopacity{0.000000}%
\pgfsetdash{}{0pt}%
\pgfpathmoveto{\pgfqpoint{0.500000in}{0.375000in}}%
\pgfpathlineto{\pgfqpoint{3.600000in}{0.375000in}}%
\pgfpathlineto{\pgfqpoint{3.600000in}{2.640000in}}%
\pgfpathlineto{\pgfqpoint{0.500000in}{2.640000in}}%
\pgfpathclose%
\pgfusepath{fill}%
\end{pgfscope}%
\begin{pgfscope}%
\pgfsetbuttcap%
\pgfsetroundjoin%
\definecolor{currentfill}{rgb}{0.000000,0.000000,0.000000}%
\pgfsetfillcolor{currentfill}%
\pgfsetlinewidth{0.803000pt}%
\definecolor{currentstroke}{rgb}{0.000000,0.000000,0.000000}%
\pgfsetstrokecolor{currentstroke}%
\pgfsetdash{}{0pt}%
\pgfsys@defobject{currentmarker}{\pgfqpoint{0.000000in}{-0.048611in}}{\pgfqpoint{0.000000in}{0.000000in}}{%
\pgfpathmoveto{\pgfqpoint{0.000000in}{0.000000in}}%
\pgfpathlineto{\pgfqpoint{0.000000in}{-0.048611in}}%
\pgfusepath{stroke,fill}%
}%
\begin{pgfscope}%
\pgfsys@transformshift{0.503228in}{0.375000in}%
\pgfsys@useobject{currentmarker}{}%
\end{pgfscope}%
\end{pgfscope}%
\begin{pgfscope}%
\definecolor{textcolor}{rgb}{0.000000,0.000000,0.000000}%
\pgfsetstrokecolor{textcolor}%
\pgfsetfillcolor{textcolor}%
\pgftext[x=0.503228in,y=0.277778in,,top]{\color{textcolor}\rmfamily\fontsize{10.000000}{12.000000}\selectfont \(\displaystyle {0}\)}%
\end{pgfscope}%
\begin{pgfscope}%
\pgfsetbuttcap%
\pgfsetroundjoin%
\definecolor{currentfill}{rgb}{0.000000,0.000000,0.000000}%
\pgfsetfillcolor{currentfill}%
\pgfsetlinewidth{0.803000pt}%
\definecolor{currentstroke}{rgb}{0.000000,0.000000,0.000000}%
\pgfsetstrokecolor{currentstroke}%
\pgfsetdash{}{0pt}%
\pgfsys@defobject{currentmarker}{\pgfqpoint{0.000000in}{-0.048611in}}{\pgfqpoint{0.000000in}{0.000000in}}{%
\pgfpathmoveto{\pgfqpoint{0.000000in}{0.000000in}}%
\pgfpathlineto{\pgfqpoint{0.000000in}{-0.048611in}}%
\pgfusepath{stroke,fill}%
}%
\begin{pgfscope}%
\pgfsys@transformshift{1.285585in}{0.375000in}%
\pgfsys@useobject{currentmarker}{}%
\end{pgfscope}%
\end{pgfscope}%
\begin{pgfscope}%
\definecolor{textcolor}{rgb}{0.000000,0.000000,0.000000}%
\pgfsetstrokecolor{textcolor}%
\pgfsetfillcolor{textcolor}%
\pgftext[x=1.285585in,y=0.277778in,,top]{\color{textcolor}\rmfamily\fontsize{10.000000}{12.000000}\selectfont \(\displaystyle {10}\)}%
\end{pgfscope}%
\begin{pgfscope}%
\pgfsetbuttcap%
\pgfsetroundjoin%
\definecolor{currentfill}{rgb}{0.000000,0.000000,0.000000}%
\pgfsetfillcolor{currentfill}%
\pgfsetlinewidth{0.803000pt}%
\definecolor{currentstroke}{rgb}{0.000000,0.000000,0.000000}%
\pgfsetstrokecolor{currentstroke}%
\pgfsetdash{}{0pt}%
\pgfsys@defobject{currentmarker}{\pgfqpoint{0.000000in}{-0.048611in}}{\pgfqpoint{0.000000in}{0.000000in}}{%
\pgfpathmoveto{\pgfqpoint{0.000000in}{0.000000in}}%
\pgfpathlineto{\pgfqpoint{0.000000in}{-0.048611in}}%
\pgfusepath{stroke,fill}%
}%
\begin{pgfscope}%
\pgfsys@transformshift{2.067942in}{0.375000in}%
\pgfsys@useobject{currentmarker}{}%
\end{pgfscope}%
\end{pgfscope}%
\begin{pgfscope}%
\definecolor{textcolor}{rgb}{0.000000,0.000000,0.000000}%
\pgfsetstrokecolor{textcolor}%
\pgfsetfillcolor{textcolor}%
\pgftext[x=2.067942in,y=0.277778in,,top]{\color{textcolor}\rmfamily\fontsize{10.000000}{12.000000}\selectfont \(\displaystyle {20}\)}%
\end{pgfscope}%
\begin{pgfscope}%
\pgfsetbuttcap%
\pgfsetroundjoin%
\definecolor{currentfill}{rgb}{0.000000,0.000000,0.000000}%
\pgfsetfillcolor{currentfill}%
\pgfsetlinewidth{0.803000pt}%
\definecolor{currentstroke}{rgb}{0.000000,0.000000,0.000000}%
\pgfsetstrokecolor{currentstroke}%
\pgfsetdash{}{0pt}%
\pgfsys@defobject{currentmarker}{\pgfqpoint{0.000000in}{-0.048611in}}{\pgfqpoint{0.000000in}{0.000000in}}{%
\pgfpathmoveto{\pgfqpoint{0.000000in}{0.000000in}}%
\pgfpathlineto{\pgfqpoint{0.000000in}{-0.048611in}}%
\pgfusepath{stroke,fill}%
}%
\begin{pgfscope}%
\pgfsys@transformshift{2.850299in}{0.375000in}%
\pgfsys@useobject{currentmarker}{}%
\end{pgfscope}%
\end{pgfscope}%
\begin{pgfscope}%
\definecolor{textcolor}{rgb}{0.000000,0.000000,0.000000}%
\pgfsetstrokecolor{textcolor}%
\pgfsetfillcolor{textcolor}%
\pgftext[x=2.850299in,y=0.277778in,,top]{\color{textcolor}\rmfamily\fontsize{10.000000}{12.000000}\selectfont \(\displaystyle {30}\)}%
\end{pgfscope}%
\begin{pgfscope}%
\definecolor{textcolor}{rgb}{0.000000,0.000000,0.000000}%
\pgfsetstrokecolor{textcolor}%
\pgfsetfillcolor{textcolor}%
\pgftext[x=2.050000in,y=0.098766in,,top]{\color{textcolor}\rmfamily\fontsize{10.000000}{12.000000}\selectfont SNR (dB)}%
\end{pgfscope}%
\begin{pgfscope}%
\pgfsetbuttcap%
\pgfsetroundjoin%
\definecolor{currentfill}{rgb}{0.000000,0.000000,0.000000}%
\pgfsetfillcolor{currentfill}%
\pgfsetlinewidth{0.803000pt}%
\definecolor{currentstroke}{rgb}{0.000000,0.000000,0.000000}%
\pgfsetstrokecolor{currentstroke}%
\pgfsetdash{}{0pt}%
\pgfsys@defobject{currentmarker}{\pgfqpoint{-0.048611in}{0.000000in}}{\pgfqpoint{-0.000000in}{0.000000in}}{%
\pgfpathmoveto{\pgfqpoint{-0.000000in}{0.000000in}}%
\pgfpathlineto{\pgfqpoint{-0.048611in}{0.000000in}}%
\pgfusepath{stroke,fill}%
}%
\begin{pgfscope}%
\pgfsys@transformshift{0.500000in}{0.375000in}%
\pgfsys@useobject{currentmarker}{}%
\end{pgfscope}%
\end{pgfscope}%
\begin{pgfscope}%
\definecolor{textcolor}{rgb}{0.000000,0.000000,0.000000}%
\pgfsetstrokecolor{textcolor}%
\pgfsetfillcolor{textcolor}%
\pgftext[x=0.225308in, y=0.326775in, left, base]{\color{textcolor}\rmfamily\fontsize{10.000000}{12.000000}\selectfont \(\displaystyle {0.0}\)}%
\end{pgfscope}%
\begin{pgfscope}%
\pgfsetbuttcap%
\pgfsetroundjoin%
\definecolor{currentfill}{rgb}{0.000000,0.000000,0.000000}%
\pgfsetfillcolor{currentfill}%
\pgfsetlinewidth{0.803000pt}%
\definecolor{currentstroke}{rgb}{0.000000,0.000000,0.000000}%
\pgfsetstrokecolor{currentstroke}%
\pgfsetdash{}{0pt}%
\pgfsys@defobject{currentmarker}{\pgfqpoint{-0.048611in}{0.000000in}}{\pgfqpoint{-0.000000in}{0.000000in}}{%
\pgfpathmoveto{\pgfqpoint{-0.000000in}{0.000000in}}%
\pgfpathlineto{\pgfqpoint{-0.048611in}{0.000000in}}%
\pgfusepath{stroke,fill}%
}%
\begin{pgfscope}%
\pgfsys@transformshift{0.500000in}{0.828000in}%
\pgfsys@useobject{currentmarker}{}%
\end{pgfscope}%
\end{pgfscope}%
\begin{pgfscope}%
\definecolor{textcolor}{rgb}{0.000000,0.000000,0.000000}%
\pgfsetstrokecolor{textcolor}%
\pgfsetfillcolor{textcolor}%
\pgftext[x=0.225308in, y=0.779775in, left, base]{\color{textcolor}\rmfamily\fontsize{10.000000}{12.000000}\selectfont \(\displaystyle {0.2}\)}%
\end{pgfscope}%
\begin{pgfscope}%
\pgfsetbuttcap%
\pgfsetroundjoin%
\definecolor{currentfill}{rgb}{0.000000,0.000000,0.000000}%
\pgfsetfillcolor{currentfill}%
\pgfsetlinewidth{0.803000pt}%
\definecolor{currentstroke}{rgb}{0.000000,0.000000,0.000000}%
\pgfsetstrokecolor{currentstroke}%
\pgfsetdash{}{0pt}%
\pgfsys@defobject{currentmarker}{\pgfqpoint{-0.048611in}{0.000000in}}{\pgfqpoint{-0.000000in}{0.000000in}}{%
\pgfpathmoveto{\pgfqpoint{-0.000000in}{0.000000in}}%
\pgfpathlineto{\pgfqpoint{-0.048611in}{0.000000in}}%
\pgfusepath{stroke,fill}%
}%
\begin{pgfscope}%
\pgfsys@transformshift{0.500000in}{1.281000in}%
\pgfsys@useobject{currentmarker}{}%
\end{pgfscope}%
\end{pgfscope}%
\begin{pgfscope}%
\definecolor{textcolor}{rgb}{0.000000,0.000000,0.000000}%
\pgfsetstrokecolor{textcolor}%
\pgfsetfillcolor{textcolor}%
\pgftext[x=0.225308in, y=1.232775in, left, base]{\color{textcolor}\rmfamily\fontsize{10.000000}{12.000000}\selectfont \(\displaystyle {0.4}\)}%
\end{pgfscope}%
\begin{pgfscope}%
\pgfsetbuttcap%
\pgfsetroundjoin%
\definecolor{currentfill}{rgb}{0.000000,0.000000,0.000000}%
\pgfsetfillcolor{currentfill}%
\pgfsetlinewidth{0.803000pt}%
\definecolor{currentstroke}{rgb}{0.000000,0.000000,0.000000}%
\pgfsetstrokecolor{currentstroke}%
\pgfsetdash{}{0pt}%
\pgfsys@defobject{currentmarker}{\pgfqpoint{-0.048611in}{0.000000in}}{\pgfqpoint{-0.000000in}{0.000000in}}{%
\pgfpathmoveto{\pgfqpoint{-0.000000in}{0.000000in}}%
\pgfpathlineto{\pgfqpoint{-0.048611in}{0.000000in}}%
\pgfusepath{stroke,fill}%
}%
\begin{pgfscope}%
\pgfsys@transformshift{0.500000in}{1.734000in}%
\pgfsys@useobject{currentmarker}{}%
\end{pgfscope}%
\end{pgfscope}%
\begin{pgfscope}%
\definecolor{textcolor}{rgb}{0.000000,0.000000,0.000000}%
\pgfsetstrokecolor{textcolor}%
\pgfsetfillcolor{textcolor}%
\pgftext[x=0.225308in, y=1.685775in, left, base]{\color{textcolor}\rmfamily\fontsize{10.000000}{12.000000}\selectfont \(\displaystyle {0.6}\)}%
\end{pgfscope}%
\begin{pgfscope}%
\pgfsetbuttcap%
\pgfsetroundjoin%
\definecolor{currentfill}{rgb}{0.000000,0.000000,0.000000}%
\pgfsetfillcolor{currentfill}%
\pgfsetlinewidth{0.803000pt}%
\definecolor{currentstroke}{rgb}{0.000000,0.000000,0.000000}%
\pgfsetstrokecolor{currentstroke}%
\pgfsetdash{}{0pt}%
\pgfsys@defobject{currentmarker}{\pgfqpoint{-0.048611in}{0.000000in}}{\pgfqpoint{-0.000000in}{0.000000in}}{%
\pgfpathmoveto{\pgfqpoint{-0.000000in}{0.000000in}}%
\pgfpathlineto{\pgfqpoint{-0.048611in}{0.000000in}}%
\pgfusepath{stroke,fill}%
}%
\begin{pgfscope}%
\pgfsys@transformshift{0.500000in}{2.187000in}%
\pgfsys@useobject{currentmarker}{}%
\end{pgfscope}%
\end{pgfscope}%
\begin{pgfscope}%
\definecolor{textcolor}{rgb}{0.000000,0.000000,0.000000}%
\pgfsetstrokecolor{textcolor}%
\pgfsetfillcolor{textcolor}%
\pgftext[x=0.225308in, y=2.138775in, left, base]{\color{textcolor}\rmfamily\fontsize{10.000000}{12.000000}\selectfont \(\displaystyle {0.8}\)}%
\end{pgfscope}%
\begin{pgfscope}%
\pgfsetbuttcap%
\pgfsetroundjoin%
\definecolor{currentfill}{rgb}{0.000000,0.000000,0.000000}%
\pgfsetfillcolor{currentfill}%
\pgfsetlinewidth{0.803000pt}%
\definecolor{currentstroke}{rgb}{0.000000,0.000000,0.000000}%
\pgfsetstrokecolor{currentstroke}%
\pgfsetdash{}{0pt}%
\pgfsys@defobject{currentmarker}{\pgfqpoint{-0.048611in}{0.000000in}}{\pgfqpoint{-0.000000in}{0.000000in}}{%
\pgfpathmoveto{\pgfqpoint{-0.000000in}{0.000000in}}%
\pgfpathlineto{\pgfqpoint{-0.048611in}{0.000000in}}%
\pgfusepath{stroke,fill}%
}%
\begin{pgfscope}%
\pgfsys@transformshift{0.500000in}{2.640000in}%
\pgfsys@useobject{currentmarker}{}%
\end{pgfscope}%
\end{pgfscope}%
\begin{pgfscope}%
\definecolor{textcolor}{rgb}{0.000000,0.000000,0.000000}%
\pgfsetstrokecolor{textcolor}%
\pgfsetfillcolor{textcolor}%
\pgftext[x=0.225308in, y=2.591775in, left, base]{\color{textcolor}\rmfamily\fontsize{10.000000}{12.000000}\selectfont \(\displaystyle {1.0}\)}%
\end{pgfscope}%
\begin{pgfscope}%
\definecolor{textcolor}{rgb}{0.000000,0.000000,0.000000}%
\pgfsetstrokecolor{textcolor}%
\pgfsetfillcolor{textcolor}%
\pgftext[x=0.169753in,y=1.507500in,,bottom,rotate=90.000000]{\color{textcolor}\rmfamily\fontsize{10.000000}{12.000000}\selectfont Accuracy (\%)}%
\end{pgfscope}%
\begin{pgfscope}%
\pgfpathrectangle{\pgfqpoint{0.500000in}{0.375000in}}{\pgfqpoint{3.100000in}{2.265000in}}%
\pgfusepath{clip}%
\pgfsetrectcap%
\pgfsetroundjoin%
\pgfsetlinewidth{1.505625pt}%
\definecolor{currentstroke}{rgb}{1.000000,0.000000,0.000000}%
\pgfsetstrokecolor{currentstroke}%
\pgfsetdash{}{0pt}%
\pgfpathmoveto{\pgfqpoint{3.459091in}{2.497808in}}%
\pgfpathlineto{\pgfqpoint{2.988065in}{2.496131in}}%
\pgfpathlineto{\pgfqpoint{2.752552in}{2.494033in}}%
\pgfpathlineto{\pgfqpoint{2.614786in}{2.490258in}}%
\pgfpathlineto{\pgfqpoint{2.517039in}{2.479772in}}%
\pgfpathlineto{\pgfqpoint{2.379273in}{2.464253in}}%
\pgfpathlineto{\pgfqpoint{2.326897in}{2.449992in}}%
\pgfpathlineto{\pgfqpoint{2.441221in}{2.478094in}}%
\pgfpathlineto{\pgfqpoint{2.281526in}{2.437408in}}%
\pgfpathlineto{\pgfqpoint{2.241507in}{2.439086in}}%
\pgfpathlineto{\pgfqpoint{2.197318in}{2.421889in}}%
\pgfpathlineto{\pgfqpoint{1.965974in}{2.334644in}}%
\pgfpathlineto{\pgfqpoint{1.829609in}{2.262081in}}%
\pgfpathlineto{\pgfqpoint{1.732565in}{2.241108in}}%
\pgfpathlineto{\pgfqpoint{1.657169in}{2.181547in}}%
\pgfpathlineto{\pgfqpoint{1.595503in}{2.100594in}}%
\pgfpathlineto{\pgfqpoint{1.543328in}{2.080042in}}%
\pgfpathlineto{\pgfqpoint{1.498109in}{1.961339in}}%
\pgfpathlineto{\pgfqpoint{1.458207in}{1.926106in}}%
\pgfpathlineto{\pgfqpoint{1.422503in}{1.848928in}}%
\pgfpathlineto{\pgfqpoint{1.390196in}{1.753294in}}%
\pgfpathlineto{\pgfqpoint{1.360696in}{1.703800in}}%
\pgfpathlineto{\pgfqpoint{1.333554in}{1.588453in}}%
\pgfpathlineto{\pgfqpoint{1.308421in}{1.543992in}}%
\pgfpathlineto{\pgfqpoint{1.285019in}{1.494078in}}%
\pgfpathlineto{\pgfqpoint{1.263126in}{1.447100in}}%
\pgfpathlineto{\pgfqpoint{1.242558in}{1.408931in}}%
\pgfpathlineto{\pgfqpoint{1.223165in}{1.320008in}}%
\pgfpathlineto{\pgfqpoint{1.204819in}{1.131258in}}%
\pgfpathlineto{\pgfqpoint{1.187414in}{1.169428in}}%
\pgfpathlineto{\pgfqpoint{1.170856in}{1.170267in}}%
\pgfpathlineto{\pgfqpoint{1.155068in}{0.989486in}}%
\pgfpathlineto{\pgfqpoint{1.139982in}{0.941250in}}%
\pgfpathlineto{\pgfqpoint{1.125536in}{0.799478in}}%
\pgfpathlineto{\pgfqpoint{1.111680in}{0.768858in}}%
\pgfpathlineto{\pgfqpoint{1.098367in}{0.852328in}}%
\pgfpathlineto{\pgfqpoint{1.085556in}{0.697133in}}%
\pgfpathlineto{\pgfqpoint{1.073211in}{0.601081in}}%
\pgfpathlineto{\pgfqpoint{1.061298in}{0.772633in}}%
\pgfpathlineto{\pgfqpoint{1.049789in}{0.724397in}}%
\pgfpathlineto{\pgfqpoint{1.038657in}{0.583464in}}%
\pgfpathlineto{\pgfqpoint{1.027878in}{0.558717in}}%
\pgfpathlineto{\pgfqpoint{1.017431in}{0.543617in}}%
\pgfpathlineto{\pgfqpoint{1.007295in}{0.539003in}}%
\pgfpathlineto{\pgfqpoint{0.997453in}{0.536067in}}%
\pgfpathlineto{\pgfqpoint{0.987888in}{0.494542in}}%
\pgfpathlineto{\pgfqpoint{0.978585in}{0.488250in}}%
\pgfpathlineto{\pgfqpoint{0.969530in}{0.514256in}}%
\pgfpathlineto{\pgfqpoint{0.960710in}{0.511739in}}%
\pgfpathlineto{\pgfqpoint{0.952113in}{0.462244in}}%
\pgfpathlineto{\pgfqpoint{0.943728in}{0.466019in}}%
\pgfpathlineto{\pgfqpoint{0.935545in}{0.440014in}}%
\pgfpathlineto{\pgfqpoint{0.927555in}{0.443369in}}%
\pgfpathlineto{\pgfqpoint{0.919748in}{0.451339in}}%
\pgfpathlineto{\pgfqpoint{0.912117in}{0.430367in}}%
\pgfpathlineto{\pgfqpoint{0.904653in}{0.403942in}}%
\pgfpathlineto{\pgfqpoint{0.897350in}{0.420300in}}%
\pgfpathlineto{\pgfqpoint{0.890200in}{0.404781in}}%
\pgfpathlineto{\pgfqpoint{0.883198in}{0.408975in}}%
\pgfpathlineto{\pgfqpoint{0.876337in}{0.402264in}}%
\pgfpathlineto{\pgfqpoint{0.869612in}{0.396392in}}%
\pgfpathlineto{\pgfqpoint{0.863017in}{0.390519in}}%
\pgfpathlineto{\pgfqpoint{0.856548in}{0.393036in}}%
\pgfpathlineto{\pgfqpoint{0.850200in}{0.393036in}}%
\pgfpathlineto{\pgfqpoint{0.843969in}{0.393036in}}%
\pgfpathlineto{\pgfqpoint{0.837849in}{0.398069in}}%
\pgfpathlineto{\pgfqpoint{0.831838in}{0.386744in}}%
\pgfpathlineto{\pgfqpoint{0.825931in}{0.387583in}}%
\pgfpathlineto{\pgfqpoint{0.820126in}{0.392197in}}%
\pgfpathlineto{\pgfqpoint{0.814417in}{0.389261in}}%
\pgfpathlineto{\pgfqpoint{0.808804in}{0.389681in}}%
\pgfpathlineto{\pgfqpoint{0.803281in}{0.385906in}}%
\pgfpathlineto{\pgfqpoint{0.797847in}{0.385906in}}%
\pgfpathlineto{\pgfqpoint{0.792498in}{0.388003in}}%
\pgfpathlineto{\pgfqpoint{0.787232in}{0.386325in}}%
\pgfpathlineto{\pgfqpoint{0.782046in}{0.386744in}}%
\pgfpathlineto{\pgfqpoint{0.776939in}{0.385486in}}%
\pgfpathlineto{\pgfqpoint{0.771907in}{0.384647in}}%
\pgfpathlineto{\pgfqpoint{0.766948in}{0.386325in}}%
\pgfpathlineto{\pgfqpoint{0.762061in}{0.385067in}}%
\pgfpathlineto{\pgfqpoint{0.757243in}{0.384228in}}%
\pgfpathlineto{\pgfqpoint{0.752493in}{0.384647in}}%
\pgfpathlineto{\pgfqpoint{0.747808in}{0.384228in}}%
\pgfpathlineto{\pgfqpoint{0.743187in}{0.387164in}}%
\pgfpathlineto{\pgfqpoint{0.738627in}{0.384647in}}%
\pgfpathlineto{\pgfqpoint{0.734129in}{0.384228in}}%
\pgfpathlineto{\pgfqpoint{0.729689in}{0.384228in}}%
\pgfpathlineto{\pgfqpoint{0.725306in}{0.384228in}}%
\pgfpathlineto{\pgfqpoint{0.720979in}{0.384228in}}%
\pgfpathlineto{\pgfqpoint{0.716706in}{0.384647in}}%
\pgfpathlineto{\pgfqpoint{0.712487in}{0.384228in}}%
\pgfpathlineto{\pgfqpoint{0.708319in}{0.384228in}}%
\pgfpathlineto{\pgfqpoint{0.704202in}{0.384228in}}%
\pgfpathlineto{\pgfqpoint{0.700134in}{0.384228in}}%
\pgfpathlineto{\pgfqpoint{0.696114in}{0.384647in}}%
\pgfpathlineto{\pgfqpoint{0.692141in}{0.384228in}}%
\pgfpathlineto{\pgfqpoint{0.688214in}{0.384228in}}%
\pgfpathlineto{\pgfqpoint{0.684332in}{0.384228in}}%
\pgfpathlineto{\pgfqpoint{0.680494in}{0.384228in}}%
\pgfpathlineto{\pgfqpoint{0.676698in}{0.384228in}}%
\pgfpathlineto{\pgfqpoint{0.672945in}{0.384228in}}%
\pgfpathlineto{\pgfqpoint{0.669233in}{0.384228in}}%
\pgfpathlineto{\pgfqpoint{0.665560in}{0.384228in}}%
\pgfpathlineto{\pgfqpoint{0.661927in}{0.384228in}}%
\pgfpathlineto{\pgfqpoint{0.658333in}{0.384228in}}%
\pgfpathlineto{\pgfqpoint{0.654776in}{0.384228in}}%
\pgfpathlineto{\pgfqpoint{0.651256in}{0.384228in}}%
\pgfpathlineto{\pgfqpoint{0.647772in}{0.384228in}}%
\pgfpathlineto{\pgfqpoint{0.644323in}{0.384228in}}%
\pgfpathlineto{\pgfqpoint{0.640909in}{0.384228in}}%
\pgfusepath{stroke}%
\end{pgfscope}%
\begin{pgfscope}%
\pgfpathrectangle{\pgfqpoint{0.500000in}{0.375000in}}{\pgfqpoint{3.100000in}{2.265000in}}%
\pgfusepath{clip}%
\pgfsetrectcap%
\pgfsetroundjoin%
\pgfsetlinewidth{1.505625pt}%
\definecolor{currentstroke}{rgb}{0.000000,0.000000,1.000000}%
\pgfsetstrokecolor{currentstroke}%
\pgfsetdash{}{0pt}%
\pgfpathmoveto{\pgfqpoint{3.459091in}{2.498228in}}%
\pgfpathlineto{\pgfqpoint{2.988065in}{2.495711in}}%
\pgfpathlineto{\pgfqpoint{2.752552in}{2.494453in}}%
\pgfpathlineto{\pgfqpoint{2.614786in}{2.489419in}}%
\pgfpathlineto{\pgfqpoint{2.517039in}{2.487322in}}%
\pgfpathlineto{\pgfqpoint{2.379273in}{2.468447in}}%
\pgfpathlineto{\pgfqpoint{2.326897in}{2.457122in}}%
\pgfpathlineto{\pgfqpoint{2.441221in}{2.478514in}}%
\pgfpathlineto{\pgfqpoint{2.281526in}{2.453347in}}%
\pgfpathlineto{\pgfqpoint{2.241507in}{2.442861in}}%
\pgfpathlineto{\pgfqpoint{2.197318in}{2.436989in}}%
\pgfpathlineto{\pgfqpoint{1.965974in}{2.353939in}}%
\pgfpathlineto{\pgfqpoint{1.829609in}{2.303186in}}%
\pgfpathlineto{\pgfqpoint{1.732565in}{2.237333in}}%
\pgfpathlineto{\pgfqpoint{1.657169in}{2.181547in}}%
\pgfpathlineto{\pgfqpoint{1.595503in}{2.107306in}}%
\pgfpathlineto{\pgfqpoint{1.543328in}{2.059489in}}%
\pgfpathlineto{\pgfqpoint{1.498109in}{2.006219in}}%
\pgfpathlineto{\pgfqpoint{1.458207in}{1.910167in}}%
\pgfpathlineto{\pgfqpoint{1.422503in}{1.853122in}}%
\pgfpathlineto{\pgfqpoint{1.390196in}{1.802789in}}%
\pgfpathlineto{\pgfqpoint{1.360696in}{1.730644in}}%
\pgfpathlineto{\pgfqpoint{1.333554in}{1.655564in}}%
\pgfpathlineto{\pgfqpoint{1.308421in}{1.640044in}}%
\pgfpathlineto{\pgfqpoint{1.285019in}{1.543572in}}%
\pgfpathlineto{\pgfqpoint{1.263126in}{1.454650in}}%
\pgfpathlineto{\pgfqpoint{1.242558in}{1.433258in}}%
\pgfpathlineto{\pgfqpoint{1.223165in}{1.362792in}}%
\pgfpathlineto{\pgfqpoint{1.204819in}{1.377053in}}%
\pgfpathlineto{\pgfqpoint{1.187414in}{1.364889in}}%
\pgfpathlineto{\pgfqpoint{1.170856in}{1.244928in}}%
\pgfpathlineto{\pgfqpoint{1.155068in}{1.219342in}}%
\pgfpathlineto{\pgfqpoint{1.139982in}{1.151392in}}%
\pgfpathlineto{\pgfqpoint{1.125536in}{1.132936in}}%
\pgfpathlineto{\pgfqpoint{1.111680in}{1.160200in}}%
\pgfpathlineto{\pgfqpoint{1.098367in}{1.096864in}}%
\pgfpathlineto{\pgfqpoint{1.085556in}{1.089733in}}%
\pgfpathlineto{\pgfqpoint{1.073211in}{1.093089in}}%
\pgfpathlineto{\pgfqpoint{1.061298in}{0.942508in}}%
\pgfpathlineto{\pgfqpoint{1.049789in}{0.936217in}}%
\pgfpathlineto{\pgfqpoint{1.038657in}{0.933281in}}%
\pgfpathlineto{\pgfqpoint{1.027878in}{0.882108in}}%
\pgfpathlineto{\pgfqpoint{1.017431in}{0.882947in}}%
\pgfpathlineto{\pgfqpoint{1.007295in}{0.814997in}}%
\pgfpathlineto{\pgfqpoint{0.997453in}{0.775150in}}%
\pgfpathlineto{\pgfqpoint{0.987888in}{0.742853in}}%
\pgfpathlineto{\pgfqpoint{0.978585in}{0.785636in}}%
\pgfpathlineto{\pgfqpoint{0.969530in}{0.716428in}}%
\pgfpathlineto{\pgfqpoint{0.960710in}{0.712653in}}%
\pgfpathlineto{\pgfqpoint{0.952113in}{0.701328in}}%
\pgfpathlineto{\pgfqpoint{0.943728in}{0.656867in}}%
\pgfpathlineto{\pgfqpoint{0.935545in}{0.671967in}}%
\pgfpathlineto{\pgfqpoint{0.927555in}{0.585561in}}%
\pgfpathlineto{\pgfqpoint{0.919748in}{0.571300in}}%
\pgfpathlineto{\pgfqpoint{0.912117in}{0.578850in}}%
\pgfpathlineto{\pgfqpoint{0.904653in}{0.603597in}}%
\pgfpathlineto{\pgfqpoint{0.897350in}{0.601500in}}%
\pgfpathlineto{\pgfqpoint{0.890200in}{0.563750in}}%
\pgfpathlineto{\pgfqpoint{0.883198in}{0.546133in}}%
\pgfpathlineto{\pgfqpoint{0.876337in}{0.554103in}}%
\pgfpathlineto{\pgfqpoint{0.869612in}{0.552425in}}%
\pgfpathlineto{\pgfqpoint{0.863017in}{0.502092in}}%
\pgfpathlineto{\pgfqpoint{0.856548in}{0.505867in}}%
\pgfpathlineto{\pgfqpoint{0.850200in}{0.500833in}}%
\pgfpathlineto{\pgfqpoint{0.843969in}{0.510900in}}%
\pgfpathlineto{\pgfqpoint{0.837849in}{0.459308in}}%
\pgfpathlineto{\pgfqpoint{0.831838in}{0.491186in}}%
\pgfpathlineto{\pgfqpoint{0.825931in}{0.475247in}}%
\pgfpathlineto{\pgfqpoint{0.820126in}{0.476506in}}%
\pgfpathlineto{\pgfqpoint{0.814417in}{0.488250in}}%
\pgfpathlineto{\pgfqpoint{0.808804in}{0.444628in}}%
\pgfpathlineto{\pgfqpoint{0.803281in}{0.414847in}}%
\pgfpathlineto{\pgfqpoint{0.797847in}{0.463503in}}%
\pgfpathlineto{\pgfqpoint{0.792498in}{0.431206in}}%
\pgfpathlineto{\pgfqpoint{0.787232in}{0.463922in}}%
\pgfpathlineto{\pgfqpoint{0.782046in}{0.429108in}}%
\pgfpathlineto{\pgfqpoint{0.776939in}{0.441272in}}%
\pgfpathlineto{\pgfqpoint{0.771907in}{0.441272in}}%
\pgfpathlineto{\pgfqpoint{0.766948in}{0.414847in}}%
\pgfpathlineto{\pgfqpoint{0.762061in}{0.427431in}}%
\pgfpathlineto{\pgfqpoint{0.757243in}{0.427011in}}%
\pgfpathlineto{\pgfqpoint{0.752493in}{0.416944in}}%
\pgfpathlineto{\pgfqpoint{0.747808in}{0.424914in}}%
\pgfpathlineto{\pgfqpoint{0.743187in}{0.421978in}}%
\pgfpathlineto{\pgfqpoint{0.738627in}{0.420300in}}%
\pgfpathlineto{\pgfqpoint{0.734129in}{0.406878in}}%
\pgfpathlineto{\pgfqpoint{0.729689in}{0.406878in}}%
\pgfpathlineto{\pgfqpoint{0.725306in}{0.420719in}}%
\pgfpathlineto{\pgfqpoint{0.720979in}{0.406039in}}%
\pgfpathlineto{\pgfqpoint{0.716706in}{0.408975in}}%
\pgfpathlineto{\pgfqpoint{0.712487in}{0.403103in}}%
\pgfpathlineto{\pgfqpoint{0.708319in}{0.395133in}}%
\pgfpathlineto{\pgfqpoint{0.704202in}{0.407297in}}%
\pgfpathlineto{\pgfqpoint{0.700134in}{0.398069in}}%
\pgfpathlineto{\pgfqpoint{0.696114in}{0.393036in}}%
\pgfpathlineto{\pgfqpoint{0.692141in}{0.396392in}}%
\pgfpathlineto{\pgfqpoint{0.688214in}{0.392197in}}%
\pgfpathlineto{\pgfqpoint{0.684332in}{0.386325in}}%
\pgfpathlineto{\pgfqpoint{0.680494in}{0.392617in}}%
\pgfpathlineto{\pgfqpoint{0.676698in}{0.389681in}}%
\pgfpathlineto{\pgfqpoint{0.672945in}{0.396392in}}%
\pgfpathlineto{\pgfqpoint{0.669233in}{0.395133in}}%
\pgfpathlineto{\pgfqpoint{0.665560in}{0.389261in}}%
\pgfpathlineto{\pgfqpoint{0.661927in}{0.392617in}}%
\pgfpathlineto{\pgfqpoint{0.658333in}{0.391358in}}%
\pgfpathlineto{\pgfqpoint{0.654776in}{0.388422in}}%
\pgfpathlineto{\pgfqpoint{0.651256in}{0.385067in}}%
\pgfpathlineto{\pgfqpoint{0.647772in}{0.393456in}}%
\pgfpathlineto{\pgfqpoint{0.644323in}{0.386744in}}%
\pgfpathlineto{\pgfqpoint{0.640909in}{0.386744in}}%
\pgfusepath{stroke}%
\end{pgfscope}%
\begin{pgfscope}%
\pgfpathrectangle{\pgfqpoint{0.500000in}{0.375000in}}{\pgfqpoint{3.100000in}{2.265000in}}%
\pgfusepath{clip}%
\pgfsetbuttcap%
\pgfsetroundjoin%
\pgfsetlinewidth{0.501875pt}%
\definecolor{currentstroke}{rgb}{0.000000,0.000000,0.000000}%
\pgfsetstrokecolor{currentstroke}%
\pgfsetdash{{1.850000pt}{0.800000pt}}{0.000000pt}%
\pgfpathmoveto{\pgfqpoint{1.073211in}{0.375000in}}%
\pgfpathlineto{\pgfqpoint{1.073211in}{2.653889in}}%
\pgfusepath{stroke}%
\end{pgfscope}%
\begin{pgfscope}%
\pgfsetrectcap%
\pgfsetmiterjoin%
\pgfsetlinewidth{0.803000pt}%
\definecolor{currentstroke}{rgb}{0.000000,0.000000,0.000000}%
\pgfsetstrokecolor{currentstroke}%
\pgfsetdash{}{0pt}%
\pgfpathmoveto{\pgfqpoint{0.500000in}{0.375000in}}%
\pgfpathlineto{\pgfqpoint{0.500000in}{2.640000in}}%
\pgfusepath{stroke}%
\end{pgfscope}%
\begin{pgfscope}%
\pgfsetrectcap%
\pgfsetmiterjoin%
\pgfsetlinewidth{0.803000pt}%
\definecolor{currentstroke}{rgb}{0.000000,0.000000,0.000000}%
\pgfsetstrokecolor{currentstroke}%
\pgfsetdash{}{0pt}%
\pgfpathmoveto{\pgfqpoint{3.600000in}{0.375000in}}%
\pgfpathlineto{\pgfqpoint{3.600000in}{2.640000in}}%
\pgfusepath{stroke}%
\end{pgfscope}%
\begin{pgfscope}%
\pgfsetrectcap%
\pgfsetmiterjoin%
\pgfsetlinewidth{0.803000pt}%
\definecolor{currentstroke}{rgb}{0.000000,0.000000,0.000000}%
\pgfsetstrokecolor{currentstroke}%
\pgfsetdash{}{0pt}%
\pgfpathmoveto{\pgfqpoint{0.500000in}{0.375000in}}%
\pgfpathlineto{\pgfqpoint{3.600000in}{0.375000in}}%
\pgfusepath{stroke}%
\end{pgfscope}%
\begin{pgfscope}%
\pgfsetrectcap%
\pgfsetmiterjoin%
\pgfsetlinewidth{0.803000pt}%
\definecolor{currentstroke}{rgb}{0.000000,0.000000,0.000000}%
\pgfsetstrokecolor{currentstroke}%
\pgfsetdash{}{0pt}%
\pgfpathmoveto{\pgfqpoint{0.500000in}{2.640000in}}%
\pgfpathlineto{\pgfqpoint{3.600000in}{2.640000in}}%
\pgfusepath{stroke}%
\end{pgfscope}%
\begin{pgfscope}%
\pgfsetbuttcap%
\pgfsetmiterjoin%
\definecolor{currentfill}{rgb}{1.000000,1.000000,1.000000}%
\pgfsetfillcolor{currentfill}%
\pgfsetfillopacity{0.800000}%
\pgfsetlinewidth{1.003750pt}%
\definecolor{currentstroke}{rgb}{0.800000,0.800000,0.800000}%
\pgfsetstrokecolor{currentstroke}%
\pgfsetstrokeopacity{0.800000}%
\pgfsetdash{}{0pt}%
\pgfpathmoveto{\pgfqpoint{1.777466in}{0.444444in}}%
\pgfpathlineto{\pgfqpoint{3.502778in}{0.444444in}}%
\pgfpathquadraticcurveto{\pgfqpoint{3.530556in}{0.444444in}}{\pgfqpoint{3.530556in}{0.472222in}}%
\pgfpathlineto{\pgfqpoint{3.530556in}{1.039352in}}%
\pgfpathquadraticcurveto{\pgfqpoint{3.530556in}{1.067129in}}{\pgfqpoint{3.502778in}{1.067129in}}%
\pgfpathlineto{\pgfqpoint{1.777466in}{1.067129in}}%
\pgfpathquadraticcurveto{\pgfqpoint{1.749688in}{1.067129in}}{\pgfqpoint{1.749688in}{1.039352in}}%
\pgfpathlineto{\pgfqpoint{1.749688in}{0.472222in}}%
\pgfpathquadraticcurveto{\pgfqpoint{1.749688in}{0.444444in}}{\pgfqpoint{1.777466in}{0.444444in}}%
\pgfpathclose%
\pgfusepath{stroke,fill}%
\end{pgfscope}%
\begin{pgfscope}%
\pgfsetrectcap%
\pgfsetroundjoin%
\pgfsetlinewidth{1.505625pt}%
\definecolor{currentstroke}{rgb}{1.000000,0.000000,0.000000}%
\pgfsetstrokecolor{currentstroke}%
\pgfsetdash{}{0pt}%
\pgfpathmoveto{\pgfqpoint{1.805244in}{0.962963in}}%
\pgfpathlineto{\pgfqpoint{2.083022in}{0.962963in}}%
\pgfusepath{stroke}%
\end{pgfscope}%
\begin{pgfscope}%
\definecolor{textcolor}{rgb}{0.000000,0.000000,0.000000}%
\pgfsetstrokecolor{textcolor}%
\pgfsetfillcolor{textcolor}%
\pgftext[x=2.194133in,y=0.914352in,left,base]{\color{textcolor}\rmfamily\fontsize{10.000000}{12.000000}\selectfont freq attack on SW}%
\end{pgfscope}%
\begin{pgfscope}%
\pgfsetrectcap%
\pgfsetroundjoin%
\pgfsetlinewidth{1.505625pt}%
\definecolor{currentstroke}{rgb}{0.000000,0.000000,1.000000}%
\pgfsetstrokecolor{currentstroke}%
\pgfsetdash{}{0pt}%
\pgfpathmoveto{\pgfqpoint{1.805244in}{0.769290in}}%
\pgfpathlineto{\pgfqpoint{2.083022in}{0.769290in}}%
\pgfusepath{stroke}%
\end{pgfscope}%
\begin{pgfscope}%
\definecolor{textcolor}{rgb}{0.000000,0.000000,0.000000}%
\pgfsetstrokecolor{textcolor}%
\pgfsetfillcolor{textcolor}%
\pgftext[x=2.194133in,y=0.720679in,left,base]{\color{textcolor}\rmfamily\fontsize{10.000000}{12.000000}\selectfont wav attack on SW}%
\end{pgfscope}%
\begin{pgfscope}%
\pgfsetbuttcap%
\pgfsetroundjoin%
\pgfsetlinewidth{0.501875pt}%
\definecolor{currentstroke}{rgb}{0.000000,0.000000,0.000000}%
\pgfsetstrokecolor{currentstroke}%
\pgfsetdash{{1.850000pt}{0.800000pt}}{0.000000pt}%
\pgfpathmoveto{\pgfqpoint{1.805244in}{0.575617in}}%
\pgfpathlineto{\pgfqpoint{2.083022in}{0.575617in}}%
\pgfusepath{stroke}%
\end{pgfscope}%
\begin{pgfscope}%
\definecolor{textcolor}{rgb}{0.000000,0.000000,0.000000}%
\pgfsetstrokecolor{textcolor}%
\pgfsetfillcolor{textcolor}%
\pgftext[x=2.194133in,y=0.527006in,left,base]{\color{textcolor}\rmfamily\fontsize{10.000000}{12.000000}\selectfont Max Diff. at 7.29 dB}%
\end{pgfscope}%
\end{pgfpicture}%
\makeatother%
\endgroup%

%% file: figures/foolrate_shift.pgf
\begingroup%
\makeatletter%
\begin{pgfpicture}%
\pgfpathrectangle{\pgfpointorigin}{\pgfqpoint{4.000000in}{3.000000in}}%
\pgfusepath{use as bounding box, clip}%
\begin{pgfscope}%
\pgfsetbuttcap%
\pgfsetmiterjoin%
\pgfsetlinewidth{0.000000pt}%
\definecolor{currentstroke}{rgb}{1.000000,1.000000,1.000000}%
\pgfsetstrokecolor{currentstroke}%
\pgfsetstrokeopacity{0.000000}%
\pgfsetdash{}{0pt}%
\pgfpathmoveto{\pgfqpoint{0.000000in}{0.000000in}}%
\pgfpathlineto{\pgfqpoint{4.000000in}{0.000000in}}%
\pgfpathlineto{\pgfqpoint{4.000000in}{3.000000in}}%
\pgfpathlineto{\pgfqpoint{0.000000in}{3.000000in}}%
\pgfpathclose%
\pgfusepath{}%
\end{pgfscope}%
\begin{pgfscope}%
\pgfsetbuttcap%
\pgfsetmiterjoin%
\definecolor{currentfill}{rgb}{1.000000,1.000000,1.000000}%
\pgfsetfillcolor{currentfill}%
\pgfsetlinewidth{0.000000pt}%
\definecolor{currentstroke}{rgb}{0.000000,0.000000,0.000000}%
\pgfsetstrokecolor{currentstroke}%
\pgfsetstrokeopacity{0.000000}%
\pgfsetdash{}{0pt}%
\pgfpathmoveto{\pgfqpoint{0.500000in}{0.375000in}}%
\pgfpathlineto{\pgfqpoint{3.600000in}{0.375000in}}%
\pgfpathlineto{\pgfqpoint{3.600000in}{2.640000in}}%
\pgfpathlineto{\pgfqpoint{0.500000in}{2.640000in}}%
\pgfpathclose%
\pgfusepath{fill}%
\end{pgfscope}%
\begin{pgfscope}%
\pgfpathrectangle{\pgfqpoint{0.500000in}{0.375000in}}{\pgfqpoint{3.100000in}{2.265000in}}%
\pgfusepath{clip}%
\pgfsetbuttcap%
\pgfsetroundjoin%
\definecolor{currentfill}{rgb}{0.121569,0.466667,0.705882}%
\pgfsetfillcolor{currentfill}%
\pgfsetfillopacity{0.200000}%
\pgfsetlinewidth{0.000000pt}%
\definecolor{currentstroke}{rgb}{0.000000,0.000000,0.000000}%
\pgfsetstrokecolor{currentstroke}%
\pgfsetdash{}{0pt}%
\pgfpathmoveto{\pgfqpoint{0.640909in}{1.232288in}}%
\pgfpathlineto{\pgfqpoint{0.640909in}{1.876977in}}%
\pgfpathlineto{\pgfqpoint{1.043506in}{1.929576in}}%
\pgfpathlineto{\pgfqpoint{1.446104in}{1.808042in}}%
\pgfpathlineto{\pgfqpoint{1.848701in}{1.919152in}}%
\pgfpathlineto{\pgfqpoint{2.251299in}{1.875590in}}%
\pgfpathlineto{\pgfqpoint{2.653896in}{1.929225in}}%
\pgfpathlineto{\pgfqpoint{3.056494in}{1.863465in}}%
\pgfpathlineto{\pgfqpoint{3.459091in}{1.849058in}}%
\pgfpathlineto{\pgfqpoint{3.459091in}{1.201416in}}%
\pgfpathlineto{\pgfqpoint{3.459091in}{1.201416in}}%
\pgfpathlineto{\pgfqpoint{3.056494in}{1.174109in}}%
\pgfpathlineto{\pgfqpoint{2.653896in}{1.165199in}}%
\pgfpathlineto{\pgfqpoint{2.251299in}{1.313698in}}%
\pgfpathlineto{\pgfqpoint{1.848701in}{1.208149in}}%
\pgfpathlineto{\pgfqpoint{1.446104in}{1.229532in}}%
\pgfpathlineto{\pgfqpoint{1.043506in}{1.242588in}}%
\pgfpathlineto{\pgfqpoint{0.640909in}{1.232288in}}%
\pgfpathclose%
\pgfusepath{fill}%
\end{pgfscope}%
\begin{pgfscope}%
\pgfpathrectangle{\pgfqpoint{0.500000in}{0.375000in}}{\pgfqpoint{3.100000in}{2.265000in}}%
\pgfusepath{clip}%
\pgfsetbuttcap%
\pgfsetroundjoin%
\definecolor{currentfill}{rgb}{1.000000,0.498039,0.054902}%
\pgfsetfillcolor{currentfill}%
\pgfsetfillopacity{0.200000}%
\pgfsetlinewidth{0.000000pt}%
\definecolor{currentstroke}{rgb}{0.000000,0.000000,0.000000}%
\pgfsetstrokecolor{currentstroke}%
\pgfsetdash{}{0pt}%
\pgfpathmoveto{\pgfqpoint{0.640909in}{1.147052in}}%
\pgfpathlineto{\pgfqpoint{0.640909in}{1.801024in}}%
\pgfpathlineto{\pgfqpoint{1.043506in}{1.814243in}}%
\pgfpathlineto{\pgfqpoint{1.446104in}{1.784968in}}%
\pgfpathlineto{\pgfqpoint{1.848701in}{1.909134in}}%
\pgfpathlineto{\pgfqpoint{2.251299in}{1.764470in}}%
\pgfpathlineto{\pgfqpoint{2.653896in}{1.813514in}}%
\pgfpathlineto{\pgfqpoint{3.056494in}{1.714806in}}%
\pgfpathlineto{\pgfqpoint{3.459091in}{1.773648in}}%
\pgfpathlineto{\pgfqpoint{3.459091in}{1.016664in}}%
\pgfpathlineto{\pgfqpoint{3.459091in}{1.016664in}}%
\pgfpathlineto{\pgfqpoint{3.056494in}{1.088063in}}%
\pgfpathlineto{\pgfqpoint{2.653896in}{1.188901in}}%
\pgfpathlineto{\pgfqpoint{2.251299in}{1.056665in}}%
\pgfpathlineto{\pgfqpoint{1.848701in}{1.163143in}}%
\pgfpathlineto{\pgfqpoint{1.446104in}{1.043244in}}%
\pgfpathlineto{\pgfqpoint{1.043506in}{1.053467in}}%
\pgfpathlineto{\pgfqpoint{0.640909in}{1.147052in}}%
\pgfpathclose%
\pgfusepath{fill}%
\end{pgfscope}%
\begin{pgfscope}%
\pgfsetbuttcap%
\pgfsetroundjoin%
\definecolor{currentfill}{rgb}{0.000000,0.000000,0.000000}%
\pgfsetfillcolor{currentfill}%
\pgfsetlinewidth{0.803000pt}%
\definecolor{currentstroke}{rgb}{0.000000,0.000000,0.000000}%
\pgfsetstrokecolor{currentstroke}%
\pgfsetdash{}{0pt}%
\pgfsys@defobject{currentmarker}{\pgfqpoint{0.000000in}{-0.048611in}}{\pgfqpoint{0.000000in}{0.000000in}}{%
\pgfpathmoveto{\pgfqpoint{0.000000in}{0.000000in}}%
\pgfpathlineto{\pgfqpoint{0.000000in}{-0.048611in}}%
\pgfusepath{stroke,fill}%
}%
\begin{pgfscope}%
\pgfsys@transformshift{0.640909in}{0.375000in}%
\pgfsys@useobject{currentmarker}{}%
\end{pgfscope}%
\end{pgfscope}%
\begin{pgfscope}%
\definecolor{textcolor}{rgb}{0.000000,0.000000,0.000000}%
\pgfsetstrokecolor{textcolor}%
\pgfsetfillcolor{textcolor}%
\pgftext[x=0.640909in,y=0.277778in,,top]{\color{textcolor}\rmfamily\fontsize{10.000000}{12.000000}\selectfont \(\displaystyle {0}\)}%
\end{pgfscope}%
\begin{pgfscope}%
\pgfsetbuttcap%
\pgfsetroundjoin%
\definecolor{currentfill}{rgb}{0.000000,0.000000,0.000000}%
\pgfsetfillcolor{currentfill}%
\pgfsetlinewidth{0.803000pt}%
\definecolor{currentstroke}{rgb}{0.000000,0.000000,0.000000}%
\pgfsetstrokecolor{currentstroke}%
\pgfsetdash{}{0pt}%
\pgfsys@defobject{currentmarker}{\pgfqpoint{0.000000in}{-0.048611in}}{\pgfqpoint{0.000000in}{0.000000in}}{%
\pgfpathmoveto{\pgfqpoint{0.000000in}{0.000000in}}%
\pgfpathlineto{\pgfqpoint{0.000000in}{-0.048611in}}%
\pgfusepath{stroke,fill}%
}%
\begin{pgfscope}%
\pgfsys@transformshift{1.446104in}{0.375000in}%
\pgfsys@useobject{currentmarker}{}%
\end{pgfscope}%
\end{pgfscope}%
\begin{pgfscope}%
\definecolor{textcolor}{rgb}{0.000000,0.000000,0.000000}%
\pgfsetstrokecolor{textcolor}%
\pgfsetfillcolor{textcolor}%
\pgftext[x=1.446104in,y=0.277778in,,top]{\color{textcolor}\rmfamily\fontsize{10.000000}{12.000000}\selectfont \(\displaystyle {2000}\)}%
\end{pgfscope}%
\begin{pgfscope}%
\pgfsetbuttcap%
\pgfsetroundjoin%
\definecolor{currentfill}{rgb}{0.000000,0.000000,0.000000}%
\pgfsetfillcolor{currentfill}%
\pgfsetlinewidth{0.803000pt}%
\definecolor{currentstroke}{rgb}{0.000000,0.000000,0.000000}%
\pgfsetstrokecolor{currentstroke}%
\pgfsetdash{}{0pt}%
\pgfsys@defobject{currentmarker}{\pgfqpoint{0.000000in}{-0.048611in}}{\pgfqpoint{0.000000in}{0.000000in}}{%
\pgfpathmoveto{\pgfqpoint{0.000000in}{0.000000in}}%
\pgfpathlineto{\pgfqpoint{0.000000in}{-0.048611in}}%
\pgfusepath{stroke,fill}%
}%
\begin{pgfscope}%
\pgfsys@transformshift{2.251299in}{0.375000in}%
\pgfsys@useobject{currentmarker}{}%
\end{pgfscope}%
\end{pgfscope}%
\begin{pgfscope}%
\definecolor{textcolor}{rgb}{0.000000,0.000000,0.000000}%
\pgfsetstrokecolor{textcolor}%
\pgfsetfillcolor{textcolor}%
\pgftext[x=2.251299in,y=0.277778in,,top]{\color{textcolor}\rmfamily\fontsize{10.000000}{12.000000}\selectfont \(\displaystyle {4000}\)}%
\end{pgfscope}%
\begin{pgfscope}%
\pgfsetbuttcap%
\pgfsetroundjoin%
\definecolor{currentfill}{rgb}{0.000000,0.000000,0.000000}%
\pgfsetfillcolor{currentfill}%
\pgfsetlinewidth{0.803000pt}%
\definecolor{currentstroke}{rgb}{0.000000,0.000000,0.000000}%
\pgfsetstrokecolor{currentstroke}%
\pgfsetdash{}{0pt}%
\pgfsys@defobject{currentmarker}{\pgfqpoint{0.000000in}{-0.048611in}}{\pgfqpoint{0.000000in}{0.000000in}}{%
\pgfpathmoveto{\pgfqpoint{0.000000in}{0.000000in}}%
\pgfpathlineto{\pgfqpoint{0.000000in}{-0.048611in}}%
\pgfusepath{stroke,fill}%
}%
\begin{pgfscope}%
\pgfsys@transformshift{3.056494in}{0.375000in}%
\pgfsys@useobject{currentmarker}{}%
\end{pgfscope}%
\end{pgfscope}%
\begin{pgfscope}%
\definecolor{textcolor}{rgb}{0.000000,0.000000,0.000000}%
\pgfsetstrokecolor{textcolor}%
\pgfsetfillcolor{textcolor}%
\pgftext[x=3.056494in,y=0.277778in,,top]{\color{textcolor}\rmfamily\fontsize{10.000000}{12.000000}\selectfont \(\displaystyle {6000}\)}%
\end{pgfscope}%
\begin{pgfscope}%
\definecolor{textcolor}{rgb}{0.000000,0.000000,0.000000}%
\pgfsetstrokecolor{textcolor}%
\pgfsetfillcolor{textcolor}%
\pgftext[x=2.050000in,y=0.098766in,,top]{\color{textcolor}\rmfamily\fontsize{10.000000}{12.000000}\selectfont Shift}%
\end{pgfscope}%
\begin{pgfscope}%
\pgfsetbuttcap%
\pgfsetroundjoin%
\definecolor{currentfill}{rgb}{0.000000,0.000000,0.000000}%
\pgfsetfillcolor{currentfill}%
\pgfsetlinewidth{0.803000pt}%
\definecolor{currentstroke}{rgb}{0.000000,0.000000,0.000000}%
\pgfsetstrokecolor{currentstroke}%
\pgfsetdash{}{0pt}%
\pgfsys@defobject{currentmarker}{\pgfqpoint{-0.048611in}{0.000000in}}{\pgfqpoint{-0.000000in}{0.000000in}}{%
\pgfpathmoveto{\pgfqpoint{-0.000000in}{0.000000in}}%
\pgfpathlineto{\pgfqpoint{-0.048611in}{0.000000in}}%
\pgfusepath{stroke,fill}%
}%
\begin{pgfscope}%
\pgfsys@transformshift{0.500000in}{0.375000in}%
\pgfsys@useobject{currentmarker}{}%
\end{pgfscope}%
\end{pgfscope}%
\begin{pgfscope}%
\definecolor{textcolor}{rgb}{0.000000,0.000000,0.000000}%
\pgfsetstrokecolor{textcolor}%
\pgfsetfillcolor{textcolor}%
\pgftext[x=0.155863in, y=0.326775in, left, base]{\color{textcolor}\rmfamily\fontsize{10.000000}{12.000000}\selectfont \(\displaystyle {0.65}\)}%
\end{pgfscope}%
\begin{pgfscope}%
\pgfsetbuttcap%
\pgfsetroundjoin%
\definecolor{currentfill}{rgb}{0.000000,0.000000,0.000000}%
\pgfsetfillcolor{currentfill}%
\pgfsetlinewidth{0.803000pt}%
\definecolor{currentstroke}{rgb}{0.000000,0.000000,0.000000}%
\pgfsetstrokecolor{currentstroke}%
\pgfsetdash{}{0pt}%
\pgfsys@defobject{currentmarker}{\pgfqpoint{-0.048611in}{0.000000in}}{\pgfqpoint{-0.000000in}{0.000000in}}{%
\pgfpathmoveto{\pgfqpoint{-0.000000in}{0.000000in}}%
\pgfpathlineto{\pgfqpoint{-0.048611in}{0.000000in}}%
\pgfusepath{stroke,fill}%
}%
\begin{pgfscope}%
\pgfsys@transformshift{0.500000in}{0.677000in}%
\pgfsys@useobject{currentmarker}{}%
\end{pgfscope}%
\end{pgfscope}%
\begin{pgfscope}%
\definecolor{textcolor}{rgb}{0.000000,0.000000,0.000000}%
\pgfsetstrokecolor{textcolor}%
\pgfsetfillcolor{textcolor}%
\pgftext[x=0.155863in, y=0.628775in, left, base]{\color{textcolor}\rmfamily\fontsize{10.000000}{12.000000}\selectfont \(\displaystyle {0.66}\)}%
\end{pgfscope}%
\begin{pgfscope}%
\pgfsetbuttcap%
\pgfsetroundjoin%
\definecolor{currentfill}{rgb}{0.000000,0.000000,0.000000}%
\pgfsetfillcolor{currentfill}%
\pgfsetlinewidth{0.803000pt}%
\definecolor{currentstroke}{rgb}{0.000000,0.000000,0.000000}%
\pgfsetstrokecolor{currentstroke}%
\pgfsetdash{}{0pt}%
\pgfsys@defobject{currentmarker}{\pgfqpoint{-0.048611in}{0.000000in}}{\pgfqpoint{-0.000000in}{0.000000in}}{%
\pgfpathmoveto{\pgfqpoint{-0.000000in}{0.000000in}}%
\pgfpathlineto{\pgfqpoint{-0.048611in}{0.000000in}}%
\pgfusepath{stroke,fill}%
}%
\begin{pgfscope}%
\pgfsys@transformshift{0.500000in}{0.979000in}%
\pgfsys@useobject{currentmarker}{}%
\end{pgfscope}%
\end{pgfscope}%
\begin{pgfscope}%
\definecolor{textcolor}{rgb}{0.000000,0.000000,0.000000}%
\pgfsetstrokecolor{textcolor}%
\pgfsetfillcolor{textcolor}%
\pgftext[x=0.155863in, y=0.930775in, left, base]{\color{textcolor}\rmfamily\fontsize{10.000000}{12.000000}\selectfont \(\displaystyle {0.67}\)}%
\end{pgfscope}%
\begin{pgfscope}%
\pgfsetbuttcap%
\pgfsetroundjoin%
\definecolor{currentfill}{rgb}{0.000000,0.000000,0.000000}%
\pgfsetfillcolor{currentfill}%
\pgfsetlinewidth{0.803000pt}%
\definecolor{currentstroke}{rgb}{0.000000,0.000000,0.000000}%
\pgfsetstrokecolor{currentstroke}%
\pgfsetdash{}{0pt}%
\pgfsys@defobject{currentmarker}{\pgfqpoint{-0.048611in}{0.000000in}}{\pgfqpoint{-0.000000in}{0.000000in}}{%
\pgfpathmoveto{\pgfqpoint{-0.000000in}{0.000000in}}%
\pgfpathlineto{\pgfqpoint{-0.048611in}{0.000000in}}%
\pgfusepath{stroke,fill}%
}%
\begin{pgfscope}%
\pgfsys@transformshift{0.500000in}{1.281000in}%
\pgfsys@useobject{currentmarker}{}%
\end{pgfscope}%
\end{pgfscope}%
\begin{pgfscope}%
\definecolor{textcolor}{rgb}{0.000000,0.000000,0.000000}%
\pgfsetstrokecolor{textcolor}%
\pgfsetfillcolor{textcolor}%
\pgftext[x=0.155863in, y=1.232775in, left, base]{\color{textcolor}\rmfamily\fontsize{10.000000}{12.000000}\selectfont \(\displaystyle {0.68}\)}%
\end{pgfscope}%
\begin{pgfscope}%
\pgfsetbuttcap%
\pgfsetroundjoin%
\definecolor{currentfill}{rgb}{0.000000,0.000000,0.000000}%
\pgfsetfillcolor{currentfill}%
\pgfsetlinewidth{0.803000pt}%
\definecolor{currentstroke}{rgb}{0.000000,0.000000,0.000000}%
\pgfsetstrokecolor{currentstroke}%
\pgfsetdash{}{0pt}%
\pgfsys@defobject{currentmarker}{\pgfqpoint{-0.048611in}{0.000000in}}{\pgfqpoint{-0.000000in}{0.000000in}}{%
\pgfpathmoveto{\pgfqpoint{-0.000000in}{0.000000in}}%
\pgfpathlineto{\pgfqpoint{-0.048611in}{0.000000in}}%
\pgfusepath{stroke,fill}%
}%
\begin{pgfscope}%
\pgfsys@transformshift{0.500000in}{1.583000in}%
\pgfsys@useobject{currentmarker}{}%
\end{pgfscope}%
\end{pgfscope}%
\begin{pgfscope}%
\definecolor{textcolor}{rgb}{0.000000,0.000000,0.000000}%
\pgfsetstrokecolor{textcolor}%
\pgfsetfillcolor{textcolor}%
\pgftext[x=0.155863in, y=1.534775in, left, base]{\color{textcolor}\rmfamily\fontsize{10.000000}{12.000000}\selectfont \(\displaystyle {0.69}\)}%
\end{pgfscope}%
\begin{pgfscope}%
\pgfsetbuttcap%
\pgfsetroundjoin%
\definecolor{currentfill}{rgb}{0.000000,0.000000,0.000000}%
\pgfsetfillcolor{currentfill}%
\pgfsetlinewidth{0.803000pt}%
\definecolor{currentstroke}{rgb}{0.000000,0.000000,0.000000}%
\pgfsetstrokecolor{currentstroke}%
\pgfsetdash{}{0pt}%
\pgfsys@defobject{currentmarker}{\pgfqpoint{-0.048611in}{0.000000in}}{\pgfqpoint{-0.000000in}{0.000000in}}{%
\pgfpathmoveto{\pgfqpoint{-0.000000in}{0.000000in}}%
\pgfpathlineto{\pgfqpoint{-0.048611in}{0.000000in}}%
\pgfusepath{stroke,fill}%
}%
\begin{pgfscope}%
\pgfsys@transformshift{0.500000in}{1.885000in}%
\pgfsys@useobject{currentmarker}{}%
\end{pgfscope}%
\end{pgfscope}%
\begin{pgfscope}%
\definecolor{textcolor}{rgb}{0.000000,0.000000,0.000000}%
\pgfsetstrokecolor{textcolor}%
\pgfsetfillcolor{textcolor}%
\pgftext[x=0.155863in, y=1.836775in, left, base]{\color{textcolor}\rmfamily\fontsize{10.000000}{12.000000}\selectfont \(\displaystyle {0.70}\)}%
\end{pgfscope}%
\begin{pgfscope}%
\pgfsetbuttcap%
\pgfsetroundjoin%
\definecolor{currentfill}{rgb}{0.000000,0.000000,0.000000}%
\pgfsetfillcolor{currentfill}%
\pgfsetlinewidth{0.803000pt}%
\definecolor{currentstroke}{rgb}{0.000000,0.000000,0.000000}%
\pgfsetstrokecolor{currentstroke}%
\pgfsetdash{}{0pt}%
\pgfsys@defobject{currentmarker}{\pgfqpoint{-0.048611in}{0.000000in}}{\pgfqpoint{-0.000000in}{0.000000in}}{%
\pgfpathmoveto{\pgfqpoint{-0.000000in}{0.000000in}}%
\pgfpathlineto{\pgfqpoint{-0.048611in}{0.000000in}}%
\pgfusepath{stroke,fill}%
}%
\begin{pgfscope}%
\pgfsys@transformshift{0.500000in}{2.187000in}%
\pgfsys@useobject{currentmarker}{}%
\end{pgfscope}%
\end{pgfscope}%
\begin{pgfscope}%
\definecolor{textcolor}{rgb}{0.000000,0.000000,0.000000}%
\pgfsetstrokecolor{textcolor}%
\pgfsetfillcolor{textcolor}%
\pgftext[x=0.155863in, y=2.138775in, left, base]{\color{textcolor}\rmfamily\fontsize{10.000000}{12.000000}\selectfont \(\displaystyle {0.71}\)}%
\end{pgfscope}%
\begin{pgfscope}%
\pgfsetbuttcap%
\pgfsetroundjoin%
\definecolor{currentfill}{rgb}{0.000000,0.000000,0.000000}%
\pgfsetfillcolor{currentfill}%
\pgfsetlinewidth{0.803000pt}%
\definecolor{currentstroke}{rgb}{0.000000,0.000000,0.000000}%
\pgfsetstrokecolor{currentstroke}%
\pgfsetdash{}{0pt}%
\pgfsys@defobject{currentmarker}{\pgfqpoint{-0.048611in}{0.000000in}}{\pgfqpoint{-0.000000in}{0.000000in}}{%
\pgfpathmoveto{\pgfqpoint{-0.000000in}{0.000000in}}%
\pgfpathlineto{\pgfqpoint{-0.048611in}{0.000000in}}%
\pgfusepath{stroke,fill}%
}%
\begin{pgfscope}%
\pgfsys@transformshift{0.500000in}{2.489000in}%
\pgfsys@useobject{currentmarker}{}%
\end{pgfscope}%
\end{pgfscope}%
\begin{pgfscope}%
\definecolor{textcolor}{rgb}{0.000000,0.000000,0.000000}%
\pgfsetstrokecolor{textcolor}%
\pgfsetfillcolor{textcolor}%
\pgftext[x=0.155863in, y=2.440775in, left, base]{\color{textcolor}\rmfamily\fontsize{10.000000}{12.000000}\selectfont \(\displaystyle {0.72}\)}%
\end{pgfscope}%
\begin{pgfscope}%
\definecolor{textcolor}{rgb}{0.000000,0.000000,0.000000}%
\pgfsetstrokecolor{textcolor}%
\pgfsetfillcolor{textcolor}%
\pgftext[x=0.12in,y=1.507500in,,bottom,rotate=90.000000]{\color{textcolor}\rmfamily\fontsize{10.000000}{12.000000}\selectfont Accuracy}%
\end{pgfscope}%
\begin{pgfscope}%
\pgfpathrectangle{\pgfqpoint{0.500000in}{0.375000in}}{\pgfqpoint{3.100000in}{2.265000in}}%
\pgfusepath{clip}%
\pgfsetrectcap%
\pgfsetroundjoin%
\pgfsetlinewidth{1.505625pt}%
\definecolor{currentstroke}{rgb}{0.121569,0.466667,0.705882}%
\pgfsetstrokecolor{currentstroke}%
\pgfsetdash{}{0pt}%
\pgfpathmoveto{\pgfqpoint{0.640909in}{1.554632in}}%
\pgfpathlineto{\pgfqpoint{1.043506in}{1.586082in}}%
\pgfpathlineto{\pgfqpoint{1.446104in}{1.518787in}}%
\pgfpathlineto{\pgfqpoint{1.848701in}{1.563651in}}%
\pgfpathlineto{\pgfqpoint{2.251299in}{1.594644in}}%
\pgfpathlineto{\pgfqpoint{2.653896in}{1.547212in}}%
\pgfpathlineto{\pgfqpoint{3.056494in}{1.518787in}}%
\pgfpathlineto{\pgfqpoint{3.459091in}{1.525237in}}%
\pgfusepath{stroke}%
\end{pgfscope}%
\begin{pgfscope}%
\pgfpathrectangle{\pgfqpoint{0.500000in}{0.375000in}}{\pgfqpoint{3.100000in}{2.265000in}}%
\pgfusepath{clip}%
\pgfsetrectcap%
\pgfsetroundjoin%
\pgfsetlinewidth{1.505625pt}%
\definecolor{currentstroke}{rgb}{1.000000,0.498039,0.054902}%
\pgfsetstrokecolor{currentstroke}%
\pgfsetdash{}{0pt}%
\pgfpathmoveto{\pgfqpoint{0.640909in}{1.474038in}}%
\pgfpathlineto{\pgfqpoint{1.043506in}{1.433855in}}%
\pgfpathlineto{\pgfqpoint{1.446104in}{1.414106in}}%
\pgfpathlineto{\pgfqpoint{1.848701in}{1.536139in}}%
\pgfpathlineto{\pgfqpoint{2.251299in}{1.410567in}}%
\pgfpathlineto{\pgfqpoint{2.653896in}{1.501207in}}%
\pgfpathlineto{\pgfqpoint{3.056494in}{1.401435in}}%
\pgfpathlineto{\pgfqpoint{3.459091in}{1.395156in}}%
\pgfusepath{stroke}%
\end{pgfscope}%
\begin{pgfscope}%
\pgfsetrectcap%
\pgfsetmiterjoin%
\pgfsetlinewidth{0.803000pt}%
\definecolor{currentstroke}{rgb}{0.000000,0.000000,0.000000}%
\pgfsetstrokecolor{currentstroke}%
\pgfsetdash{}{0pt}%
\pgfpathmoveto{\pgfqpoint{0.500000in}{0.375000in}}%
\pgfpathlineto{\pgfqpoint{0.500000in}{2.640000in}}%
\pgfusepath{stroke}%
\end{pgfscope}%
\begin{pgfscope}%
\pgfsetrectcap%
\pgfsetmiterjoin%
\pgfsetlinewidth{0.803000pt}%
\definecolor{currentstroke}{rgb}{0.000000,0.000000,0.000000}%
\pgfsetstrokecolor{currentstroke}%
\pgfsetdash{}{0pt}%
\pgfpathmoveto{\pgfqpoint{3.600000in}{0.375000in}}%
\pgfpathlineto{\pgfqpoint{3.600000in}{2.640000in}}%
\pgfusepath{stroke}%
\end{pgfscope}%
\begin{pgfscope}%
\pgfsetrectcap%
\pgfsetmiterjoin%
\pgfsetlinewidth{0.803000pt}%
\definecolor{currentstroke}{rgb}{0.000000,0.000000,0.000000}%
\pgfsetstrokecolor{currentstroke}%
\pgfsetdash{}{0pt}%
\pgfpathmoveto{\pgfqpoint{0.500000in}{0.375000in}}%
\pgfpathlineto{\pgfqpoint{3.600000in}{0.375000in}}%
\pgfusepath{stroke}%
\end{pgfscope}%
\begin{pgfscope}%
\pgfsetrectcap%
\pgfsetmiterjoin%
\pgfsetlinewidth{0.803000pt}%
\definecolor{currentstroke}{rgb}{0.000000,0.000000,0.000000}%
\pgfsetstrokecolor{currentstroke}%
\pgfsetdash{}{0pt}%
\pgfpathmoveto{\pgfqpoint{0.500000in}{2.640000in}}%
\pgfpathlineto{\pgfqpoint{3.600000in}{2.640000in}}%
\pgfusepath{stroke}%
\end{pgfscope}%
\begin{pgfscope}%
\pgfsetbuttcap%
\pgfsetmiterjoin%
\definecolor{currentfill}{rgb}{1.000000,1.000000,1.000000}%
\pgfsetfillcolor{currentfill}%
\pgfsetfillopacity{0.800000}%
\pgfsetlinewidth{1.003750pt}%
\definecolor{currentstroke}{rgb}{0.800000,0.800000,0.800000}%
\pgfsetstrokecolor{currentstroke}%
\pgfsetstrokeopacity{0.800000}%
\pgfsetdash{}{0pt}%
\pgfpathmoveto{\pgfqpoint{2.475385in}{2.141543in}}%
\pgfpathlineto{\pgfqpoint{3.502778in}{2.141543in}}%
\pgfpathquadraticcurveto{\pgfqpoint{3.530556in}{2.141543in}}{\pgfqpoint{3.530556in}{2.169321in}}%
\pgfpathlineto{\pgfqpoint{3.530556in}{2.542778in}}%
\pgfpathquadraticcurveto{\pgfqpoint{3.530556in}{2.570556in}}{\pgfqpoint{3.502778in}{2.570556in}}%
\pgfpathlineto{\pgfqpoint{2.475385in}{2.570556in}}%
\pgfpathquadraticcurveto{\pgfqpoint{2.447607in}{2.570556in}}{\pgfqpoint{2.447607in}{2.542778in}}%
\pgfpathlineto{\pgfqpoint{2.447607in}{2.169321in}}%
\pgfpathquadraticcurveto{\pgfqpoint{2.447607in}{2.141543in}}{\pgfqpoint{2.475385in}{2.141543in}}%
\pgfpathclose%
\pgfusepath{stroke,fill}%
\end{pgfscope}%
\begin{pgfscope}%
\pgfsetrectcap%
\pgfsetroundjoin%
\pgfsetlinewidth{1.505625pt}%
\definecolor{currentstroke}{rgb}{0.121569,0.466667,0.705882}%
\pgfsetstrokecolor{currentstroke}%
\pgfsetdash{}{0pt}%
\pgfpathmoveto{\pgfqpoint{2.503162in}{2.466389in}}%
\pgfpathlineto{\pgfqpoint{2.780940in}{2.466389in}}%
\pgfusepath{stroke}%
\end{pgfscope}%
\begin{pgfscope}%
\definecolor{textcolor}{rgb}{0.000000,0.000000,0.000000}%
\pgfsetstrokecolor{textcolor}%
\pgfsetfillcolor{textcolor}%
\pgftext[x=2.892051in,y=2.417778in,left,base]{\color{textcolor}\rmfamily\fontsize{10.000000}{12.000000}\selectfont waveform}%
\end{pgfscope}%
\begin{pgfscope}%
\pgfsetrectcap%
\pgfsetroundjoin%
\pgfsetlinewidth{1.505625pt}%
\definecolor{currentstroke}{rgb}{1.000000,0.498039,0.054902}%
\pgfsetstrokecolor{currentstroke}%
\pgfsetdash{}{0pt}%
\pgfpathmoveto{\pgfqpoint{2.503162in}{2.272716in}}%
\pgfpathlineto{\pgfqpoint{2.780940in}{2.272716in}}%
\pgfusepath{stroke}%
\end{pgfscope}%
\begin{pgfscope}%
\definecolor{textcolor}{rgb}{0.000000,0.000000,0.000000}%
\pgfsetstrokecolor{textcolor}%
\pgfsetfillcolor{textcolor}%
\pgftext[x=2.892051in,y=2.224105in,left,base]{\color{textcolor}\rmfamily\fontsize{10.000000}{12.000000}\selectfont frequency}%
\end{pgfscope}%
\end{pgfpicture}%
\makeatother%
\endgroup%

%% file: figures/accuracy_vs_iterations.pgf
\begingroup%
\makeatletter%
\begin{pgfpicture}%
\pgfpathrectangle{\pgfpointorigin}{\pgfqpoint{4.000000in}{3.000000in}}%
\pgfusepath{use as bounding box, clip}%
\begin{pgfscope}%
\pgfsetbuttcap%
\pgfsetmiterjoin%
\pgfsetlinewidth{0.000000pt}%
\definecolor{currentstroke}{rgb}{1.000000,1.000000,1.000000}%
\pgfsetstrokecolor{currentstroke}%
\pgfsetstrokeopacity{0.000000}%
\pgfsetdash{}{0pt}%
\pgfpathmoveto{\pgfqpoint{0.000000in}{0.000000in}}%
\pgfpathlineto{\pgfqpoint{4.000000in}{0.000000in}}%
\pgfpathlineto{\pgfqpoint{4.000000in}{3.000000in}}%
\pgfpathlineto{\pgfqpoint{0.000000in}{3.000000in}}%
\pgfpathclose%
\pgfusepath{}%
\end{pgfscope}%
\begin{pgfscope}%
\pgfsetbuttcap%
\pgfsetmiterjoin%
\definecolor{currentfill}{rgb}{1.000000,1.000000,1.000000}%
\pgfsetfillcolor{currentfill}%
\pgfsetlinewidth{0.000000pt}%
\definecolor{currentstroke}{rgb}{0.000000,0.000000,0.000000}%
\pgfsetstrokecolor{currentstroke}%
\pgfsetstrokeopacity{0.000000}%
\pgfsetdash{}{0pt}%
\pgfpathmoveto{\pgfqpoint{0.500000in}{0.375000in}}%
\pgfpathlineto{\pgfqpoint{3.600000in}{0.375000in}}%
\pgfpathlineto{\pgfqpoint{3.600000in}{2.640000in}}%
\pgfpathlineto{\pgfqpoint{0.500000in}{2.640000in}}%
\pgfpathclose%
\pgfusepath{fill}%
\end{pgfscope}%
\begin{pgfscope}%
\pgfpathrectangle{\pgfqpoint{0.500000in}{0.375000in}}{\pgfqpoint{3.100000in}{2.265000in}}%
\pgfusepath{clip}%
\pgfsetbuttcap%
\pgfsetroundjoin%
\definecolor{currentfill}{rgb}{0.121569,0.466667,0.705882}%
\pgfsetfillcolor{currentfill}%
\pgfsetfillopacity{0.200000}%
\pgfsetlinewidth{0.000000pt}%
\definecolor{currentstroke}{rgb}{0.000000,0.000000,0.000000}%
\pgfsetstrokecolor{currentstroke}%
\pgfsetdash{}{0pt}%
\pgfpathmoveto{\pgfqpoint{0.640909in}{2.537045in}}%
\pgfpathlineto{\pgfqpoint{0.640909in}{1.733159in}}%
\pgfpathlineto{\pgfqpoint{0.993182in}{1.326108in}}%
\pgfpathlineto{\pgfqpoint{1.345455in}{1.143515in}}%
\pgfpathlineto{\pgfqpoint{1.697727in}{0.995463in}}%
\pgfpathlineto{\pgfqpoint{2.050000in}{0.846392in}}%
\pgfpathlineto{\pgfqpoint{2.402273in}{0.749792in}}%
\pgfpathlineto{\pgfqpoint{2.754545in}{0.673493in}}%
\pgfpathlineto{\pgfqpoint{3.106818in}{0.730877in}}%
\pgfpathlineto{\pgfqpoint{3.459091in}{0.666649in}}%
\pgfpathlineto{\pgfqpoint{3.459091in}{1.003255in}}%
\pgfpathlineto{\pgfqpoint{3.459091in}{1.003255in}}%
\pgfpathlineto{\pgfqpoint{3.106818in}{1.029088in}}%
\pgfpathlineto{\pgfqpoint{2.754545in}{1.138542in}}%
\pgfpathlineto{\pgfqpoint{2.402273in}{1.198404in}}%
\pgfpathlineto{\pgfqpoint{2.050000in}{1.294665in}}%
\pgfpathlineto{\pgfqpoint{1.697727in}{1.506495in}}%
\pgfpathlineto{\pgfqpoint{1.345455in}{1.613666in}}%
\pgfpathlineto{\pgfqpoint{0.993182in}{1.916593in}}%
\pgfpathlineto{\pgfqpoint{0.640909in}{2.537045in}}%
\pgfpathclose%
\pgfusepath{fill}%
\end{pgfscope}%
\begin{pgfscope}%
\pgfpathrectangle{\pgfqpoint{0.500000in}{0.375000in}}{\pgfqpoint{3.100000in}{2.265000in}}%
\pgfusepath{clip}%
\pgfsetbuttcap%
\pgfsetroundjoin%
\definecolor{currentfill}{rgb}{1.000000,0.498039,0.054902}%
\pgfsetfillcolor{currentfill}%
\pgfsetfillopacity{0.200000}%
\pgfsetlinewidth{0.000000pt}%
\definecolor{currentstroke}{rgb}{0.000000,0.000000,0.000000}%
\pgfsetstrokecolor{currentstroke}%
\pgfsetdash{}{0pt}%
\pgfpathmoveto{\pgfqpoint{0.640909in}{1.172464in}}%
\pgfpathlineto{\pgfqpoint{0.640909in}{0.617145in}}%
\pgfpathlineto{\pgfqpoint{0.993182in}{0.553296in}}%
\pgfpathlineto{\pgfqpoint{1.345455in}{0.477955in}}%
\pgfpathlineto{\pgfqpoint{1.697727in}{0.561468in}}%
\pgfpathlineto{\pgfqpoint{2.050000in}{0.564808in}}%
\pgfpathlineto{\pgfqpoint{2.402273in}{0.540375in}}%
\pgfpathlineto{\pgfqpoint{2.754545in}{0.561007in}}%
\pgfpathlineto{\pgfqpoint{3.106818in}{0.558859in}}%
\pgfpathlineto{\pgfqpoint{3.459091in}{0.526752in}}%
\pgfpathlineto{\pgfqpoint{3.459091in}{1.112328in}}%
\pgfpathlineto{\pgfqpoint{3.459091in}{1.112328in}}%
\pgfpathlineto{\pgfqpoint{3.106818in}{1.095802in}}%
\pgfpathlineto{\pgfqpoint{2.754545in}{1.389602in}}%
\pgfpathlineto{\pgfqpoint{2.402273in}{1.419633in}}%
\pgfpathlineto{\pgfqpoint{2.050000in}{1.264291in}}%
\pgfpathlineto{\pgfqpoint{1.697727in}{1.259627in}}%
\pgfpathlineto{\pgfqpoint{1.345455in}{1.318302in}}%
\pgfpathlineto{\pgfqpoint{0.993182in}{1.257505in}}%
\pgfpathlineto{\pgfqpoint{0.640909in}{1.172464in}}%
\pgfpathclose%
\pgfusepath{fill}%
\end{pgfscope}%
\begin{pgfscope}%
\pgfsetbuttcap%
\pgfsetroundjoin%
\definecolor{currentfill}{rgb}{0.000000,0.000000,0.000000}%
\pgfsetfillcolor{currentfill}%
\pgfsetlinewidth{0.803000pt}%
\definecolor{currentstroke}{rgb}{0.000000,0.000000,0.000000}%
\pgfsetstrokecolor{currentstroke}%
\pgfsetdash{}{0pt}%
\pgfsys@defobject{currentmarker}{\pgfqpoint{0.000000in}{-0.048611in}}{\pgfqpoint{0.000000in}{0.000000in}}{%
\pgfpathmoveto{\pgfqpoint{0.000000in}{0.000000in}}%
\pgfpathlineto{\pgfqpoint{0.000000in}{-0.048611in}}%
\pgfusepath{stroke,fill}%
}%
\begin{pgfscope}%
\pgfsys@transformshift{0.993182in}{0.375000in}%
\pgfsys@useobject{currentmarker}{}%
\end{pgfscope}%
\end{pgfscope}%
\begin{pgfscope}%
\definecolor{textcolor}{rgb}{0.000000,0.000000,0.000000}%
\pgfsetstrokecolor{textcolor}%
\pgfsetfillcolor{textcolor}%
\pgftext[x=0.993182in,y=0.277778in,,top]{\color{textcolor}\rmfamily\fontsize{10.000000}{12.000000}\selectfont \(\displaystyle {2}\)}%
\end{pgfscope}%
\begin{pgfscope}%
\pgfsetbuttcap%
\pgfsetroundjoin%
\definecolor{currentfill}{rgb}{0.000000,0.000000,0.000000}%
\pgfsetfillcolor{currentfill}%
\pgfsetlinewidth{0.803000pt}%
\definecolor{currentstroke}{rgb}{0.000000,0.000000,0.000000}%
\pgfsetstrokecolor{currentstroke}%
\pgfsetdash{}{0pt}%
\pgfsys@defobject{currentmarker}{\pgfqpoint{0.000000in}{-0.048611in}}{\pgfqpoint{0.000000in}{0.000000in}}{%
\pgfpathmoveto{\pgfqpoint{0.000000in}{0.000000in}}%
\pgfpathlineto{\pgfqpoint{0.000000in}{-0.048611in}}%
\pgfusepath{stroke,fill}%
}%
\begin{pgfscope}%
\pgfsys@transformshift{1.697727in}{0.375000in}%
\pgfsys@useobject{currentmarker}{}%
\end{pgfscope}%
\end{pgfscope}%
\begin{pgfscope}%
\definecolor{textcolor}{rgb}{0.000000,0.000000,0.000000}%
\pgfsetstrokecolor{textcolor}%
\pgfsetfillcolor{textcolor}%
\pgftext[x=1.697727in,y=0.277778in,,top]{\color{textcolor}\rmfamily\fontsize{10.000000}{12.000000}\selectfont \(\displaystyle {4}\)}%
\end{pgfscope}%
\begin{pgfscope}%
\pgfsetbuttcap%
\pgfsetroundjoin%
\definecolor{currentfill}{rgb}{0.000000,0.000000,0.000000}%
\pgfsetfillcolor{currentfill}%
\pgfsetlinewidth{0.803000pt}%
\definecolor{currentstroke}{rgb}{0.000000,0.000000,0.000000}%
\pgfsetstrokecolor{currentstroke}%
\pgfsetdash{}{0pt}%
\pgfsys@defobject{currentmarker}{\pgfqpoint{0.000000in}{-0.048611in}}{\pgfqpoint{0.000000in}{0.000000in}}{%
\pgfpathmoveto{\pgfqpoint{0.000000in}{0.000000in}}%
\pgfpathlineto{\pgfqpoint{0.000000in}{-0.048611in}}%
\pgfusepath{stroke,fill}%
}%
\begin{pgfscope}%
\pgfsys@transformshift{2.402273in}{0.375000in}%
\pgfsys@useobject{currentmarker}{}%
\end{pgfscope}%
\end{pgfscope}%
\begin{pgfscope}%
\definecolor{textcolor}{rgb}{0.000000,0.000000,0.000000}%
\pgfsetstrokecolor{textcolor}%
\pgfsetfillcolor{textcolor}%
\pgftext[x=2.402273in,y=0.277778in,,top]{\color{textcolor}\rmfamily\fontsize{10.000000}{12.000000}\selectfont \(\displaystyle {6}\)}%
\end{pgfscope}%
\begin{pgfscope}%
\pgfsetbuttcap%
\pgfsetroundjoin%
\definecolor{currentfill}{rgb}{0.000000,0.000000,0.000000}%
\pgfsetfillcolor{currentfill}%
\pgfsetlinewidth{0.803000pt}%
\definecolor{currentstroke}{rgb}{0.000000,0.000000,0.000000}%
\pgfsetstrokecolor{currentstroke}%
\pgfsetdash{}{0pt}%
\pgfsys@defobject{currentmarker}{\pgfqpoint{0.000000in}{-0.048611in}}{\pgfqpoint{0.000000in}{0.000000in}}{%
\pgfpathmoveto{\pgfqpoint{0.000000in}{0.000000in}}%
\pgfpathlineto{\pgfqpoint{0.000000in}{-0.048611in}}%
\pgfusepath{stroke,fill}%
}%
\begin{pgfscope}%
\pgfsys@transformshift{3.106818in}{0.375000in}%
\pgfsys@useobject{currentmarker}{}%
\end{pgfscope}%
\end{pgfscope}%
\begin{pgfscope}%
\definecolor{textcolor}{rgb}{0.000000,0.000000,0.000000}%
\pgfsetstrokecolor{textcolor}%
\pgfsetfillcolor{textcolor}%
\pgftext[x=3.106818in,y=0.277778in,,top]{\color{textcolor}\rmfamily\fontsize{10.000000}{12.000000}\selectfont \(\displaystyle {8}\)}%
\end{pgfscope}%
\begin{pgfscope}%
\definecolor{textcolor}{rgb}{0.000000,0.000000,0.000000}%
\pgfsetstrokecolor{textcolor}%
\pgfsetfillcolor{textcolor}%
\pgftext[x=2.050000in,y=0.098766in,,top]{\color{textcolor}\rmfamily\fontsize{10.000000}{12.000000}\selectfont Iteration}%
\end{pgfscope}%
\begin{pgfscope}%
\pgfsetbuttcap%
\pgfsetroundjoin%
\definecolor{currentfill}{rgb}{0.000000,0.000000,0.000000}%
\pgfsetfillcolor{currentfill}%
\pgfsetlinewidth{0.803000pt}%
\definecolor{currentstroke}{rgb}{0.000000,0.000000,0.000000}%
\pgfsetstrokecolor{currentstroke}%
\pgfsetdash{}{0pt}%
\pgfsys@defobject{currentmarker}{\pgfqpoint{-0.048611in}{0.000000in}}{\pgfqpoint{-0.000000in}{0.000000in}}{%
\pgfpathmoveto{\pgfqpoint{-0.000000in}{0.000000in}}%
\pgfpathlineto{\pgfqpoint{-0.048611in}{0.000000in}}%
\pgfusepath{stroke,fill}%
}%
\begin{pgfscope}%
\pgfsys@transformshift{0.500000in}{0.620461in}%
\pgfsys@useobject{currentmarker}{}%
\end{pgfscope}%
\end{pgfscope}%
\begin{pgfscope}%
\definecolor{textcolor}{rgb}{0.000000,0.000000,0.000000}%
\pgfsetstrokecolor{textcolor}%
\pgfsetfillcolor{textcolor}%
\pgftext[x=0.225308in, y=0.572236in, left, base]{\color{textcolor}\rmfamily\fontsize{10.000000}{12.000000}\selectfont \(\displaystyle {0.2}\)}%
\end{pgfscope}%
\begin{pgfscope}%
\pgfsetbuttcap%
\pgfsetroundjoin%
\definecolor{currentfill}{rgb}{0.000000,0.000000,0.000000}%
\pgfsetfillcolor{currentfill}%
\pgfsetlinewidth{0.803000pt}%
\definecolor{currentstroke}{rgb}{0.000000,0.000000,0.000000}%
\pgfsetstrokecolor{currentstroke}%
\pgfsetdash{}{0pt}%
\pgfsys@defobject{currentmarker}{\pgfqpoint{-0.048611in}{0.000000in}}{\pgfqpoint{-0.000000in}{0.000000in}}{%
\pgfpathmoveto{\pgfqpoint{-0.000000in}{0.000000in}}%
\pgfpathlineto{\pgfqpoint{-0.048611in}{0.000000in}}%
\pgfusepath{stroke,fill}%
}%
\begin{pgfscope}%
\pgfsys@transformshift{0.500000in}{1.093186in}%
\pgfsys@useobject{currentmarker}{}%
\end{pgfscope}%
\end{pgfscope}%
\begin{pgfscope}%
\definecolor{textcolor}{rgb}{0.000000,0.000000,0.000000}%
\pgfsetstrokecolor{textcolor}%
\pgfsetfillcolor{textcolor}%
\pgftext[x=0.225308in, y=1.044961in, left, base]{\color{textcolor}\rmfamily\fontsize{10.000000}{12.000000}\selectfont \(\displaystyle {0.4}\)}%
\end{pgfscope}%
\begin{pgfscope}%
\pgfsetbuttcap%
\pgfsetroundjoin%
\definecolor{currentfill}{rgb}{0.000000,0.000000,0.000000}%
\pgfsetfillcolor{currentfill}%
\pgfsetlinewidth{0.803000pt}%
\definecolor{currentstroke}{rgb}{0.000000,0.000000,0.000000}%
\pgfsetstrokecolor{currentstroke}%
\pgfsetdash{}{0pt}%
\pgfsys@defobject{currentmarker}{\pgfqpoint{-0.048611in}{0.000000in}}{\pgfqpoint{-0.000000in}{0.000000in}}{%
\pgfpathmoveto{\pgfqpoint{-0.000000in}{0.000000in}}%
\pgfpathlineto{\pgfqpoint{-0.048611in}{0.000000in}}%
\pgfusepath{stroke,fill}%
}%
\begin{pgfscope}%
\pgfsys@transformshift{0.500000in}{1.565911in}%
\pgfsys@useobject{currentmarker}{}%
\end{pgfscope}%
\end{pgfscope}%
\begin{pgfscope}%
\definecolor{textcolor}{rgb}{0.000000,0.000000,0.000000}%
\pgfsetstrokecolor{textcolor}%
\pgfsetfillcolor{textcolor}%
\pgftext[x=0.225308in, y=1.517686in, left, base]{\color{textcolor}\rmfamily\fontsize{10.000000}{12.000000}\selectfont \(\displaystyle {0.6}\)}%
\end{pgfscope}%
\begin{pgfscope}%
\pgfsetbuttcap%
\pgfsetroundjoin%
\definecolor{currentfill}{rgb}{0.000000,0.000000,0.000000}%
\pgfsetfillcolor{currentfill}%
\pgfsetlinewidth{0.803000pt}%
\definecolor{currentstroke}{rgb}{0.000000,0.000000,0.000000}%
\pgfsetstrokecolor{currentstroke}%
\pgfsetdash{}{0pt}%
\pgfsys@defobject{currentmarker}{\pgfqpoint{-0.048611in}{0.000000in}}{\pgfqpoint{-0.000000in}{0.000000in}}{%
\pgfpathmoveto{\pgfqpoint{-0.000000in}{0.000000in}}%
\pgfpathlineto{\pgfqpoint{-0.048611in}{0.000000in}}%
\pgfusepath{stroke,fill}%
}%
\begin{pgfscope}%
\pgfsys@transformshift{0.500000in}{2.038637in}%
\pgfsys@useobject{currentmarker}{}%
\end{pgfscope}%
\end{pgfscope}%
\begin{pgfscope}%
\definecolor{textcolor}{rgb}{0.000000,0.000000,0.000000}%
\pgfsetstrokecolor{textcolor}%
\pgfsetfillcolor{textcolor}%
\pgftext[x=0.225308in, y=1.990411in, left, base]{\color{textcolor}\rmfamily\fontsize{10.000000}{12.000000}\selectfont \(\displaystyle {0.8}\)}%
\end{pgfscope}%
\begin{pgfscope}%
\pgfsetbuttcap%
\pgfsetroundjoin%
\definecolor{currentfill}{rgb}{0.000000,0.000000,0.000000}%
\pgfsetfillcolor{currentfill}%
\pgfsetlinewidth{0.803000pt}%
\definecolor{currentstroke}{rgb}{0.000000,0.000000,0.000000}%
\pgfsetstrokecolor{currentstroke}%
\pgfsetdash{}{0pt}%
\pgfsys@defobject{currentmarker}{\pgfqpoint{-0.048611in}{0.000000in}}{\pgfqpoint{-0.000000in}{0.000000in}}{%
\pgfpathmoveto{\pgfqpoint{-0.000000in}{0.000000in}}%
\pgfpathlineto{\pgfqpoint{-0.048611in}{0.000000in}}%
\pgfusepath{stroke,fill}%
}%
\begin{pgfscope}%
\pgfsys@transformshift{0.500000in}{2.511362in}%
\pgfsys@useobject{currentmarker}{}%
\end{pgfscope}%
\end{pgfscope}%
\begin{pgfscope}%
\definecolor{textcolor}{rgb}{0.000000,0.000000,0.000000}%
\pgfsetstrokecolor{textcolor}%
\pgfsetfillcolor{textcolor}%
\pgftext[x=0.225308in, y=2.463136in, left, base]{\color{textcolor}\rmfamily\fontsize{10.000000}{12.000000}\selectfont \(\displaystyle {1.0}\)}%
\end{pgfscope}%
\begin{pgfscope}%
\definecolor{textcolor}{rgb}{0.000000,0.000000,0.000000}%
\pgfsetstrokecolor{textcolor}%
\pgfsetfillcolor{textcolor}%
\pgftext[x=0.169753in,y=1.507500in,,bottom,rotate=90.000000]{\color{textcolor}\rmfamily\fontsize{10.000000}{12.000000}\selectfont Accuracy (\%)}%
\end{pgfscope}%
\begin{pgfscope}%
\pgfpathrectangle{\pgfqpoint{0.500000in}{0.375000in}}{\pgfqpoint{3.100000in}{2.265000in}}%
\pgfusepath{clip}%
\pgfsetrectcap%
\pgfsetroundjoin%
\pgfsetlinewidth{1.505625pt}%
\definecolor{currentstroke}{rgb}{0.121569,0.466667,0.705882}%
\pgfsetstrokecolor{currentstroke}%
\pgfsetdash{}{0pt}%
\pgfpathmoveto{\pgfqpoint{0.640909in}{2.135102in}}%
\pgfpathlineto{\pgfqpoint{0.993182in}{1.621350in}}%
\pgfpathlineto{\pgfqpoint{1.345455in}{1.378590in}}%
\pgfpathlineto{\pgfqpoint{1.697727in}{1.250979in}}%
\pgfpathlineto{\pgfqpoint{2.050000in}{1.070528in}}%
\pgfpathlineto{\pgfqpoint{2.402273in}{0.974098in}}%
\pgfpathlineto{\pgfqpoint{2.754545in}{0.906017in}}%
\pgfpathlineto{\pgfqpoint{3.106818in}{0.879982in}}%
\pgfpathlineto{\pgfqpoint{3.459091in}{0.834952in}}%
\pgfusepath{stroke}%
\end{pgfscope}%
\begin{pgfscope}%
\pgfpathrectangle{\pgfqpoint{0.500000in}{0.375000in}}{\pgfqpoint{3.100000in}{2.265000in}}%
\pgfusepath{clip}%
\pgfsetrectcap%
\pgfsetroundjoin%
\pgfsetlinewidth{1.505625pt}%
\definecolor{currentstroke}{rgb}{1.000000,0.498039,0.054902}%
\pgfsetstrokecolor{currentstroke}%
\pgfsetdash{}{0pt}%
\pgfpathmoveto{\pgfqpoint{0.640909in}{0.894804in}}%
\pgfpathlineto{\pgfqpoint{0.993182in}{0.905401in}}%
\pgfpathlineto{\pgfqpoint{1.345455in}{0.898128in}}%
\pgfpathlineto{\pgfqpoint{1.697727in}{0.910547in}}%
\pgfpathlineto{\pgfqpoint{2.050000in}{0.914550in}}%
\pgfpathlineto{\pgfqpoint{2.402273in}{0.980004in}}%
\pgfpathlineto{\pgfqpoint{2.754545in}{0.975304in}}%
\pgfpathlineto{\pgfqpoint{3.106818in}{0.827331in}}%
\pgfpathlineto{\pgfqpoint{3.459091in}{0.819540in}}%
\pgfusepath{stroke}%
\end{pgfscope}%
\begin{pgfscope}%
\pgfsetrectcap%
\pgfsetmiterjoin%
\pgfsetlinewidth{0.803000pt}%
\definecolor{currentstroke}{rgb}{0.000000,0.000000,0.000000}%
\pgfsetstrokecolor{currentstroke}%
\pgfsetdash{}{0pt}%
\pgfpathmoveto{\pgfqpoint{0.500000in}{0.375000in}}%
\pgfpathlineto{\pgfqpoint{0.500000in}{2.640000in}}%
\pgfusepath{stroke}%
\end{pgfscope}%
\begin{pgfscope}%
\pgfsetrectcap%
\pgfsetmiterjoin%
\pgfsetlinewidth{0.803000pt}%
\definecolor{currentstroke}{rgb}{0.000000,0.000000,0.000000}%
\pgfsetstrokecolor{currentstroke}%
\pgfsetdash{}{0pt}%
\pgfpathmoveto{\pgfqpoint{3.600000in}{0.375000in}}%
\pgfpathlineto{\pgfqpoint{3.600000in}{2.640000in}}%
\pgfusepath{stroke}%
\end{pgfscope}%
\begin{pgfscope}%
\pgfsetrectcap%
\pgfsetmiterjoin%
\pgfsetlinewidth{0.803000pt}%
\definecolor{currentstroke}{rgb}{0.000000,0.000000,0.000000}%
\pgfsetstrokecolor{currentstroke}%
\pgfsetdash{}{0pt}%
\pgfpathmoveto{\pgfqpoint{0.500000in}{0.375000in}}%
\pgfpathlineto{\pgfqpoint{3.600000in}{0.375000in}}%
\pgfusepath{stroke}%
\end{pgfscope}%
\begin{pgfscope}%
\pgfsetrectcap%
\pgfsetmiterjoin%
\pgfsetlinewidth{0.803000pt}%
\definecolor{currentstroke}{rgb}{0.000000,0.000000,0.000000}%
\pgfsetstrokecolor{currentstroke}%
\pgfsetdash{}{0pt}%
\pgfpathmoveto{\pgfqpoint{0.500000in}{2.640000in}}%
\pgfpathlineto{\pgfqpoint{3.600000in}{2.640000in}}%
\pgfusepath{stroke}%
\end{pgfscope}%
\begin{pgfscope}%
\definecolor{textcolor}{rgb}{0.000000,0.000000,0.000000}%
\pgfsetstrokecolor{textcolor}%
\pgfsetfillcolor{textcolor}%
\pgftext[x=2.050000in,y=2.723333in,,base]{\color{textcolor}\rmfamily\fontsize{12.000000}{14.400000}\selectfont Accuracy vs. Iterations of proposed algorithm}%
\end{pgfscope}%
\begin{pgfscope}%
\pgfsetbuttcap%
\pgfsetmiterjoin%
\definecolor{currentfill}{rgb}{1.000000,1.000000,1.000000}%
\pgfsetfillcolor{currentfill}%
\pgfsetfillopacity{0.800000}%
\pgfsetlinewidth{1.003750pt}%
\definecolor{currentstroke}{rgb}{0.800000,0.800000,0.800000}%
\pgfsetstrokecolor{currentstroke}%
\pgfsetstrokeopacity{0.800000}%
\pgfsetdash{}{0pt}%
\pgfpathmoveto{\pgfqpoint{2.822993in}{2.141543in}}%
\pgfpathlineto{\pgfqpoint{3.502778in}{2.141543in}}%
\pgfpathquadraticcurveto{\pgfqpoint{3.530556in}{2.141543in}}{\pgfqpoint{3.530556in}{2.169321in}}%
\pgfpathlineto{\pgfqpoint{3.530556in}{2.542778in}}%
\pgfpathquadraticcurveto{\pgfqpoint{3.530556in}{2.570556in}}{\pgfqpoint{3.502778in}{2.570556in}}%
\pgfpathlineto{\pgfqpoint{2.822993in}{2.570556in}}%
\pgfpathquadraticcurveto{\pgfqpoint{2.795216in}{2.570556in}}{\pgfqpoint{2.795216in}{2.542778in}}%
\pgfpathlineto{\pgfqpoint{2.795216in}{2.169321in}}%
\pgfpathquadraticcurveto{\pgfqpoint{2.795216in}{2.141543in}}{\pgfqpoint{2.822993in}{2.141543in}}%
\pgfpathclose%
\pgfusepath{stroke,fill}%
\end{pgfscope}%
\begin{pgfscope}%
\pgfsetrectcap%
\pgfsetroundjoin%
\pgfsetlinewidth{1.505625pt}%
\definecolor{currentstroke}{rgb}{0.121569,0.466667,0.705882}%
\pgfsetstrokecolor{currentstroke}%
\pgfsetdash{}{0pt}%
\pgfpathmoveto{\pgfqpoint{2.850771in}{2.466389in}}%
\pgfpathlineto{\pgfqpoint{3.128549in}{2.466389in}}%
\pgfusepath{stroke}%
\end{pgfscope}%
\begin{pgfscope}%
\definecolor{textcolor}{rgb}{0.000000,0.000000,0.000000}%
\pgfsetstrokecolor{textcolor}%
\pgfsetfillcolor{textcolor}%
\pgftext[x=3.239660in,y=2.417778in,left,base]{\color{textcolor}\rmfamily\fontsize{10.000000}{12.000000}\selectfont wav}%
\end{pgfscope}%
\begin{pgfscope}%
\pgfsetrectcap%
\pgfsetroundjoin%
\pgfsetlinewidth{1.505625pt}%
\definecolor{currentstroke}{rgb}{1.000000,0.498039,0.054902}%
\pgfsetstrokecolor{currentstroke}%
\pgfsetdash{}{0pt}%
\pgfpathmoveto{\pgfqpoint{2.850771in}{2.272716in}}%
\pgfpathlineto{\pgfqpoint{3.128549in}{2.272716in}}%
\pgfusepath{stroke}%
\end{pgfscope}%
\begin{pgfscope}%
\definecolor{textcolor}{rgb}{0.000000,0.000000,0.000000}%
\pgfsetstrokecolor{textcolor}%
\pgfsetfillcolor{textcolor}%
\pgftext[x=3.239660in,y=2.224105in,left,base]{\color{textcolor}\rmfamily\fontsize{10.000000}{12.000000}\selectfont freq}%
\end{pgfscope}%
\end{pgfpicture}%
\makeatother%
\endgroup%